Knowledge Management in Medium-Sized Software Consulting Companies

An investigation of Intranet-based Knowledge Management Tools for Knowledge Cartography and Knowledge Repositories for Learning Software Organisations

Torgeir Dingsøyr

# Knowledge Management in Medium-Sized Software Consulting Companies

An Investigation of Intranet-based Knowledge Management Tools for Knowledge Cartography and Knowledge Repositories for Learning Software Organisations

# Abstract

Companies that develop software have a pressure from customers to deliver better solutions, and to deliver solutions faster and cheaper. Many researchers have worked with suggestions on how to improve the development process; software process improvement. As software development is a very knowledge intensive task, both researchers and industry have recently turned their attention to knowledge management as a means to improve software development. This often involves developing technical tools, which many companies have spent resources on. But the tools are often not used in practise by developers and managers in the companies, and it is often unknown if the tools improve how knowledge is managed.

In order to build efficient knowledge management tools, we need a better understanding of how the tools that exist are applied and used in software development.

We present and analyse eight case studies of knowledge management initiatives from the literature. We found evidence of improved software quality, reduced development costs and evidence of a better working environment for developers as a result of these initiatives.

Further, we examine success criteria in knowledge management codification initiatives, based on Intranet tools in medium-sized software companies. We found four factors that we consider important: Having a culture for sharing knowledge, having a stable focus on knowledge management, developing knowledge management tools incrementally, and coupling knowledge management initiatives well to business goals. This research was based on participation with software companies in improvement projects.

In addition, we investigate how knowledge management tools are used for different purposes by different groups of users in two software consulting companies. They use tools both as support for personalization and codification strategies. The consulting companies are two medium-sized Norwegian companies with 40 and 150 employees, which work in development projects that lasts from a few weeks to several years. We used semi-structured interviews with developers, project managers and managers, examined logs of tool usage, and company-internal minutes from development meetings, as well as handbooks, project plans and annual reports.

The frequency of usage varied between the two companies: in one, most employees used tools on a daily basis, whilst in the other, employees used tools weekly. We find that tools for codification are in use for transferring knowledge from projects in order to solve technical problems, get an overview of technical problem areas, avoiding rework in having to explain many people about the same technical solution, improving the employees' work situation by tips on better configuration of technical tools, and also for finding who knows what in the organisation. The tools for personalization are in use for searching for competence to solve technical problems, resource allocation, finding projects and external marketing, and for competence development. In all, we found a variety of uses of a variety of tools by several groups of employees in a company.

*«The maturation of the information technology revolution in the 1990s has transformed the work process, introducing new forms of social and technical division of labor.»*

Manuel Castells in The Rise of the Network Society

# Contents

# Acknowledgements

During the work on this thesis, I have benefited from a lot of communication with many people, who have provided inspiration and motivation. First of all, I would like to thank my supervisor, Reidar Conradi, who has commented on an enormous number of drafts through the last four years. Also, thanks to present and former members of the software engineering group at the Department of Computer and Information Science, NTNU, for providing a good work environment: Elisabeth Bayegan, Roxana Diaconescu, Monica Divitini, Ekaterina Prasolova-Førland, Letizia Jaccheri, Jens-Otto Larsen, Øystein Nytrø, Tor Stålhane, Sivert Sørumgård, Carl-Fredrik Sørensen, Marco Torchiano and Alf Inge Wang.

Another source of inspiration has been the possibility to participate in two research projects during my PhD work: The Software Process Improvement for Better Quality (SPIQ) and Process Improvement for IT-industry (PROFIT) projects. I would like to thank participants from Sintef Telecom and Informatics: Tore Dybå, Geir Kjetil Hanssen, Nils Brede Moe and Kari Juul Wedde (now Clustra) as well as from the University of Oslo: Erik Arisholm, Dag Sjøberg, Magne Jørgensen and project manager Tor Ulsund from Bravida Geomatikk. Thanks are also due to the contact persons from companies that participated in these projects, which provided a pragmatic research environment that greatly influenced the focus of this PhD. I am also grateful to the Norwegian Research Council who financed the PhD work.

A part of the work for the thesis was done as fieldwork in two companies, and I am deeply grateful to these companies and the contact persons and the people I interviewed for their willingness to share experience about their knowledge management efforts.

During the last phase of the work, I was able to stay at the Fraunhofer Institute for Experimental Software Engineering in Kaiserslautern, Germany. Many people deserve thanks for both social and professional inclusion. It was especially nice to be able to work on the COIN Experience Factory project with Björn Decker and Markus Nick, which made me see knowledge management from another perspective. And I am very grateful to Klaus-Dieter Althoff for organising the stay, and also to the rest of the Systematic Learning and Improvement group at the institute for numerous discussions.

Through the years that I have been working as a doctoral candidate, I have benefited greatly from discussions with students that I have supervised in project and diploma work: Arne Bakkebø, Terje Nygaard, Bjørgulv O. Sandanger, Thies Schrader, Torkel Westgaard, Helge Jenssen and Bjørn-Ovin Wivestad. In the early phase of thesis work, I also had many discussions with Bent Ingebretsen, who had written his masters thesis in this field.

There are also other people that I would like to thank for collaboration during the thesis work: Emil Røyrvik on the Kunne project at Sintef Technology Management for discussions and writings on skills management. Trond Knudsen at Norsk Regnesentral for letting me present an early version of research approach which provoked many new thoughts. Petter Gottschalk at the Norwegian School of Management BI, for giving me an early version of his book on knowledge management which directed me to new references. Also, I would like to thank Frank Maurer at the University of Calgary and Mirjam Minor at Humboldt Universität Berlin for organising guest lectures where some of the material in the thesis was presented and discussed.

I am further grateful to the people who have commented on parts of the thesis: Of course, Reidar Conradi as my supervisor, but also Stefan Biffl at the Technical University of Vienna, Magne Jørgensen at the University of Oslo, and Monica Divitini, Letizia Jaccheri, Roxana Diaconescu and Alf Inge Wang at the Department of Computer and Information Science at NTNU. I am also grateful to Terje Brasethvik at the Information Systems group at NTNU for a number of discussions on research topics and comments on drafts. Also thanks to Gavin Gaudet for tips on how to improve the oral English in the Empirical Investigations chapter. And thanks to Preben Randhol for help with scripts to handle usage logs.

Finally, I would like to thank family and friends for inspiraton and encouragement, and especially Sissel for her patience.

Trondheim, January 21st
Torgeir Dingsøyr

# List of Figures

# List of Tables

# 1 Introduction

This thesis is about how Intranet-based Knowledge Management Tools can be used to support what has been called a «Learning Software Organisation». An Intranet-based tool is a software program that provides help for software developers. We will define what we mean by a tool more precisely later.

Software development usually takes place in team-based projects where the participants work towards a shared goal. Many companies have problems with transferring what people learn in one project to other projects in the same company. Knowledge Management is a set of strategies and techniques to increase the transfer and use of different types of knowledge in a company or organisation.

We find many knowledge management tools and methods in companies and in the research literature, but most of the scientific work on tools is concentrating on technology to build such tools; on the structure of knowledge and technical work on retrieval mechanisms. Also, work on knowledge management methods is usually describing an ideal way of collecting and sharing knowledge, which is often difficult to reproduce in practise. There is little work on how tools and methods for knowledge management are actually applied in the software engineering domain. Also, many tools that are introduced in companies are abandoned later. This is often because they turned out not to be so useful as people thought before they were introduced.

We think that we would be able to design better tools and methods, if we knew more about how the existing tools are used - or why they are not used.

In this thesis we discuss how companies can improve their knowledge management by adjusting Intranet-based knowledge management tools, and thus become more of a learning organisation. We will base this discussion on an examination of tools and initiatives that are used in medium-sized companies that develop software. These medium-sized companies are four case companies in a prestudy, and a main and a contrast case in a main study - as well as reports of knowledge management tools from the literature.

Now, we go on to define a problem outline for this thesis that will be further narrowed later and state the main contributions of this thesis. Then, we briefly state what main choices we have made for carrying out research. Further, we narrow the scope of this work, and finally give an overview of the structure of the thesis.

## 1.1   Problem Outline

In this thesis, we are interested in studying how tools for knowledge management are used in medium-sized companies that develop software. The specific tools are Intranet-based tools that companies have produced themselves. There are, however, many such tools, and we will only be concerned with Knowledge Repository and Library, and Knowledge Cartography Tools. We will introduce these types later. and argue why these are particularly interesting to examine.

The type of companies where we have studied this phenomenon is in medium-sized companies in Norway that develop software. By medium-sized we will mean companies with from 50 to 500 employees.

Many knowledge management tools are in use in the software industry. But there has been done little work on how these tools actually work in practise. Also, many research prototypes for knowledge management tools exist in the research literature. But not many of them has made it into industrial practise.

We are then asking the following research question:

- How can Intranet-based knowledge management tools be used in medium-sized software consulting companies to facilitate a «learning software organisation»?

This research question will be further discussed and elaborated after we have introduced more theory. We will also elaborate what we mean by a «Learning Software Organisation».

The critical reader might already now ask: But do these knowledge management tools help solve the problems that the software industry has (which will be described in the next chapter)? The answer is: we are not sure. But we think we need to know more about the tools in use, and about how they are used before we can begin to answer the question of whether they are solving problems or not, and of how cost-effective they are.

But is it really any use in studying such tools? The technology is changing so fast. When we have completed this study, the tools will be completely different! Although we think that developing tools for knowledge management is a long process, and the ones we will study are by no means «completed» - we still think it is important to study how they work, before moving on to something else. It has been claimed that it is a general problem in software engineering, that we do not systematically study the effect of technology and methods, before we jump on to newer technologies. We think it is a sound scientific task to analyse the impact of «new» tools. Yet, we acknowledge that the results might be a bit «old» when we finish.

Then, when we examine such tools, what is the relation between their usage and the potential improvement of the productivity or quality of the software that is developed? It is a long chain of events from the effects of a knowledge management tool, to this knowledge being learned and used by employees, which should then finally affect the quality of the developed software or the productivity of the software development team. We do not intend to show a causal relationship between these factors, but we think a knowledge management tool is one of many factors in a good work environment that can stimulate learning, creativity, and employee motivation, which will affect the quality of the output. But we limit ourselves here to study how knowledge management tools can be used, leaving more «hard measurements» for further work.

## 1.2   Claimed Contributions

The work in this thesis can be divided into four major phases, where we claim to have some contributions in each, related to the field of studying Learning Software Organisations by empirical methods. Some work in the thesis has been published before, and we give references to these papers for each phase:

- Literature study: We present literature on knowledge management in software engineering, and have made a taxonomy of knowledge management tools based on findings from the literature. We have also surveyed existing case studies of how knowledge management tools are applied in companies that develop software, and present, and discuss these approaches. This work can be found in chapter 3 in the thesis, and in papers 1 and 8.

- Method for experience capture: We have contributed in developing a method to capture experience from completed software projects though a group process: lightweight postmortem reviews. This method is given as an example of experience capture methods in chapter 3, and is described in further detail in papers 2, 5 and 6.

- Four cases studies on knowledge management in software engineering companies: Here, we studied four companies that have applied different knowledge management initiatives, and discuss success factors. The cases are presented in chapter 5.1, and discussed in chapter 6. This analysis has also been published as paper 4.

- Deep case study and a contrast case: We examine further what kind of knowledge management tools that exist in two companies, and describe how different groups of users apply them. The cases are presented in chapter 5.2 and 5.3, and are discussed in chapter 6. Some of the work here on Skills Management has been published in papers 3 and 10.

We have further published paper 7 as a first discussion on the selected research topic and research questions in this thesis, that can be found in chapter 4. Finally, paper 9 gives a further description of knowledge management tools than the ones that can be found in chapter 3.

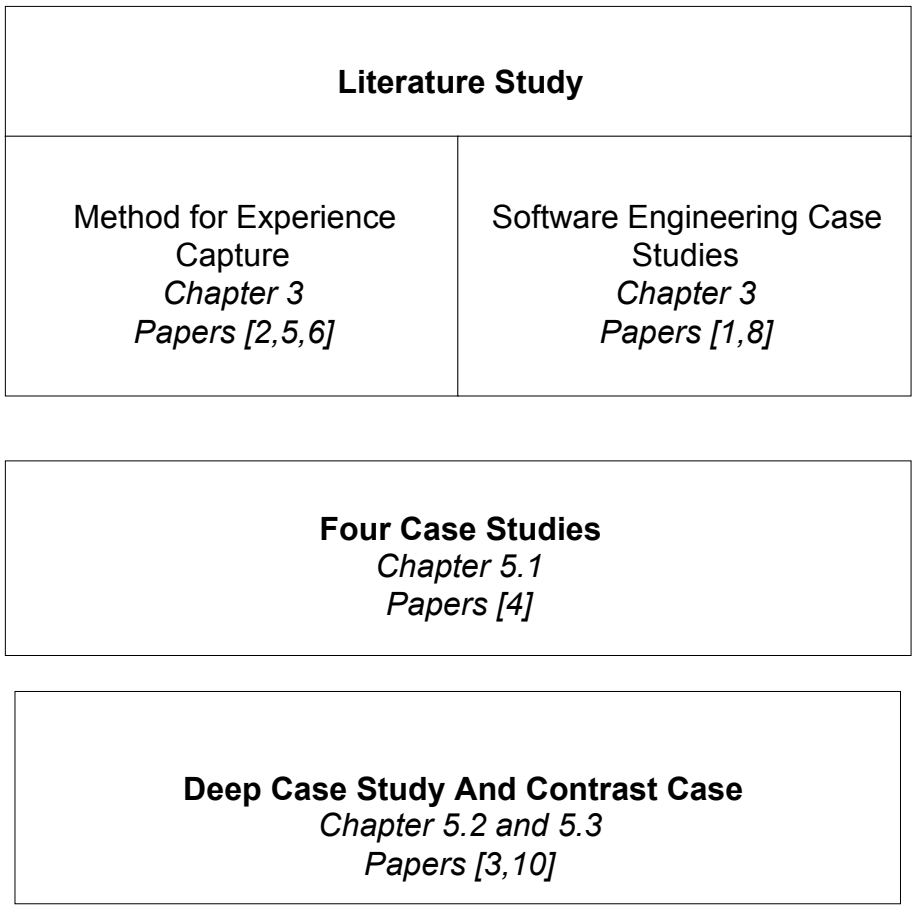

*Figure 1.1: The main contributions in this thesis, with references to thesis chapters and published papers.*

The following are the papers that has been published or are undergoing a publication process:

### Journal articles

1.　　Dingsøyr, Torgeir, Conradi, Reidar: A Survey of Case Studies of Knowledge Management in Software Engineering, submitted to International Journal of Software Engineering and Knowledge Engineering. A previous version of this paper was published as paper 9.

### Book chapter

### Conference papers

### Workshop papers

Workshop on Programming Environment Research, 28-30 May, Lille-hammer, Norway.

8.      Dingsøyr, Torgeir (2000) An evaluation of Research on Experience Factory, Workshop on Learning Software Organisations at the international conference on Product-Focused Software Process Improvement, Oulo, Finland, University of Oulu, VTT Electronics, Fraunhofer IESE, pp. 55 - 66.

9.      Dingsøyr, Torgeir (2000) An Analysis of Process Support in Knowledge Management Tools for Software Engineering, Workshop on Flexible Strategies for Maintaining Knowledge Containers,14th European Conference on Artificial Intelligence, 20-25. August, Berlin, Germany, Humboldt-Universität zu Berlin, ECAI Workshop Notes, pp. 6 - 13.

10.      Dingsøyr, Torgeir and Røyrvik, Emil (2001) Skills Management as Knowledge Technology in a Software Consultancy Company, Learning Software Organizations Workshop, 12 - 13 September, Kaiserslautern, Germany, Springer Verlag, Lecture Notes in Computer Science, vol. 2176, pp. 96-107.

## 1.3   Chosen Research Strategy

In researching the question outlined in section 1.1, we have chosen to investigate it in a real environment. That is, to go into a real organisation, and study tools in «vivo». We will discuss this further in the Research Methods and Design chapter. The main reasons for choosing to study real organisations, and doing case and field studies, are that:

- Many prototype knowledge management tools are already developed in research institutions, so the need for making more prototypes is small.
- Few studies exist on how knowledge management tools are used in software companies.

In software engineering, several environments have expressed the need for a more empirical basis of software engineering, promoting what has been called *empirical software engineering*.

In empirical software engineering, it is necessary to use different research methods than normally applied in software engineering. This is because we have no strict control of the environment. Also, in our case, there is relatively little information to find about the usage of knowledge management systems in the research literature.

In studying organisations, we have used research methods that are common in social science, but not in technology-oriented disciplines such as software engineering. A common problem when using such methods is that: «technologists regard sociologists as, apparently, merely wishing to observe and give an account of what they observe, with no interest necessarily in this leading to social action». While on the other hand, «sociologists regard technologists as simply wanting plans of action to make their technology more 'effective'» (Low et al., 1996). Here, we hope that our proposed theory will be seen as a contribution to better understand the tools, and then be useful for anyone wanting to improve the design or usage of such tools later.

Much of the work in software engineering has been done in the spirit of modernity; with a rational view that the problems at hand can be solved if we just establish good enough work methods and tools. The search for a silver bullet (which will be discussed further in the next chapter) is evidence of such a view.

However, many people now have a more post-modern view of software development. That is, it is futile to «solve problems» related to organizational, human and technological factors by say technology alone. Instead of looking for a silver bullet, we can only hope to find a set of «weapons» - that will help us to reduce the impact of some problems as they appear to some people.

In fact, the whole idea of software «engineering» is questioned by some environments (see an interesting discussion on the engineering metaphor in (Bryant, 2000)). Engineering is often associated with words like «science», «mathematics», as well as «practical methods». But we could also see software development as a creative task (Glass, 1995), where for example improvisation (Dybå, 2000) is more important than rigour.

In this thesis we will adopt a subjectivist, or postmodern view, that different people might have different goals, and they do not necessarily always act in a pre-planned or even rational manner. In studying how people use knowledge management tools, we consider the software practitioners (or «community of practise») to be the true, skilled, experts to judge what kind of tools they find useful or not. Therefore, we have opted for a research strategy with a close interaction with developers, project managers and management in the field. We will discuss this further in our chapter about research goals, method and design.

## 1.4   Research Context

The work which was performed in this thesis was a part of two larger research projects on software process improvement (SPI) which involved many Norwegian companies that develop software.

The Software Process Improvement for Better Quality (SPIQ) project aimed to increase the competitiveness of 12 participating Norwegian software companies, by creating an improvement environment in the companies, and introducing ideas from an American context, like the Experience Factory, and adopting it for small and medium-sized enterprises in Norway (Conradi, 1996). It also included pilot projects for improvement in companies, as well as discussion forums for issues related to process improvement. Further, it contained dissemination activities like conferences and the writing of a method handbook for process improvement in Norwegian (Dybå et al., 2000). This project lasted from 1997 to 1999.

This project was followed by the Process Improvement for IT industry (PROFIT) project, which focused more on software process improvement in companies with frequent changes in technology and market. Can such companies benefit from the same improvement initiatives as more stable organisations? This was one of the major questions in this project, which is still ongoing, and involved eight companies from the start. This project lasts from 2000 to 2002.

## 1.5   Scope

In this thesis, we are concerned with how Intranet-based tools are used for knowledge management in medium-sized organisations that develop software. We have thus limited the field of knowledge management to those processes that can be supported by computer tools, and specifically tools with a web-interface on a company-internal Intranet. We also concentrate on a specific set of tools that will be discussed later. Further, we have limited the usage of these tools to the domain of software development and maintenance, and specifically in medium-sized companies, where most of the development is done «in-house», and where most of the staff spends much of their working day in front of a computer.

When we examined the knowledge management tools, we have only looked at how they are used. We have not looked at issued in developing such tools, and not on economical issues - whether they are cost-effective or not.

We have neither looked at specific tools for reusing code or other software artifacts, but at tools that operate on a higher level of abstraction. But these tools may be linked to code, like a system that help you solve problems by showing example code. In the companies where we have been working, reuse of code is usually organised through development of software libraries of «baseline products» that get input from all people in the organisation.

To introduce knowledge management as an «improvement» in an organisation is of course not without problems. What some people in the organisation see as «improvements» might be seen as «deteriorating» efforts by other people. For example, some employees might think that their knowledge is ignored by a company, because it is not included in a computer tool. This, and other political issues in deploying knowledge management tools are not issues that we will discuss here.

## 1.6   Structure of the Thesis

The structure of the rest of this thesis is as follows:

*Chapter 2: Software Development; Problems and Remedies.* In this chapter, we discuss what software development is about, and some of the challenges the field is concerned with. We also discuss some of the main improvement initiatives that have been in the field, and discuss one of them, namely knowledge management and learning organisations in more detail. Finally, we give an overview of research methods in software engineering.

*Chapter 3: Knowledge Management:* In General and in Software Engineering. Here we first discuss knowledge management in general, and then specifically its application in software engineering. We discuss terms like experience, information and knowledge, and other common terms in the knowledge management field, like organisational memory, corporate memory, and experience factory. We also examine how knowledge is transferred in an organisation, and introduce a knowledge management program as a strategy, a set of processes and a set of tools. We present case studies on knowledge management tools in companies that develop software, found in the literature.

*Chapter 4: Research Goals, Method and Design.* Here we further specify our research goals, using concepts from chapter 3. We list the topics of interest in the form of research questions. We present the research method that we selected, with arguments for why this approach is suiting the topics under study.

*Chapter 5: Empirical Investigation.* First, we present a prestudy of four case studies of knowledge management programs in Norwegian companies. Then, we present two companies where we did case studies, together with projects that we followed in each of them. We present the infrastructure for knowledge management that exist in the companies, and our findings on the usage of them.

*Chapter 6: Discussion and Analysis.* We discuss the findings from the literature, our prestudy and main study cases in light of the theory which is given in chapter 3.

*Chapter 7: Conclusion and Further Work.* We sum up the main findings from the discussion, and outline possible further work in the field of learning software organisations.

*Appendix A: Interview guides* - here we present the interview guides that was used in semi-structured interviews in the two companies in the main study.

*Appendix B: Processed Usage Logs* - Here, we list processed usage data from Knowledge Management Tools in Alpha, one of the main study companies.

This is a doctoral thesis, written for the research community. It is not the intention to come up with direct, practical aid for companies on how to improve their knowledge management, but more to bring forward theory about how knowledge management is used. This will hopefully make it into practise, but it is out of the scope to concentrate on that issue. That is the responsibility of the research field as a whole.

Reading this thesis requires knowledge of software engineering and specifically software process improvement, and what has been called learning software organisations (knowledge management in software engineering). It also requires knowledge of research methods in general.

# 2 Software Development; Problems and Remedies

Now, we first discuss what software development is about, and future trends. Then, we describe some of the main challenges for the software engineering field, and some solutions that have been suggested in the literature. Further, we present knowledge management and learning organisations as an interesting new field in improving software development practise, before we briefly discuss research methods in the software engineering domain: What methods have we got to scientifically examine the problems and remedies at hand?

## 2.1  Software development

To develop and maintain software is often referred to as «software engineering». One definition is that software engineering «is concerned with theories, methods and tools which are needed to develop software... for computers». It differs from other types of engineering because it is «not constrained by materials governed by physical laws or by manufacturing processes» (Sommerville, 1996). Of course, we also have constraints in software, for example due to manpower, organisation or skills.

With this definition of «software engineering» (see (Bryant, 2000) for an interesting discussion on the use of this word), we include everything from eliciting requirements for a software system from a customer, via specifying architectural details of the software system, to actual implementation in one or more programming languages. We also include activities to check or improve the quality of the software, like testing and inspection. Usually, the developers use a wide range of tools, from tools for handling documents to design tools and editors, compilers, debuggers and tools for version control and project management.

A common way to develop software, is to divide the work into several phases. A much referenced model of such phases is the waterfall model, which comes in many variants, but most include (Sommerville 1996):

- Requirements analysis and definition - to find what the software system will be used for, that is, find the requirements.
- System and software design - make decisions on technical issues, like software architecture, database design and user interface design.
- Implementation and unit testing - write, adapt or generate the actual code in a programming language, and test each program unit.
- Integration and system testing - check that the implementation fulfils the requirements.
- Operation and maintenance - enhance the software, or correct errors that are found during usage.

The sequential waterfall model works best when the requirements for the system are stable and easy to establish. For software where the requirements are largely unknown, or frequently changing, more incremental models for software development should be used.

To develop software is a typical example of what Peter Drucker has called «knowledge work»; where «value is (...) created by 'productivity' and 'innovation'» (Drucker, 1993). Knowledge is the only scarce resource in software development - not other «means of production» like computer hardware and software, office buildings or capital.

What is software likely to be in the future? A panel that constructed scenarios for software in the future saw some major trends in development and usage (Tellioglu and Wagner, 2000):

- Because you can charge more for a service than a product, software will more and more be seen as a service. This «implies new responsibilities for developers, particularly in light of the risk of software failure». For example, we might expect software companies to pay higher penalties if their software does not work.
- Developers often take many choices that the users have no influence on, but that will have an impact on how the software system can be used later. This will probably require developers to involve the users more in the technical development to a higher degree in the future.

- Software will interact more with users through natural forms, using speech and multimedia technology.
- Software will adapt to the users and their working styles.

In all, we can expect software to be more complex than today, and used in even more situations than today.

Other trends in software development that we have seen is that storage and memory has become practically free. Now the limitations are bandwidth. Some think that technical limitations will decrease in the future, and the limitations we will meet is in our own creativity and ability to design software solutions.

We have now given a broad overview of software development. Let us then go on to discuss some problems, which particularly are related to the quality of software, and the quality of the software development process.

## 2.2 Problems in Software Engineering: Overruns and Unfulfilled Requirements

To develop software is challenging. There are many examples of software projects that have failed. The much cited Standish report on software projects (1995) «shows a staggering 31.1% of projects will be cancelled before they ever get completed. Further results indicate 52.7% of projects will cost 189% of their original estimates. The cost of these failures and overruns are just the tip of the proverbial iceberg. The lost opportunity costs are not measurable, but could easily be in the trillions of dollars...». Now, we should be a bit careful in thinking that the percentages given in the report will be true for all projects, but they may give a good indication. The view that the software systems we use today are not very mature is also supported by the American «President's Information Technology Advisory Committee», that writes: «The Nation needs robust systems, but the software our systems depend on is often fragile. Software fragility is its tendency not to work properly - or at all. Fragility is manifested as unreliability, lack of security, performance lapses, errors and difficulty in upgrading» (Joy and Kennedy, 1999).

So what are the consequences of this kind of problems? In september 2001, the US bank Citigroup had severe problems as their «systems began crashing on Tuesday afternoon and nearly 24 hours later, officials of the largest US banking company were still scrambling to provide a reason»[1]. This affected 2000 ATMs and 750.000 online banking customers.

We can say that we have problems in software engineering that is related to low quality of software, and a long time to market. That is, developing a product can often take much more time than predicted, and this can be very critical in emerging markets like the Internet industry, where we saw companies offering a service first got all the attention of media and customers.

But before continuing - what do we mean by quality when we speak of software? Many define quality simply as «satisfied customers», but we have several other types of «quality», for example in requirement specifications, we can talk of syntactic quality (that the requirements are stated syntactically correct) and semantic quality (that the meaning of the requirement is correct) (Krogstie, 2001). Also, different user groups of software systems can have different perceptions of quality (See (Wong and Jeffery, 2001)).

So why are there so many problems related to software development projects? Software is an immaterial product, and it can be difficult to get an overview of a total program system, which can be millions of lines of code, to identify all possible error sources. Also, a very small defect might have a lot of influence in critical systems, like the European Space Agency's Ariane 5 satellite launcher, that ended in a failure in 1996. About 40 seconds after initiation, the launcher «veered off its flight path, broke up and exploded» according to the report by the inquiry board (Lions, 1996). The error was «caused by an internal variable related to the horizontal velocity of the launcher exceeding a limit which existed in the software». Thus, just a few lines of code that was lacking, had severe consequences - a loss of around 312 million Euro. We could say that this was a process error, in that the module that failed was not tested under the right conditions.

---

1 Article «Citigroup struggles with growing pains», Financial Times, Friday 7th of September, 2001.

Other problems can be that the communication between the end-users and the software developers is lacking, or that project management is difficult in an environment where a small bug can take a very long time to correct, and where it is often difficult to estimate the schedule and amount of remaining work.

Numerous examples of problems in software development projects can be found in popular books like Crash - Learning from the World's worst Computer Disasters (Collins and Bicknell, 1997) and Software Runaways (Glass, 1998).

After listing all these problems that exist in software and its development, you may ask: are all software systems that bad? Of course, it is not so, there are a lot of software projects that deliver software that is highly usable and working. Robert Glass has argued that the software failures are the exception rather than the trend (Glass, 2000) - «we tend to focus on the unusual things that go wrong because they're more interesting or important than the run-of-the-mill things that go right». Glass argues that we should not use a word like «crisis» to describe the software development field when we know of so many well-working systems. The main reason for this argument is that problems in software are used to motivate a lot of research - which should be able to stand on it's own feet.

We acknowledge that there have been more writings about the failures than the successes in software engineering projects, and that the situation might not be as bad as it looks. But as the reports we have cited earlier show, there is at least quite a lot of processes and projects that could improve, although it is not right to use a word like «crisis».

## 2.3    Suggested Solutions: Is There A Silver Bullet?

There has been much discussion in the software engineering community about finding a «silver bullet» to end the problems, or at least reduce the impact of them. Several solutions have been suggested to improve the way software is developed. Some have tried to change the way software is produced; the «process», some by introducing new programming languages, and other have worked with supporting tools to assist in devel-opment. The goal is usually to increase productivity and/or the quality of the software that is made.

There are a lot of factors involved in developing software; we have outlined a set of major factors in Figure 2.1.

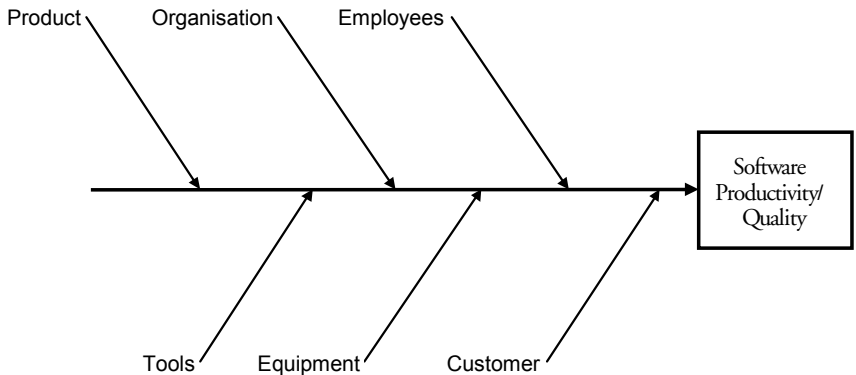

*Figure 2.1: Some factors that influence productivity and quality in software development.*

Clearly, software development depends on the software product to be developed, but also on the way software development is organised. (Organisation includes what kind of work processes that are used, and how communication in projects is organised). Other factors are: the skills and attitudes of the individual developers, what computer tools that are used, as well as what physical equipment (such as computers). And it depends on the need of, and the relations to the customer.

But let us go back to the suggested remedies for improving the productivity and quality of software. The outcome of several of these initiatives was summed up in an article in Communications of the ACM (Glass, 1999). Claims of «order of magnitude» improvements were evaluated, on different «technologies» like:

- Structured techniques - using structured analysis, design and programming.
- Fourth generation programming languages (application generation).
- Computer Aided Software Engineering - tools to support software engineering, mainly in analysis and design.

- Formal methods - formal specification and verification of software.
- Cleanroom methodologies - methods for removing software defects.
- Process models - descriptions of appropriate processes in software engineering.
- Object-oriented technology - to find «objects» in the problem to be solved, and use those in generating software solutions.

Many of the technologies show promising results, but there are few scientific studies that evaluate how the different technology and methods actually work. Also, in some studies that claim improvement, the improvement technology is confused with other changes, like changes in the programming language. So there is still a need for more systematic studies on how these technologies work in practise.

If we look at the improvement initiatives that include organizational as well as technical aspects, we find a subfield of software engineering named «software process improvement». The idea here is to change work practice to be more effective or predictive, or to develop software with higher quality. The underlying idea is that the way you produce software affects the final product. Within this field, we find what we can describe as two different positions: One that imposes «top-down standardization» to increase the quality of software, and one that imposes more «bottom up» quality improvement initiatives. To give a further overview of this area, we also present «software process modelling», which has provided important contributions to this field, and finally the topic that interests us the most; improving software by stimulating learning: «Knowledge Management and Learning Organisations».

Let us first present the standardization approach, then Total Quality Management and Process Modelling before we briefly discuss these approaches. Note that the two first approaches may require a «champion» in the organisation that is seeking improvement: A person who initiates and follows up improvement efforts - either in the top or in the bottom of the organisation.

## 2.3.1 Improving through Standardization; The Capability Maturity Model

The motivation here is, like standardization in other fields, that if we develop software in a more well-defined and predictable way, the resulting software will also be of higher quality, and it will be easier to reach goals on cost, schedule and quality for a software development project. In a statistical sense, we can say that we aim to reduce the variance in cost and quality between different teams and projects, in other words to reduce the risk for severe overruns. The most known standardization approach in the software engineering field, is the Capability Maturity Model (Humphrey, 1989) developed at the Software Engineering Institute at Carnegie Mellon University. This is a framework to evaluate the «maturity» of software developing companies, where companies that produce software in a very planned and documented way is considered to have a «higher maturity level» than other companies. The framework divides companies into five levels (Paulk et al. 1995, Pfleeger, 1998):

1. Initial - software development is done in an ad-hoc fashion, with little control on effort spent or remaining, and on the quality of the software.
2. Repeatable - inputs and outputs of different parts of the development process are defined, such as budget, schedule, resources that will be used, as well as functionality. Each project has a standard.
3. Defined - activities to produce the software («processes») are documented and standardized. The organisation has a standard.
4. Managed - measures of process and product quality make it possible to find problems and assess effects of possible solutions. The risk of overruns is reduced.
5. Optimizing - new tools and techniques are tested to find how they work before they affect processes and products, and possible faults are discovered before they appear. This means to be more efficient.

Some have criticized this way of improving because the model is not suited to the everyday problems of most software developing companies. It was originally developed for software contraction in the defence industry, but has since been applied in other fields, like in the telecommunication industry. However, CMM is not very tailorable to the situation a company might be in, it only focuses on the more managerial aspects of the development process, and does not consider that most software companies have to sell their product in a market. So one critique of this model, like with most other forms of standardization, is that a company can get a very high score without really doing well in the market. This is similar to being «ISO-certified» to make useless life jackets of concrete - as long as the production process is well documented. Anyway, many of these issues, especially in the lower levels of the CMM are issues that most companies will benefit from, like more systematic version management and project planning.

## 2.3.2 Involving everyone in Improvement; Total Quality Management and the Quality Improvement Paradigm

Another position in the software engineering community is to try to improve, but focus more on the specific needs of a company in its market situation. This is based on thoughts from Total Quality Management, TQM (Deming, 2000, Pascale, 1991), which has been a popular improvement strategy the last twenty years. Some important aspects of this approach has been (Neerland, 2000): organized improvement, involvement of every employee, increased customer support, improved performance and integration of activities. «Quality is everyone's responsibility», and «quality is satisfied customers» are slogans that gives a good description of TQM, and several technologies have been developed to help all employees in a company to focus on quality.

A central idea in TQM is to learn from the activities that you do in a company. For this purpose, and for improving your performance from what you learn, the *plan-do-check-act* cycle is a structured way of working. The idea is to first *plan* an improvement or change activity, then *do* it, then *check* whether you reached the intended goals, and finally *act*; make changes to work processes in order to do better the next time, based on what you have learned.

This kind of feedback-loop is also used in software engineering under the name Quality Improvement Paradigm (Basili, 1985), developed at the NASA Software Engineering Laboratory. Here, we find six steps to apply in improvement work: 1) characterise the environment, 2) set goals, 3) choose process, 4) execute, 5) analyse, and 6) package. This is a further breakdown of the steps from total quality management, but the fo-cus in the quality improvement paradigm has traditionally been on gathering quantitative data. A technique for focusing data collection is called the Goal Question Metric method (van Solingen and Berghout, 1999). Here, you start by defining some goals, like «to improve the quality of the user interface», then go on to find some questions that can give you an answer to whether you have reached you goal or not, like «how often does the user interface crash?» or «how much time does users spend to familiarize with the user interface?», and then you finally decide on some metrics to define the data to collect. In our example, this could be «number of user interface system crashes per 100 hours», or «average time before users claim to master the interface». When a company wants to measure how a new method or tool performs over time, the goal question metric method can be a valuable support as a part of a quality improvement program.

### 2.3.3 Making Work Practice Explicit, and Automate it

Yet other people have been working on process-sensitive tools to define and support the software process that the developer is supposed to follow, for example by assisting in tasks like planning and organisation. A software process can be defined as «the coherent set of policies, organizational structures, technologies, procedures, and artefacts that are needed to conceive, develop, deploy and maintain a software product» (Fuggetta, 2000).

An important step here is to find how software is actually developed in a company (to elicit and define the process model), and then to design tools that support this way of development (enact in the development process). Of course, it is also possible here to change the way processes are carried out in order to make development simpler or focus more on quality aspects in the development.

### 2.3.4 Summary and Discussion

Now, we have seen some different perspectives on what software process improvement can be. All the views we have presented can overlap in normal improvement activities in software companies, and the different fields loan from each other. In the Capability Maturity Model, for example, work processes should be documented when you reach level two.

This documentation is the expertise of the software process modelling field. Such process models can also be a prerequisite in more bottom-up improvement initiatives involving measurement, as in the Quality Improvement Paradigm. And finally, in a computer-supported knowledge management tool which we will discuss in the next section, can be beneficial to «tag» knowledge about issues to existing «processes» in a company.

Note that most of these improvement strategies has their main goal as «optimising» how work is done, in the spirit of scientific management. Another approach to be more productive is not to try to optimise, but to «improvise» (Dybå, 2000). We will come back to this point later.

## 2.4   Knowledge Management and Learning Organisations

Another recent improvement «trend» has been knowledge management, which is also related to creating «learning organisations», in software engineering: «learning software organisations».

Objectives of knowledge management might be «to make the enterprise act as intelligently as possible to secure its viability and overall success» (Wiig, 1997). If we look more into knowledge management, we find that some important aspects are (Wiig, 1995):

- Survey, develop, maintain and secure the intellectual and knowledge resources of the enterprise.
- Determine the knowledge and expertise required to perform work tasks, organize it, make the requisite knowledge available, «package it» and distribute it to the relevant points of action.
- Provide (...) a knowledge architecture so that the enterprise's facilities, procedures, guidelines, standards, examples, and practices facilitate and support active Knowledge management as part of the organization's practices and culture.

An example of a knowledge management tool is the COIN Experience Factory, an Intranet tool developed in the Fraunhofer Institute for Experimental Software Engineering (Tautz, 2000). This tool allows researchers to search in a database of experience gathered from previous projects. This experience has been gathered through in-depth interviews with project participants, and then structured according to topic.

We will discuss the term knowledge management in depth in chapter 3.

Another holistic approach, which includes organisations and technology in improvement, has been to create «learning organisations». A learning organisation is «an organisation skilled at creating, acquiring, and transferring knowledge, and at modifying its behaviour to reflect new knowledge and insight» (Garvin, 1993). George Huber gives some advice on what managers can do to make their organisations more «learning» (Huber, 1996):

- Learn from experience - systematically capture, store, interpret and distribute relevant experience gathered from projects; and also to investigate new ideas by carrying out experiments.
- Learn by watching and listening - make people act as «sensors» to learn on behalf of the organisation by participating in communities and reading relevant information.
- Using a computer-based organisational memory - to capture knowledge obtained from experts to spread it through the organisation.

A research area that is linked to organizational learning is research on «communities of practise» as a basis for learning. Etienne Wenger writes: «learning is an issue of sustaining the interconnected communities of practise through which an organization knows what it knows» (Wenger, 1998).

In the much-cited book on learning organisations, The Fifth Discipline (Senge, 1990), we find further characteristics of learning organisations: the ability of «systems thinking» - to see more than just parts of a system. This often means to involve people in an organisation to develop a «shared vision», some common grounds that make the work meaningful, and also serve to explain aspects that you yourself do not have hands-on experience in. Another way of improving communication in an organisation is to work on «mental models» that support action, «personal mastery»; that people make use of their creativity and abilities. And finally «group learning» - to enhance dialogue and openness in the organisation.

Many researchers use the word «best practise» in relation to knowledge management - implying that the aim is to transfer the «best practise» from some people to the whole organisation. We think that this is both difficult and unwanted in most organisations: To «capture» a «best practise» and make it applicable in different settings, you need to capture the context, and this makes it expensive. We also think that such descriptions of process models can better serve as «good examples» that is not necessarily the «best» way to develop software.

Another misconception about knowledge management is that it is synonymous to «reuse» or «replication». This can be a part of knowledge management, but then more as «getting to know which artefacts that are reusable», and probably modify it to suit a specific need than to re-apply an old piece of work.

## 2.4.1 Why is Knowledge Management a Good Approach?

After having seen some different possible solutions to some of the common problems in software engineering, why would we suggest another one like knowledge management? Let us first discuss why this approach is relevant for software companies, and why it is interesting as a research topic.

Our main argument why knowledge management is a good solution to common problems in software engineering is that software development is knowledge-intensive work, and knowledge-intensive work can be improved by managing knowledge better. We claim that software engineering is knowledge-intensive because:

1. To develop software requires deep technical knowledge in many specific domains.
2. The required knowledge is changing because of technological changes, and because the market wants new solutions.

So, it requires knowledge both to do a good job, and also to cope with rapid changes in both technology and needs in an application domain. Then we reach our second step in the argument: Knowledge intensive-work can be improved by managing knowledge better, because:

1. Work that requires knowledge can be done better if you know that the knowledge is relevant and up to date, which requires learning.

2. To ensure that you learn relevant knowledge, it is best to learn from you own environment, which is the essence of knowledge management. This also means that you «try to make the best out of the resources you have available already».

3. To improve knowledge work, we need a holistic approach with both technical and organisational aspects. People learn better when they are motivated to do so.

4. Focusing on managing knowledge will activate local knowledge that exists in a company.

Some knowledge is easier to transfer to others if it is written down, like in a (possibly) formal document. Frederik Brooks writes about this in his book The Mythical Man-Month about software development, where he recommends that «no matter how small the project, however, the manager is wise to begin immediately to formalize at least mini-documents to serve as his database» (Brooks, 1995).

Of course, many companies are interested in having knowledge from employees written down - to make it easier to replace the employees if they leave for another company, or another position internally. This is an issue that can make normal employees sceptic to knowledge management, as this can reduce their «value» in the company. However, we can also expect the contrary to be the case: that employees that are good at sharing knowledge with others become even more valuable for a company than before.

We think that knowledge management is a promising set of methods and tools, that could help knowledge workers in performing their job better, and that will probably be used in many different occupations in the future. It seems that the last years' focus on knowledge management has made a business climate for learning, and even learning «on the job». The field of knowledge management is also a truly interdisciplinary arena, where many communities including artificial intelligence, organisational development, software engineering, pedagogy and psychology meet.

Many software developers have long workdays and stress because of the complexity of software development and short time limits. We see knowledge management as one way to make the workday better for developers, by giving a better overview of the work situation, as well as helping people to be more effective.

Knowledge management is a field dominated by a lot of hype and a mixture of theory and technology from different research fields. It can be difficult to understand the different knowledge management initiatives. Especially in software engineering, where technology from artificial intelligence and software development meet with theories from consulting companies, the word «knowledge management» can have a lot of meanings. We think there is a great need to clarify which approaches exist in this domain, and relate different theories and technologies to each other.

The field of knowledge management has been criticised as the next «fad to forget people» (Swan et al., 1999). It has been criticised on four points: (we will introduce the words codify and tacit knowledge later) «overstating the codifiability of knowledge», «overemphasise the utility of new ITs for improving organizational performance improvements», making «unjustified assumptions about the willingness of employees to use such IT systems», and finally not seeing that «the codification of tacit knowledge into formal systems may generate its own pathology; the informal and locally situated practises that allow the firm to cope with uncertainty, may become rigidified by the system». We agree with this criticism in that it is vital to keep in mind what knowledge management is used for, and that it is neither wanted or economical to «downskill» all activities in a company by writing down required knowledge. We view knowledge management as a field that is open to combine lessons learned from social science as well as technology fields.

A problem with many previous studies of knowledge management in software engineering is that the work relies on work descriptions and «methods» that you find in manuals in the companies. But as it has been pointed out in ethnographic studies, there are often a lot of differences between how people really work, and how the organisation describes the work that is done (Brown and Duguid, 1991). Therefore, we think it is a great need for more empirical studies in this field.

Why did we choose to study knowledge management in the software engineering domain?

The «science» or «art» of software development is relatively immature, and so there is a huge potential for learning, as we have seen from the introductory sections. Also, software development is a large industry, where much effort is spent on improvement initiatives. Another reason for choosing this domain, is that people who work in software engineering are used to make use of computerized tools - they often spend most of their workday in front of a computer, either in their office, or at a customer. These people will probably be the first to use tools that will reach a wider audience later, both because of the high computer use, but also because this group of people are likely to write new software to help in their own work. Therefore, it is particularly interesting to study the use of knowledge management tools in this domain.

Many of the improvement initiatives in the previous sections have been tried out in real software companies. But, as we will describe later, we are mostly interested in medium-sized companies. So why this fascination with medium-sized companies? First, many companies belong to this category (see Fayad et al., 2000)), and many more such companies appear every year. Second, there have not been many studies of process improvement efforts in this type of companies, and especially efforts related to knowledge management initiatives. This might be due to that very few of these companies have an own research and development department that can cooperate in studies with academia. It is interesting to see what kind of technology and organisation that exists in this type of companies, as they are usually very «lean» and can not afford to develop large internal systems as for larger companies. A third reason is that we have actually participated in research projects with this type of companies, which we will describe later.

Another point is that medium-sized companies quite often use different technical solutions that larger ones. Many large companies use Lous Notes as a tool for knowledge management. The consultancy company Ernst & Young in the UK claims to have more than 1 million documents available on their Intranet in addition to their 5.000 internal Lotus Notes databases (Ezingeard et al., 2000). Medium- sized companies will certainly not have systems of this size, and will probably also use more low-cost solutions like Intranets than larger companies.

But could we not also have looked at small companies? After all, there are more of them than the medium-sized ones. We could have looked at them as well, but think that they do usually not have such a great need for computer systems to share knowledge, because the people working in the company can communicate directly much easier. In general, we think computer-based tools needs an environment of a certain size to be effective.

## 2.5   Research Methods in Software Engineering

Three types of validation methods in software engineering are discussed in an article by (Zelkowitz and Wallace, 1998): observational, historical and controlled. We will refer to these as research methods rather than validation methods, as they can be used also for constructing research results and not just for validating research hypothesis. We now present each of these set of methods from Table 2.1, and also add action research as a fourth group. We place special emphasis on the case study as this is a method that we will be using later, and describe a way of data collection in case studies called ethnography, and a method for data analysis called grounded theory.

*Table 2.1: Validation Methods for Software Engineering (Modified from (Zelkowitz and Wallace, 1998))*

| Category | Validation method | Description |
|---|---|---|
| Observational | Project | Collect developmend data. |
| | Case study | Monitor projects in depth. |
| | Assertion | Use ad hoc validation techniques. |
| | Field study | Monitor multiple projects. |
| Historical | Literature | Examine previously published studies. |
| | Legacy | Examine data from completed projects. |
| | Lessons | Examine qualitative data from completed projects. |
| | Static analysis | Examine structure of the product. |
| Controlled | Replicated | Use different approaches. |
| | Synthetic | Replicate one factor in laborarory setting. |
| | Dynamic | Examine developed product. |
| | Simulation | Use developed product in a simulation. |
| Action Research | - | Cooperate with industry to reach improvement goals. |

### 2.5.1  Historical Research Methods

Historical methods use secondary or indirect data sources, for example from projects that are already finished. One historical method is literature search, where scientific papers that are publicly available are analysed. Another method is to study documents from previously completed projects, like program source code and documentation. This method is called to study legacy data. Other documents for study could be lessons learned documents, which are often reported after finishing larger projects. A last method is static analysis, where we look at a finished (software) product.

## 2.5.2 Controlled Research Methods

Controlled methods imply that we have control over important aspects of the environment where the study is performed. A study is replicated if it is performed several times with changing assumptions to see the impact of different variables. If it is difficult to perform a study in a normal environment, we can make an approximation of the environment and perform a synthetic environment experiment. If the subject under study is a software product, we can use dynamic analysis to test the product under changing conditions. Another method would be to use simulation - to evaluate a product in a simulated environment. For a wider discussion on experiments in software engineering, see (Wohlin et al., 2000).

## 2.5.3 Action Research

We could also add a new category that we do not find in Zelkowitz and Wallace's overview, which is relying both on observation and on taking a more active part in forming results, namely action research. This is defined as (Greenwood and Levin, 1998) «social research carried out by a team encompassing a professional action researcher and members of an organization or community seeking to improve their situation. Action research promotes broad participation in the research process and supports actions leading to a more just or satisfying situation for the stakeholders. Together, the professional researcher and the stakeholders define the problems to be examined, cogenerate relevant knowledge about them, learn and execute social research techniques, take actions, and interpret the results of actions based on what they have learned».

This is a kind of democratic research view that gives credit to the knowledge that people in normal work have, and also ensures that the research is relevant for the software industry (Avison et al., 1999).

Potential problems with this kind of research is that it can easily be biased, in that everyone is interested in reaching the goals that are set up. Thus, we do not know if the same results would be achieved with an-other set of researchers, or with other people from the company, or with another company in the same situation. But this kind of research is a way to get interaction with companies in a way that would not be possible if it was not so much in the company's interest.

## 2.5.4 Observational Research Methods

By observational research methods we mean to collect information about our subject of study without strict control over the environment. As in most research, we have to decide what type of information or data to collect, and a proper way to collect it. Data collection methods can be through written questionnaires, visual observation, interviews, written reports, logs, etc. We can further distinguish between different observational studies when looking at how the research material is collected. If we simply study a project with no special efforts to gather data, and no interference, we call it project monitoring. If the researchers are more involved in deciding what information should be collected, we call it a case study. If there is no strong distinction between the subjects participating in the study and the researchers, we call it an assertion. This type of studies would increase the possibilities of biased results. If we collect data from several projects, we call it a field study.

Note that this usage of the word «field study» differs from normal usage in social science, where a field study would imply that the researchers actually spend time on working in the field.

## 2.5.5 A Further Description of The Case Study

A definition of a case study is: «an empirical inquiry that (Yin, 1994):

- Investigates a contemporary phenomenon within its real-life context, especially when
- The boundaries between phenomenon and context are not clearly evident.»

To gather data for a case study, we have several options (see (Seaman, 1999) for an overview of qualitative methods in empirical studies of software engineering). We can use a method called ethnography, which we will first present, and then we present another method for analysing the data called grounded theory.

## *Ethnography - Using a Variety of Data Sources in a Company Setting*

Ethnography is a method for collecting information for a case study which has been used by anthropologists. It is described as «the art and science of describing a group or culture. The description may be of a small tribal group in an exotic land or a classroom in middle- class suburbia. The task is much like the one taken on by an investigative reporter, who interviews relevant people, reviews records, weighs the credibility of one person's opinion against another's, looks for ties to special interests and organizations, and writes the story for a concerned public and for professional colleagues. A key difference, however, is that whereas the journalist seeks out the unusual - the murder, the plane crash, or the bank robbery - the ethnographer writes about the routine, daily lives of people» (Fetterman, 1998). The main element of ethnographic research is the fieldwork: the researcher should get into the environment that she is intending to study, and gradually start to collect data. Another key element in ethnography is to rely on multiple data sources: Participant observation, questionnaires, interviews, projective techniques, videos, pictures and written material.

The analysis in ethnography is usually concentrated around triangulation - to set different sources of information up against each other, to find patterns of thought or behaviour in the community of study. Other methods include drawing organizational charts, making flowcharts of processes that happen, setting up matrices to compare and contrast data, and to use statistics.

Ethnography has been used to some extent within the Computer Science subfield of Information Systems, as well as in Software Engineering. Some researchers have used ethnography and discourse analysis to investigate how quality procedures are applied by practitioners (Sharp et al., 2000). Perry et. al. report on how software developers actually spend their time in a large company (Perry et al., 1994), which contradicts the waterfall model. Others have written about applying ethnographic methods in the construction of information systems and to analyse the development itself (Beynon-Davies, 1997). This paper also gives a good introduction to ethnography.

## *Grounded Theory - Building Inductive Theory from Field Data*

Grounded theory is one type of qualitative research, which is relying heavily on the data that is collected. It aims to use the data for building theoretical constructions unlike other methods more used in natural sciences where the data is used to evaluate hypothesis that we have. A lot of emphasis is put on the process of coding, that is «The analytic processes through which data are fractured, conceptualised, and integrated to form theory» (Strauss and Corbin, 1998). By data, we here mean transcripts of interviews, observational field notes, videos, journals, memos and other written or pictorial information. Some coding procedures are «open coding» - where information is analysed to find central ideas - «concepts» - which is then used to form theory. The analysis can be word-by-word, sentence-by-sentence or on a more abstract level to find concepts in complete documents. Another much used technique is to apply «axial coding» - to place concepts along an axis. An example from Strauss and Corbin are how teenagers describe the impact of drugs, from «getting stoned» to «not having any effect».

# 3 Knowledge Management

In this chapter, we first give an overview of topics needed to understand knowledge management: Different perspectives on knowledge, and models of learning. We then discuss what we mean by knowledge management, and present a model to understand the contents of a knowledge management system in an organisation. Finally, we present case studies of such knowledge management systems that have been described in the software engineering literature.

## 3.1 What is Knowledge?

Before we discuss knowledge management, let us clarify what we mean by some common terms in this field. The term knowledge is defined in the Oxford Dictionary and Thesaurus (1995) as: «awareness or familiarity gained by experience (of a person, fact, or thing)», «persons range of information», «specific information; facts or intelligence about something», or «a theoretical or practical understanding of a subject». A more philosophical (and positivist) view of knowledge is to see it as «justified true belief» (first introduced by Plato, according to (Nonaka and Takeuchi, 1995)).

Davenport and Prusak give a broader definition of knowledge (Davenport and Prusak, 1998): «Knowledge is a fluid mix of framed experience, values, contextual information, and expert insight that provides a framework for evaluating and incorporating new experiences and information. It originates and is applied in the minds of knowers. In organizations, it often becomes embedded not only in documents or repositories but also in organizational routines, processes, practises, and norms.»

We often divide knowledge into two types, tacit and explicit knowledge (Polanyi, 1967). By tacit knowledge we mean knowledge that a human is unable to express, but is guiding the behaviour of the human. Polanyi writes in his book about «The Tacit Dimension» that humans «know more than we can tell». For example, humans can recognise people's faces from thousands of others, but we cannot usually tell how we rec-ognize a face. Another example is the struggle of Japanese engineers to make a machine that bakes tasty bread. According to Nonaka and Takeuchi (Nonaka and Takeuchi, 1995), there were several trials to construct such a machine, but the bread simply did not taste as well as bread made by normal bakers. The company NEC decided to send people to a local baker to see how the process of making bread was actually carried out. The researchers returned with new insight on the kneading process, and later were able to replicate this in their machine. This is an example of tacit knowledge that is difficult to transfer by other means than looking at someone actually baking bread.

Explicit knowledge is knowledge that we can represent, or «codify», for example in reports, books, talks, or other communication.

Some terms that are related to knowledge, are experience and information. In normal English, experience means «actual observation of or practical acquaintance with facts or events», or «knowledge or skill resulting from this» (1995). Most people see experience as a type of knowledge that you have gained from practise, what some people call «local knowledge». Information is seen as «something told; knowledge», «items of knowledge; news». In normal English, it is difficult to distinguish the words information and knowledge. As Tom Stonier writes in his book «Information and Meaning» (Stonier, 1997): «although we all have an intuitive understanding of the term 'information', our understanding is not sufficient to allow us create, for example, a theory of information which would allow us to explain manifestations such as meaning, knowledge, insight, or wisdom». Within artificial intelligence and information processing, however, information is often referred to as «data with meaning». The characters (data) «4m» does not say much in itself, but if we know that «m» stands for «meters», it can be useful information. Knowledge is then often defined as information that is used, or made operative (in an artificial intelligence-sense: in a computer system). For an interesting discussion about the terms data, information and knowledge in artificial intelligence, see (Aamodt and Nygård, 1995). This use of the term knowledge in artificial intelligence is however disputed by Dreyfus (Dreyfus, 1992), who claims that knowledge requires other processes than those in a computer system.

Nevertheless, the view of knowledge as «usable information» is also held by Peter Drucker, who in his book about the «knowledge society» de-fines knowledge as «information effective in action, information focused on results» (Drucker, 1993).

In the book «The Social Life of Information» (Brown and Duguid, 2000), we find another discussion on the difference between knowledge and information, which differs greatly from the one in the artificial intelligence field. Brown and Duguid see three distinctions between the words: First, they claim that «knowledge» entails a «knower»; it is more associated with a person than «information». Second, knowledge is harder to detach (from the «knower») than information. Information is something that people can «pick up, possess, pass around, put in a database, lose, find, write down, accumulate, compare»... while they claim that knowledge is «hard to pick up, and hard to transfer». Third, they point out that is that knowledge entails the knower's understanding and «some degree of commitment». A person can have conflicting information, but will usually not have conflicting knowledge.

To sum up this discussion, it is clearly out of scope to finish the discussion on knowledge in this thesis, but in the following we will use a pragmatic definition of knowledge, what Taylor (Taylor, 1991), who has been working with information use environments, would call «instrumental information» - information that is used so that individuals know how to do something, or «factual information» - information that is used to determine facts. We will refer to this type of «operational information» as explicit knowledge, and we will also use the word tacit knowledge.

## 3.2   Learning

The process of transferring knowledge between people is usually referred to as learning». Webster's (1989) defines it as «to acquire knowledge of or skill in by study, instruction, or experience, to become informed of or acquainted with» or «to memorize». In organisational literature, it is often defined as a «purposefully change of action».

A term that is in common use in the knowledge management field - and which is a field of its own - is «organisational learning». What does it mean to say that an organisation as a whole learns? This differs from individual learning in two respects (Stata, 1996): First, it occurs through shared insight, knowledge and shared models. Second: it is not only based on the memory of the participants in the organisation, but also on «institutional mechanisms» like policies, strategies, explicit models and defined processes (we can call this the «culture» of the organisation). These mechanisms may change over time, what we can say is a form of learning. Argyris and Schön distinguish between what they call single and double-loop learning (Argyris and Schön, 1996) in organisations. Single-loop learning implies a better understanding of how to change (or «tune»), say a process, to remove an error from a product. It is a (single) feedback- loop from observed effects to making some changes (refinements) that influence the effects. Double loop learning, on the other hand, is when you understand the factors that influence the effects, and the nature of this influence, what they call the «governing values» (Argyris, 1990). In our example above, this could be to understand why a process is usable, that is: Which premises must be satisfied for it to be worthwhile. To make changes based on this type of understanding will be more thorough, like in a political revolution.

In software engineering, a «learning software organisation» has been defined as an organisation that have to «create a culture that promotes continuous learning and fosters the exchange of experience» (Feldmann and Althoff, 2001). Tore Dybå puts more emphasis on action in his definition: «A software organisation that promotes improved actions through better knowledge and understanding» (Dybå, 2001).

Now, we will present different models from the literature on how knowledge is transferred between individuals in organisations, what we can describe as «learning» on an individual level, and «organizational learning» for a community. We do not claim to cover the whole range of theories of learning, but will focus on three approaches that we consider interesting, and has been used in the knowledge management field, namely:

- Learning through participation: communities of practise.
- Learning from experience - a theory of individual's learning modes.
- Learning as a conversion process between tacit and explicit knowledge.

### 3.2.1 Learning through Participation: Communities of Practise

The traditional view of learning has been that it best takes place in a setting where you isolate and abstract knowledge and then «teach» it to «students» in rooms free of context. Etienne Wenger describes this view of learning as an individual process where for example collaboration is considered a kind of cheating (Wenger, 1998). In his book about communities of practise, he describes a completely different view: learning as a social phenomenon. A community of practise develops its own «practises, routines, rituals, artifacts, symbols, conventions, stories and histories». This is often different from what you find in work instructions, manuals and the like. In this context, Wenger defines learning as:

- For individuals: learning takes place in engaging in and contributing to a community.
- For communities: learning is to refine the practise.
- For organisations: learning is to sustain interconnected communities of practise.

We find these communities of practise everywhere: at work, at home, in volunteer work. And it can be a challenge to sustain such networks of people, for example in turbulent organisations that undergo many reorganisation processes.

The work on communities of practise is closely linked to work on situated learning (Lave and Wenger, 1991).

### 3.2.2 Learning from Experience

We can describe learning from experience («experiential learning», see (Kolb, 1984)) as four different learning modes that we can place in two dimensions. One dimension is how people take hold of experience, with two modes, either relying on symbolic representation - called comprehension, or through «tangible, felt qualities of immediate experience», called apprehension. The other dimension is how people transform experience, with two modes, either through internal reflection, referred to as intention, or through «active external manipulation of the external world», called extension. Let us now discuss these four modes of learn-ing a bit more in detail. First the grasping dimension, then the transformation dimension:

By apprehension we mean knowing through grasping the environment; sounds, colours, how things feel. All these factors influence humans to recognise a situation where they remember something they did before. By comprehension, we mean introducing order to sensations, to abstract issues so that they can be communicated to other people.

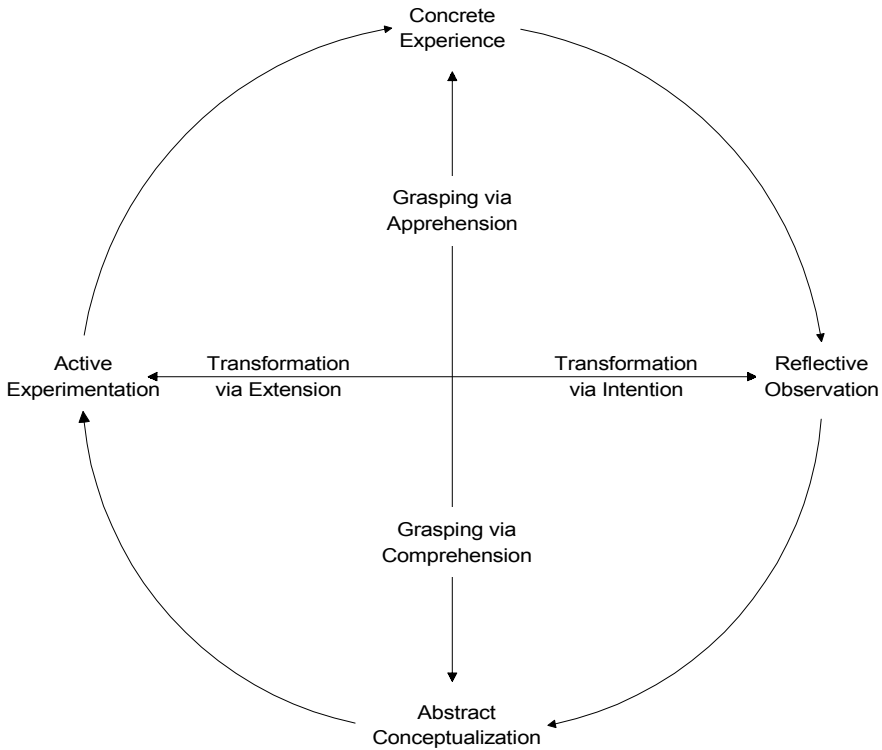

*Figure 3.1: The Four Modes of Learning in Kolb's model.*

If we turn our attention to the transformation dimension, we think of this as two learning modes that apply to knowledge that has been grasped both through comprehension and apprehension. The mode intention deals with descriptions of things, that does not necessarily exist in the real world. Characteristics of people or a «work process» can be seen as «symbols» with a meaning, that focus on some aspects and eliminate others. Extension on the other hand, deals with the real world, what symbols refer to. So, in dealing with this world and manipulating it, humans come to learn. An example of learning through extension could be to perform an experiment, although this would normally also include some abstracted learning from the results which would include intention as well.

Kolb argues that people need to take advantage of all four modes of learning to be effective, they «must be able to involve themselves fully, openly, and without bias in new experiences; reflect on and observe these experiences from many perspectives; create concepts that integrate their observations into logically sound theories; and use these theories to make decisions and solve problems» (Kolb, 1996).

### 3.2.3 Learning as a Conversion Process between Tacit and Explicit Knowledge.

In the much-cited book The Knowledge-Creating Company, where Nonaka and Takeuchi tries to explain that Japanese companies have done well because they are good at «organizational knowledge creation», they also offer a model of how knowledge is transformed and converted in an organisation (Nonaka and Takeuchi, 1995).

In the section where we discussed the word «knowledge», we divided between tacit and implicit knowledge. Nonaka and Takeuchi claims that knowledge is constantly converted from tacit to explicit and back again as it passes through an organisation. They say that knowledge can be converted from tacit to tacit, from tacit to explicit, or from explicit to either tacit or explicit knowledge as in Figure 3.2.

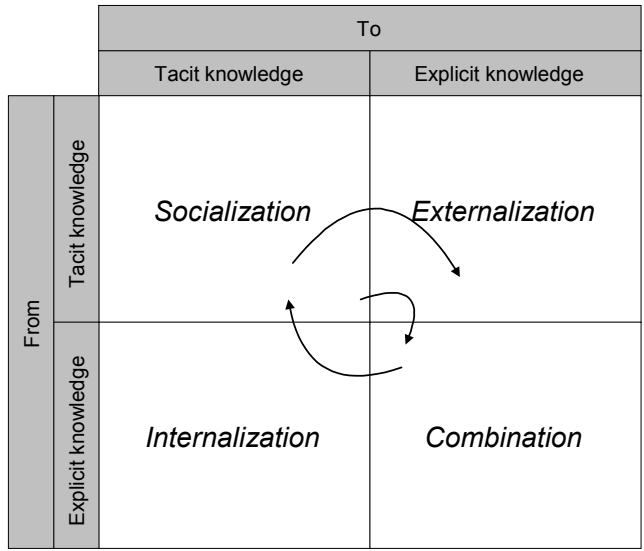

*Figure 3.2: Conversion of knowledge according to Nonaka and Takeuchi. We can imagine knowledge going through all conversion processes in a spiral form as it develops in an organisation.*

We now describe each of these four modes of conversion:

- Socialization means to transfer tacit knowledge to tacit through observation, imitation and practice, what has been referred to as «on the job» training. Craftsmanship has usually been learned in this way, where oral communication is either not used or plays a minor part.

- Internalisation is to take externalised knowledge and make it into individual tacit knowledge in the form of mental models or technical know-how. «Documents and manuals facilitate the transfer of explicit knowledge to other people, thereby helping them experience the experiences of others indirectly (i.e. 're-experience' them)».

- Externalisation means to go from tacit knowledge to explicit. Explicit knowledge can «take the shapes of metaphors, analogies, concepts, hypotheses or models». This conversion is usually triggered by dialogue or collective reflection, but can also be the result of individual reflection, for example in a writing process.

- • Combination is to go from explicit to explicit knowledge, that is, to combine and systemize knowledge from different sources such as documents, meetings, telephone conferences or bulletin boards. Systematizing this kind of explicit knowledge is to reconfigure it by sorting, adding, combining or categorizing the knowledge.

According to Nonaka and Takeuchi knowledge passes through different modes of conversion in a spiral which makes the knowledge more refined, and also spreads it across different layers in an organisation.

## 3.3    What is Knowledge Management?

Now, we will first discuss the term Knowledge Management in general and in software engineering, and then introduce a model for what a knowledge management initiative, or system, can be in a company. Finally, we discuss some success factors in working with knowledge management initiatives in companies.

### 3.3.1  Introduction

There are many interpretations of knowledge management, and of how to describe computer systems to support it in companies. In 1974, the book «The Corporate Memory» was published (Weaver and Bishop, 1974), arguing on the benefit of collecting information from different sources in a company and making it «searchable». At this time, the information was gathered on paper, and «search» would mean to submit a form to a department who would manually search through their files. The word corporate memory is still in use, but now meaning a database for storing documents from many people in a company. The word «corporate brain» is also used to describe such a database. Another related word is «organizational memory», which does not really have a clear definition, but «intuitively, organizations should be able to retrieve traces of their past activities, but the form of this memory is unclear in research literature. Early efforts assume one could consider memory as though it were a single, monolithic repository of some sort for the entire organization» (Ackerman and Halverson, 2000). Many see this term as meaning both a process of collecting and using information as well as a repository.

So what do we mean by knowledge management? We think that this term includes issues from all the terms discussed. Some goals of knowledge management can be (Wiig, 1997):

1) To make the enterprise act as intelligently as possible to secure its viability and overall success and
2) To otherwise realize the best value of its knowledge assets.

Thomas Davenport has defined it as «a method that simplifies the process of sharing, distributing, creating, capturing and understanding of a company's knowledge» (Davenport et al., 1998a). If we look a bit more into knowledge management, we find that some important aspects are to (Wiig, 1995):

- Survey, develop, maintain and secure the intellectual and knowledge resources of the enterprise.
- Determine the knowledge and expertise required to perform work tasks, organize it, make the requisite knowledge available, «package it» and distribute it to the relevant points of action.
- Provide (...) knowledge architecture so that the enterprise's facilities, procedures, guidelines, standards, examples, and practices facilitate and support active knowledge management as part of the organization's practices and culture.

This seems to be in line with what people from two different software companies that we will introduce later in this thesis see as knowledge management. We interviewed 19 managers and developers about what they meant by «knowledge management» and got answers like «manage, plan, deploy, collect and spread knowledge in an organisation, and do it in a planned manner», and «to create, store, survey, use and revise knowledge».

### 3.3.2 Knowledge Management in Software Engineering

In Software Engineering, to reuse life cycle experience, processes and products for software development is often referred to as having an «Experience Factory» (Basili et al., 1994) - a separate organisational entity with responsibility for capturing and reusing experience. This approach has been very much referenced in the software engineering field (Basili et al., 1994). Experience is collected from software development projects, and packaged and stored in an experience base. By packaging, we mean generalising, tailoring and formalising experience so that it is easy to re-use.

The Experience Factory organisation assists software developing projects with earlier experience both in upstart and during execution, and can suggest improvements in processes based on collected experience (we call this «strategic improvement management» in Figure 3.3). The interaction between the Experience Factory, the sponsoring organisation and the software development projects is shown in Figure 3.4.

These ideas were further elaborated in the PERFECT project (PERFECT Consortium, 1996). Here, we find advise on how to «implement» an Experience Factory in an organisation; on which steps to take - from «characterizing the business situation» and «setting goals», to making an «implementation proposal» and «establish an Experience Factory». It also gives advice on which roles different people in the organisation can have in this work.

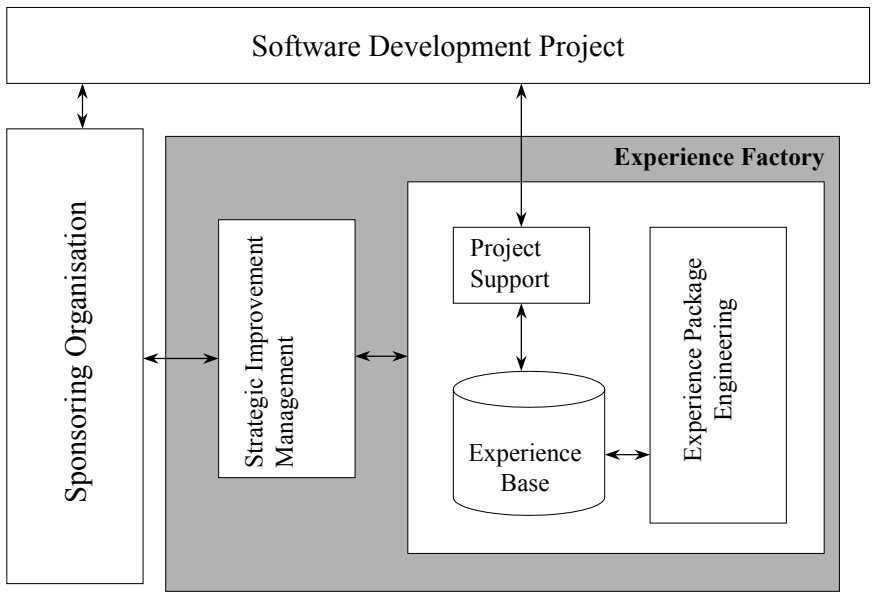

*Figure 3.3: The Experience Factory as seen in the PERFECT Project.*

Another addition to the original ideas in Experience Factory, we find in a paper from Daimler Chrysler (Houdek and Schneider, 1999), which clarifies some issues that are taken for granted in the original Experience Factory work:

- Improvement activities in a quality improvement paradigm-perspective is a long-term activity.
- For projects, process improvement and learning will require additional effort.
- Knowledge transfer between projects requires some similarity between projects.

Some of the ideas in Experience Factory would probably be implemented in a different way today, than when the ideas emerged. For example web-technology was not developed when the work started.

### 3.3.3  A Model for Knowledge Management

Now we will present a model for knowledge management systems that exist in companies, which combines strategic issues (Hansen et al., 1999) with «implementation» in a company - divided in processes and tools, taken from the process modelling field. This model will be used for discussing case studies later: A knowledge management system in a company can consist of three parts:

- An overall strategy for knowledge management. That is, what goals do the company want to achieve? And how does it proceed to do so? Usually, the goals within software engineering companies are to produce software faster, with less cost, or with a higher quality. But it can also be to improve the work situation of software developers.
- A set of processes (activities) that a company does in order to facilitate knowledge management. This will usually be methods for collecting and distributing knowledge, and can be activities a separate part of the organisation is doing, as well as project managers and software developers.
- A set of tools for knowledge management: a computer software system where operational information, or «knowledge», can be found by different groups of practitioners (like developers, project managers, quality management) in a software company, usu-ally on an Intranet. The knowledge can be represented in databases or in text or pure HTML, but the maintenance effort would be larger with the latter options. Another way to represent to make it easy retrievable is to use Case Based Reasoning, like in the COIN EF system in use at the Fraunhofer IESE (Tautz, 2000). We see knowledge as something dynamic, that might be changing over time, so a knowledge management tool, must offer possibilities for revising or discarding knowledge, as well as supplying new knowledge. For a broader view of what the artificial intelligence community see as knowledge management tools, see (Smith and Farquhar, 2000). In our definition of a tool we do not include mailinglists, Outlook folders or server folders that are available in many companies.

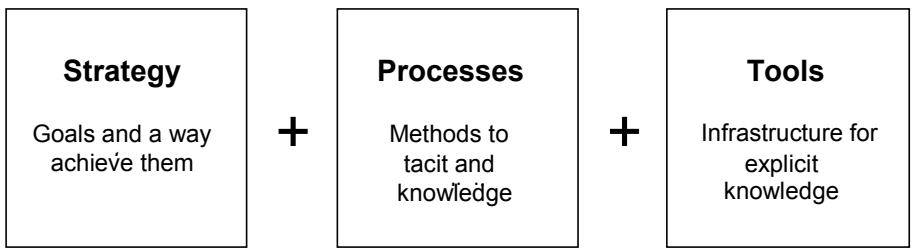

*Figure 3.4: A Model of the Components of a Knowledge Management System.*

Now, we describe different strategies, processes and tools that can exist in software developing companies.

## Two Strategies for Knowledge Management

We can divide between two different usages, or strategies for knowledge management (Hansen et al., 1999):

- Codification - to systematize and store information that represents the knowledge of the company, and make this available for the people in the company. If we look at the models for learning we presented earlier, this is what Nonaka and Takeuchi calls «externalisation» - to make tacit knowledge explicit. In Kolb's model, this is when you reason with symbolic representation, and make abstract ideas of your experience, what he refers to as intention.

- • Personalization - to support the flow of information in a company by storing information about knowledge sources, like a «yellow pages» of who knows about what in a company. Referring again to the previous subchapters on learning, we can think of a community of practise as an environment that focuses very much on person to person communication, what Nonaka and Takeuchi calls socialization. In Kolb's model, this could include both modes of the grasping and transforming dimensions.

Hansen et al. argues that companies should focus on just one of these strategies.

We should add here that the codification strategy does not fit all types of knowledge. In situations where knowledge is very context- dependent, and where the context is difficult to transfer, it can be directly dangerous to reuse knowledge without analysing it critically. For some more examples of problems with this strategy, see (Jørgensen and Sjøberg, 2000).

Another strategy than the two mentioned above could be to support the growth of knowledge - the creation of new knowledge by arranging for innovation through special learning environments or expert networks, but that is beyond the scope of this thesis.

When we go on to discuss computer systems and associated human processes that support knowledge management, we will restrict the scope to systems supporting the first two strategies.

Note that some have referred to these strategies by other names: Codification can also be called «exploitation», and personalization «exploration» (Mathiassen et al., 2002).

## Processes for Knowledge Management

What activities can an organisation perform to promote knowledge management? If we return to our three models of learning, we can say that to improve working conditions for different «communities of practise» can be one activity. This would be similar to knowledge transfer in different arenas through socialization. If we turn to Kolb, we should try to make room for reflection on experience in order to improve learning processes in a company; and understand that different people have different learning modes that they prefer. No learning recipe will suit all people. If we turn to Nonaka and Takeuchi, codifying (externalising) tacit knowledge and writing it down can be one activity, having a group of people to combine explicit knowledge a second, and finally making such externalised knowledge available for people to learn from.

As an example of a knowledge management process, we will now describe varieties of processes for «externalising» tacit knowledge, and making it explicit, what we can call «harvesting knowledge» or «knowledge acquisition». See (Eriksson, 1992) for an overview of knowledge acquisition techniques, and (Birk et al., 1999) and (Birk, 2000) for methods tailored to software engineering.

In a software engineering setting, harvesting knowledge will be mostly relevant from a project setting. We now describe two ways of capturing knowledge from projects: writing experience reports (usually written by a project manager), and a more structured method which involves as many people as possible from a project team, namely postmortem reviews. We also present a version of postmortems that requires little effort: «lightweight postmortem reviews».

### Experience Reports

A way to collect experience from a completed project is to write an «Experience Report». This document is written by the project manager after a project is finished. The report usually follows a fixed template, to make it easy to compare reports from different projects. In a Norwegian company that develops software for space satellites, the report is divided into two parts: The first part gives an overview of all the facts and numbers from the project: Start and finish date, size of contract, labour used, deviation from estimated work size, the number of source lines-of-code developed, documents produced, and the activities that contributed most to the excess consumption. The second part of the report describes problems during project execution with proposals for improvement. For each phase of the project there are Problem Descriptions and Proposals for Improvements. This information is represented as text. These reports are usually not longer than 10-15 pages in this company, and about 50% is devoted to each part.

## Postmortem Reviews

There are several ways to perform postmortem reviews. Apple has used a method (Collier et al., 1996) which includes designing a project survey, collecting objective project information, conducting a debriefing meeting, a «project history day» and finally publishing the results. At Microsoft they also devote quite much effort into writing «postmortem reports», which are more similar to what we have called «Experience Reports», except they involve more people in the writing process. These contain a discussion on «what worked well in the last project, what did not work well, and what the group should do to improve in the next project» (Cusomano and Selby, 1995). The size of the resulting documents are quite large, «groups generally take three to six months to put a postmortem document together. The documents have ranged from under 10 to more than 100 pages, and have tended to grow in length».

In a book about team software development, Watts Humphrey suggests a way to do postmortems to «learn what went right and wrong, and to see how to do the job better the next time» (Humphrey, 1999).

A problem with these approaches is that they are made for very large companies, who can spend a lot of resources on analysing completed projects. With medium-sized companies where 5-10 people usually participate in a project, ranging in size from about 8 to 50 manmonths, there are usually not enough time or resources to a thorough review. Such companies might benefit from a «lightweight» version of postmortem reviews.

## Lightweight Postmortem Reviews

A lightweight postmortem review (Dingsøyr et al., 2001) is a group process, where most of the work is done in one meeting lasting only half a day. We try to get as many as possible of the project workers to participate, together with two researchers, one in charge of the Postmortem process, the other acting as a secretary. The goal of this meeting is to collect information from the participants, make them discuss the way the project was carried out, and also to analyse causes for why things worked out well or not.

Our «requirements» for this process is that it should not take much time for the project team to participate - no individual preparation is necessary, and it should document the most important experience from the project, together with an analysis of this experience.

A description of another lightweight approach which seeks to elicit experience using interviews, and not a group process, is described by Schneider (Schneider, 2000).

We have used two techniques to carry out lightweight postmortem reviews. For a focused brainstorm on what happened in the project, we used a technique named after a Japanese ethnologist, Jiro Kawakita (Scupin, 1997) - called «the KJ Method». For each of these sessions, we give the participants a set of «post-it» notes, and ask them to write one «issue» on each. We usually hand out between three and five notes per person, depending on the number of participants. After some minutes, we ask one of them to attach one note to a whiteboard and say why this issue was important. Then the next person would present a note and so on until all the notes are on the whiteboard. The notes are then grouped and renamed.

We use a technique for Root Cause Analysis called Ishikawa or fishbone-diagrams (Straker, 1995) to analyse the causes of important issues. We draw an arrow on a whiteboard indicating the issue being discussed, and attach other arrows to this one like in a fishbone with issues the participants think cause the first issue. Sometimes, we also think about what was the subcauses for some of the causes and attached those as well.

A comparison of two methods for lightweight Postmortem Review can be found in (Stålhane et al., 2001), with information on the resources required.

## Tools for Knowledge Management

When we talk of tools for knowledge management here, we will mean tools that have several users, and are widely available for employees in an organisation. This is usually what we can call Intranet tools, that supports knowledge management in «at least three ways: 1) providing compression of time and space among the users. 2) offering the flexibility to ex-change information, and 3) supporting information transfer and organizational networking independent of direct contacts between the users» (Ruppel and Harrington, 2001).

There are many dimensions for describing knowledge management tools. Ruggles (cited in (Carlsen and Paulsen, 1999)) mentions tools that «generate knowledge», where tools for data mining can be an example - to discover new patterns in data. Further, we have «knowledge codification tools» to make knowledge available for others, and «knowledge transfer tools» that aim to decrease problems with time and space when communicating in an organisation.

Another dimension is whether the tools are «active» (Sørlie et al., 1999) or «passive». By active tools, we mean tools that notify users when it is likely that users require some kind of knowledge. Passive tools requires a user to actively seek knowledge without any system support.

Yet another way of classifying tools has been suggested by Hahn and Subramani (Hahn and Subramani, 2000), where they use two dimensions: where knowledge resides, and to what extent the knowledge is structured. If we say that knowledge can reside in individuals or as «artifacts», and that knowledge can be «structured» or «unstructured», we get the following table:

*Table 3.1: A Framework for Knowledge Management with Examples of Tools.*

| | | Locus of knowledge | |
| --- | --- | --- | --- |
| | | *Artifact* | *Individual* |
| Level of A Priori Structure | *Structured* | Document Repository. Data Warehousing | Yellow Pages of Experts, Expertise Profiles and Databases |
| | *Unstructured* | Collaborative Filtering, Intranets and Search Engines | Electronic Discussion Forums |

The artifacts are then the result of a codification process as we discussed in the section about strategies for knowledge management. And supporting the knowledge that resides in individuals corresponds to the personalization strategy. Note that Intranets are used in a more narrow sense in Table 3.1 than we have used. We see an Intranet as an underlying technology that can support both structured and unstructured knowledge as well as artifacts and transfer between individuals.

For a discussion of other dimensions like Nonaka and Tekeuchi's four modes of knowledge conversion in categorizing tools, see (Carlsen and Paulsen, 1999).

Now, we have chosen another way of categorizing the tools than the ones mentioned so far, from the book Information Technology for Knowledge Management (Borghoff and Pareschi, 1998), because this model is widely known. The authors divide technology for a «corporate memory» into four parts, shown in Figure 3.5:

- Knowledge repositories and libraries - tools for handling repositories of knowledge in the form of documents.

- Communities of knowledge workers - tools to support communities of practise in work; like organizing workspaces for communities for online discussions and distributed work.
- The flow of knowledge - here we find tools for supporting the interaction between tacit knowledge, explicit knowledge and metaknowledge; that is, that combines the three parts above.
- Knowledge cartography - tools for mapping and categorizing knowledge, from core competence in a company to individual expertise; what we can refer to as «metaknowledge».

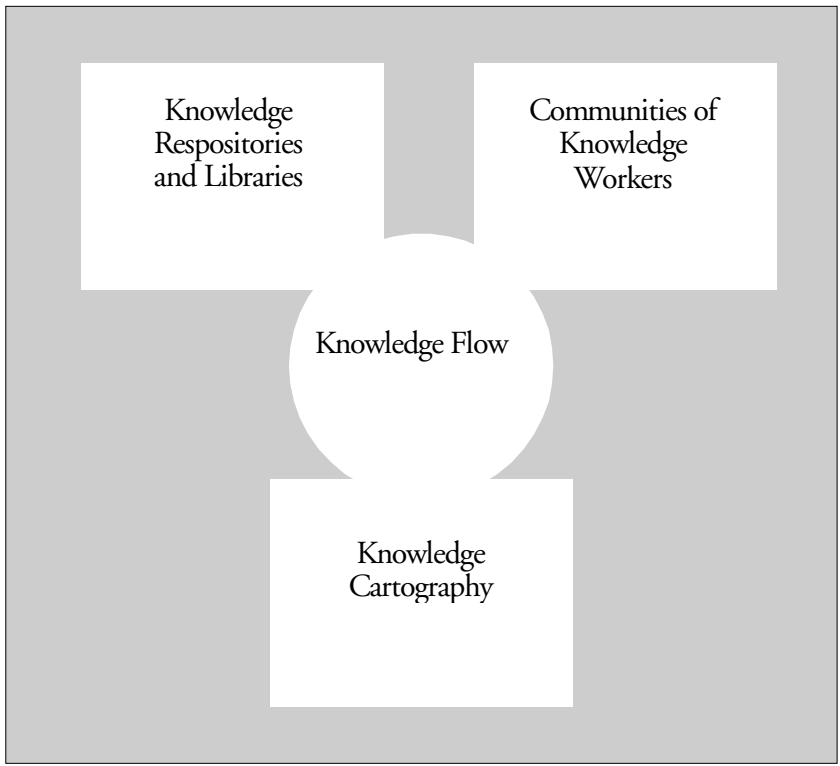

*Figure 3.5: Types of Knowledge Management Tools or Architecture (Borghoff and Pareschi).*

In Software Engineering, much work has been devoted to building knowledge repositories and libraries. In the Experience Factory concept, some examples of the contents of such repositories (called «packages») are:

1. Product Packages - information about the life cycle of a product, information on how to reuse the product, and lessons learned from reuse.
2. Process Packages - information on how to execute a method or a life cycle process, and how to reuse it.
3. Relationship Packages - used for analysis and forecasts. Can be predictive cost and defect models, resource models.
4. Tool Packages - instructions for how to use a tool and experience with it.
5. Management Packages - reference information for project managers.
6. Data Packages - data relevant for a software project or activities. Can be project databases or quality records.

If we look at technology issues in supporting repositories and libraries in software engineering, we find many tools, for example the Experience Management System (Seaman et al., 1999). Many have used Case-Based Reasoning (CBR), see (Aamodt and Plaza, 1994), for retaining and retrieving experience. For instance, (Althoff et al., 1998a) report on the benefits in using CBR technology to support experimental software engineering more generally, and (Althoff et al., 1998b) are concerned with CBR for building learning software organisations.

The University of Nebraska-Lincoln has developed a research prototype tool for knowledge management support in software development called BORE (Henniger, 1997a, Henniger, 1997b, Henniger and Schlabach): This is a tool which contains information in cases about problem solving experience, and descriptions of resources like tools, projects, people and development methods. These descriptions are used to find relevant solutions when software developers are faced with a new problem.

Another prototype system, is CODE - a general-purpose knowledge management tool which serves as a medium for knowledge capture, transfer and iteration, as well as editing or «packaging» knowledge to make it easy available (Skuce, 1995).

Yet another knowledge management tool for software engineering is developed at the University of Kaiserslautern (Feldmann, 1999). Here, a comprehensive reuse repository has been developed, with possibilities for advanced search and retrieval mechanisms.

Several technologies for experience reuse are evaluated in (Wangenheim et al., 1998), where the conclusion is that Case-Based Reasoning is suitable for reusing experience from software engineering. In (Broomé and Runeson, 1999) we find a number of technical requirements for an experience database. Yet other work has been done on using ideas from the Experience Factory in the construction of CBR systems, (Bergmann and Göker, 1999), for process improvement in developing educational software (van Aalst, 2001). Other technical approaches than CBR has also been suggested by (Feldmann et al., 1998). Additional work has been done on models for introduction of technical systems for experience reuse in an organisation (Dingsøyr, 1999).

We also find descriptions of knowledge management systems used in the industry in the literature, like the one in use in the company Computas (Carlsen et al., 1999), and at Hewlett Packard India (Bhave et al., 2001, Bhave and Narendra, 2000). Further, we find descriptions of knowledge management systems in four companies in Norway (Conradi and Dingsøyr, 2000), together with a discussion on success factors in introducing and using such systems in organisations.

### 3.3.4 Success Factors in Knowledge Management Initiatives

Davenport, Long and Beers (Davenport et al., 1998b) studied 31 knowledge management projects in 24 companies - by interviewing people in the companies. They identified eight «success factors» in these projects, which were:

- Link to economic performance or industry value (d1).
- Technical and organizational infrastructure (d2).
- Standard, flexible knowledge structure (d3).
- Knowledge-friendly culture (d4).
- Clear purpose and language (d5).
- Multiple channels for knowledge transfer (d6).
- Senior management support (d7).

We enumerate these factors d1-d7 as we will be using these again in the discussion later.

Another study about a knowledge management initiative in the Buckham laboratories (Pan and Scarbrough, 1999) also conclude that «specifically, the task for the organization is to continuously create and maintain a knowledge-enterprising culture and community whereby associated feel comfortable with knowledge and are motivated, rewarded and entrepreneurial». They further find that knowledge management systems «involve more than technology but rather a culture in which new roles and constructs are created». The importance of organisational factors is also stressed in a study from an American Consulting company. The introduction of a groupware system for sharing experience was unsuccessfull, because of a very little collaborative culture, and few structural incentives for cooperation (Orlikowski, 1992).

A fourth study that we have found, is McKinsey's survey (Kluge et al., 2001) on knowledge management in 40 companies in Europe, the US and Japan. They tried to find success in knowledge management initiatives by looking at companies «process performance» and financial success. The findings of this survey was that companies that are more «successful» focus more on the following factors (non-extensive list) in knowledge management: development efficiency, process efficiency, quality standards, product innovation. We also find factors such as «active involvement of employees in process improvement decisions» and «financial incentives for cooperation, information flow in production».

## 3.4   Case Studies of Knowledge Management in Software Engineering

If we look at work on actual use of knowledge management in an organisation, we find much less in the literature than about technology issues. We here report case studies, and examine what claims are made about knowledge management in each of them, and describe in what organisational setting each of the case studies were performed. We also place the studies in a category of scientific methods, which was discussed in section 2.5.

### 3.4.1 The NASA Software Engineering Laboratory

The first implementation of an Experience Factory was at the NASA Software Engineering Laboratory, and is reported in (Basili et al., 1992). The Experience Factory is used as described in (Basili et al., 1994).

Experience in forms of cost data, process data (as project methodology information), information on tools and technology used, product data (such as change and error information) and results on static analysis on delivered code was collected, and used to develop predictive models and to refine the software processes that is used.

The results of this activity is reported as dramatically reduced defect rates (75% from 1987-91, further by 37% from 1991-95); the cost of producing software went down by 55% from 1987-91 and 42% from 1991-95. Reuse was improved by 300% from 1987-91 and 8% from 1991-95. Finally, implemented functionality was increased five-fold from 1976-92.

The organisation produces astrophysical software for unmanned missions in NASA only. It is a bit difficult to compare this organisation with normal, more competitive companies. The article (Basili et al., 1992) reports lessons learned through 15 years of operation.

### 3.4.2 Daimler Chrysler

Daimler Chrysler has implemented three experience factories in different environments within a two-year period, in cooperation with the University of Ulm, Germany (Houdek et al., 1998). The environments were: 1) A department responsible for developing software for the aerospace area with real-time constraints. 2) A department which develops small embedded systems for cars, with special focus on keeping software portable across different micro controllers, and making sure that planned functionality was actually implemented. 3) An administrative software unit that manages internal business processes such as car sales. This unit operates only on requirements, and the software production itself is outsourced.

Other work on experience reuse from Daimler Chrysler can be found in (Landes et al., 1999, Sazama, 2000). Three case studies on experience transfer in the company can be found in (Wieser et al., 1999).

The study takes the form of a lessons learned report, and documents the following findings from the three environments (Houdek and Schneider, 1999):

- There are many sources of reusable experience, and measurement is just one of them.
- There were difficulties in finding how «packaged» users wanted the experience to be.
- Handling qualitative data was a bottleneck.
- Building predictive models from quantitative data was difficult when context information was missing.

We also find a discussion on benefits and problems of introducing an experience factory in a top-down and bottom-up manner.

### 3.4.3 Telenor Telecom Software

In an effort to reuse software development experience, Telenor Telecom Software, a company with 400 software developers in five geographical locations, decided to improve estimation of software development effort, as well as risk management (Jørgensen et al., 1998). To achieve this, they set up:

- An experience reuse process, with new and modified role descriptions.
- An experience database tool, available on the Intranet.
- Resources allocated for experience reuse and for experience database administration.

The experience database was made available like an «expert system» which would ask you questions on the nature of a new project and recommend an appropriate estimation model. It would also give you information on company experts on estimation. The database was linked to a risk management module with risk factors found from interviewing experienced project managers. This module consisted of a set of «best practise» processes, a tool to identify, assess and store risk factors, and a tool to visualize risk exposure over time. In addition to this, new roles for «experience database administrators» were set up - responsible for technical and editorial contents, as well as several roles for «process ana-

lysts», responsible for analysing information from processes such as the estimation process, project management process, and testing process.

Although the authors of the article acknowledge that the study was made too early after the initiative was introduced too draw firm conclusions, and that it was difficult to isolate the impact of their own work from other improvement initiatives in the company, they find several indications of improvement:

- The estimation accuracy improved, and estimation models were more widespread in use.
- The focus on experience-based risk management increased in the projects.
- The organization accepted the need to collect and share experience.

The study takes the form of a lessons learned report.

### 3.4.4  Ericsson Software Technology

Ericsson Software Technology in Sweden has experimented with transfer of experience on a site that develops a wide range of software applications. The site has around 1600 employees who work in business units of 20 to 30 people. The units develop software for telephone switches, base stations and mobile phone management systems. The company has formal communication channels such as meetings, e-mail and written reports, but wanted to establish a corporate culture that facilitates more oral communication of experience (Johansson et al., 1999). Two organisational roles were invented: «Experience brokers» keep track of what other people in the company know, and match people who can benefit from talking to each other. «Experience communicators» help other people solve problems, by teaching them how to solve the problems on their own. The study reports that employees are more motivated when they know that there is a working system for transferring experience.

The scientific method used in this article is a lesson learned report.

### 3.4.5 An Australian Telecom Company

Another paper (Koennecker et al., 1999) reports on the introduction of an Experience Factory in an Australian telecommunications company, done by the company in cooperation with the Centre for Advanced Empirical Software Research at the University of New South Wales, Australia. The goal was to improve the speed and quality of software development, and to enhance experience transfer of process knowledge between projects. This was done by organising information that was already documented in the company, and making it available and searchable, a kind of a «bottom up» way to start a knowledge management initiative. The article then reports the usage of this experience base over time, and classifies the searches that were made. A survey amongst the users was conducted, and the «acceptance and judgment of the product was good». The experience database is also reported to break down barriers between project environments, but this is not supported by quantitative data. Although no information is given on the research method used, it seems that the researchers involved defined the metrics to collect and we can then say that this is a case study. In a later paper, this introduction is described as a «failure» (Koennecker et al., 1999). Although an informal survey amongst users said the «acceptance and judgement of the product (possibility to search an experience base) was good», the project was abandoned by management. Some reasons for this is discussed in the paper: 1) The researchers felt that there was a lack of ongoing management support for this initiative. 2) The goals and payback-criteria for the project was not clearly defined. 3) The researchers think that a more formal approach should have been used to construct an experience-repository, because the users were physically co-located, and the number of people relatively small.

The scientific method here is an assertion.

### 3.4.6 ICL High Performance Systems

ICL High Performance Systems in the UK has developed an «Engineering Process Improvement Framework», which includes a repository for knowledge sharing (Chatters, 1999). The engineering knowledge base contains information divided into three categories (Chatters et al., 2000):

- Projects and processes - descriptions of processes.

- Topic-based instructional material to introduce new concepts.
- General background and further information.

The main objective with introducing this improvement program, was to «improve the predictability of costs and delivery dates of systems and solutions». The authors claim that there is a «perception by project members that the framework has facilitated the transfer to the new mode of working, but this perception is only backed up by anecdotal evidence». The main benefit has been to «reduce risks to achieving project deliverables within agreed budgets, on time, and with required quality». Several lessons learned are reported, like the importance of commitment from the management when introducing such a framework, and that the developers should be involved in designing the framework.

The scientific method used is a lessons learned report.

### 3.4.7  ICL Finland

ICL in Finland has also made a knowledge management system. The Finnish part of the company employs more than 800 people working with software development, in applications and services, and on Internet technology for business applications (like electronic commerce) (Markkula, 1999). ICL classifies their knowledge resources in three groups:

- External knowledge: which includes technical Internet pages, related to customers, software suppliers, tools, technical partners, journals, and research centres.
- Structured internal knowledge: includes databases for sales and marketing information and employee competence, as well as examples of frequently used documents, templates, software components, best practise information, and research reports.
- Informal internal knowledge: includes electronic discussion forums, news and «project folders». The project folders contain overviews of the projects, news and important announcements, technical documents and reusable components (for a complete list, refer to the paper cited above).

ICL did a survey about use of this «Extranet system» amongst people participating in a large project with a peak manning of 50 people, and with an estimated effort of 7800 man-days. The survey was done with a questionnaire, but it is unclear what kind of questionnaire was used, and how many people were interviewed. Based on the survey and interviews, ICL has found the following: Most of the project members say «use of the Extranet has supported the work on the project and saved time», it is also «easier to find documents and other information». Further claims are that the «use of project management and software engineering methods has been easier via the Intranet», and document templates are especially appreciated. Learning new project members about project work is also said to be easier. One interviewee estimates that project managers and other project members «save about 30 percent in time, when making a new project member familiar with the system under development». In all, the benefits of the Extranet, has been highest in «technical planning, implementation and unit testing». The most important benefit is described as the «better visibility through knowing what kind of projects are going on and have been completed at ICL, and the own unit».

The scientific method used is a lessons learned report.

## 3.4.8 sd&m

The German Software company sd&m, focuses on designing and implementing large business information systems tailored to customer needs, and used to have problems with rapid growth. In 1999, the company had 700 employees, and had grown by around 50% in some years (Brössler, 1999). Typical problems were that developers used long time to acquire programming and project management skills, and also had problems in coping with many different technological platforms and tools. It was also a problem that insights gained in one project were not applied in others, so the same «mistakes» were repeated many times in the company.

In 1997 sd&m started working with a knowledge management initiative which involved:

- A knowledge management group consisting of «knowledge brokers»; responsible for the core topics in the company. This involved maintaining a web page on the Internet relating to these topics.

- The projects are supported by the knowledge brokers, who provide pointers to internal and external knowledge resources. The brokers participate in the project kick-off and touchdown meetings.

Several databases was also made available in the company, listing employees, customers, partners, projects and acquisitions (in Lotus Notes databases), as well as a Skill Database, where all employees assess their own skills.

The company claims that these efforts on knowledge management has reduced the impact of the problems described: «it can be seen very clearly that the problems described... do not occur nearly as often as before, despite continuing double-digit growth».

The scientific method used is a lessons learned report.

## 3.4.9 Summary

We give a quick overview of what knowledge management initiatives and approaches the different companies had. We describe the approach by indicating whether the company used a personalization strategy or a codification strategy (or both). We list the companies in Table 3.2. We will discuss these initiatives in further detail in the discussion chapter.

*Table 3.2: What Knowledge Management initiatives and approaches we found in the literature*

| Company | What did they do? | Knowledge Management Approach | |
|---|---|---|---|
| | | Personalization? | Codification? |
| NASA SEL | Set up a separate organisation which collected and distributed experience. | Yes | Yes |
| Daimler Chrysler | Created three experience factories in three different company departments. | Yes | Yes |
| Telenor Telecom Software | Made an expert system based on own empirical data for effort estimation and risk management, and modified roles. | Yes | Yes |
| Ericsson Software Technology | Set up new organisational roles to increase oral communication of experience. | Yes | Yes |
| Australian Telecom Company | Collected existing explicit information regarding software development and made it searchable. | No | Yes |
| ICL High Performance Systems | Introduced an Intranet-based system with an "engineering knowledge database" | No | Yes |
| ICL Finland | Made an Intranet-based system with three structural layers. | Yes | Yes |
| sd&m | Set up a knowledge management group and Intranet system. | Yes | Yes |

# 4 Research Goals, Method and Design

Now, we first specify our research goal from the problem outline given in the introduction, and then discuss the research method we applied in the work, and detail how the research process was carried out. Then, we discuss validity of the studies that we have done, and also discuss ethical considerations in this type of studies.

## 4.1 Research Goals

Our main research question that we stated in the introduction was:

*How can Intranet-based knowledge management tools be used in medium-sized software companies to facilitate a «learning software organisation»?*

As we have discussed in chapter 3, we can distinguish between four types of Intranet-based knowledge management tools: Knowledge Repositories and Libraries, Knowledge Cartography Tools, Knowledge Flow Tools, and Tools to support Communities of Knowledge Workers. We will be concerned with the two first types of tools in this thesis.

Reasons for focusing on this type of tools are:

- Tools that Support Communities of Knowledge Workers is a very wide research are, covering for example groupware tools.
- Knowledge Flow Tools are not applied widely in companies that develop software.

We have earlier defined a «learning software organisation» as an organisation that uses such tools as mentioned above in an efficient manner to transfer knowledge between development projects, and put that knowledge in use.

That is, the main goal of this work is to investigate what characterizes a set of working or non-working knowledge management tools for com-panies that develop software. We think that this depends on technological, organizational, as well as individual factors amongst the users. What are the reasons for that? Knowledge Management tools are of course technology. The technical solutions will decide what kind of knowledge can be managed, and what type of management that is possible. The organisation of the company will influence what kind of knowledge that will pass through the tools. For example, if the people who are working with one topic sit in the same room, the need for making knowledge explicit is not pressingly large. It can also make a difference when a topic is relevant for an organisation. Work processes in the company can also be a major influence, which we can call an organisational factor. Finally, the individual users will be the ones that decide whether or not to use the available tools, and their attitude to the tools will therefore be crucial in getting them into use.

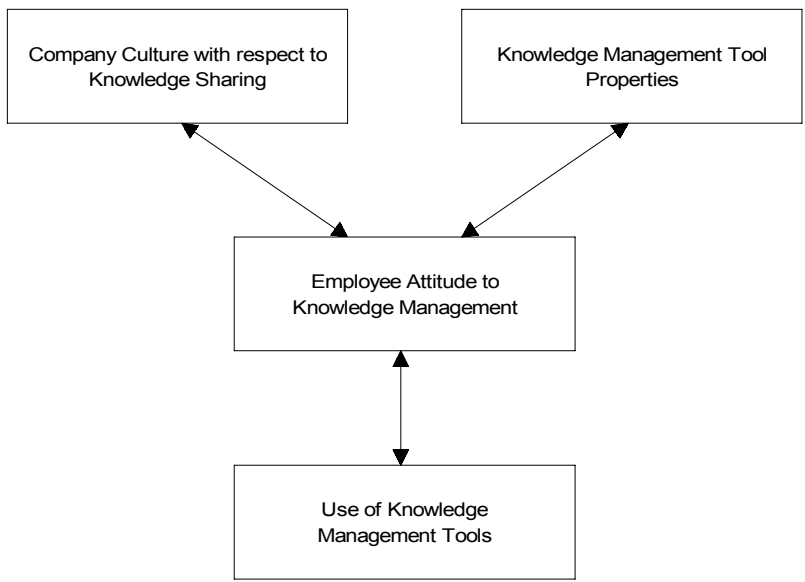

*Figure 4.1: Relationships between company culture to knowledge sharing, properties of computer tools, attitudes of employees, and actual use of knowledge management tools. Note that the arrows point both ways, indicating a relation, not a one-way causal relationship.*

We have represented the relationship between these factors in Figure 4.1.

When we were working with software companies in the Software Process Improvement for better Quality project (Conradi, 1996), we tried to introduce knowledge management programs for codification in various companies, but the projects often failed due to non-technical issues. In one company, the management decided to «hold» some improvement projects we were involved in, because they had reorganised the company and had found another «competing» improvement «strategy». This meant that an initiative that had shown promising results was abandoned. In another company, we had problems due to the change of contact persons - when one person quit, the company did not have any other «key person» to take over, and so the initiative lost momentum, as many discussions had to start again from scratch.

After this, we realized that it is a very complex task to introduce a knowledge management program, where cultural aspects can be far more important than more technical ones. In order to investigate this further, we formulate the following research question:

*R1)    Which factors influence the success of codification strategies for Intranet-based knowledge management tools in medium-sized companies that develop software?*

With this, we aim to find factors that influence the upper left box in Figure 4.1; the organisational factors.

We also want to know more about what tools for knowledge management are in use in software engineering companies. We are then interested in what kind of knowledge carriers exists in these systems, what situations are they used in, what benefits or drawbacks do the developers and management in these companies experience from their use? (This is the lower box in Figure 4.1).

We think that we should be able to design better knowledge management tools in the future, if we know more about how the ones that exist work. Or, of course, we could discover that knowledge management is no more than a marketing hype, and has no real influence in the organisations we study.

To learn more about this, we state the following research question:

*R2)     How do different groups of users in medium-sized consultancy organisations use Intranet-based knowledge management tools to transfer knowledge between software development projects?*

With this question, we want to learn about the purposes that different tools are used for, and if there are differences in tool use based on factors like the position in the organisation.

## 4.2    Research Process and Methods

Although the research reported in this thesis was done in an iterative manner, we can divide the work into three major research phases:

- A literature study.
- A prestudy of knowledge management tools in a research project setting (using action research).
- A main study of knowledge management in one company with another company as a contrast case (using etnography and grounded theory).

The second phase was aimed at research question one, while the third phase was aimed at research question two. We have tried to illustrate this process in Figure 4.2.

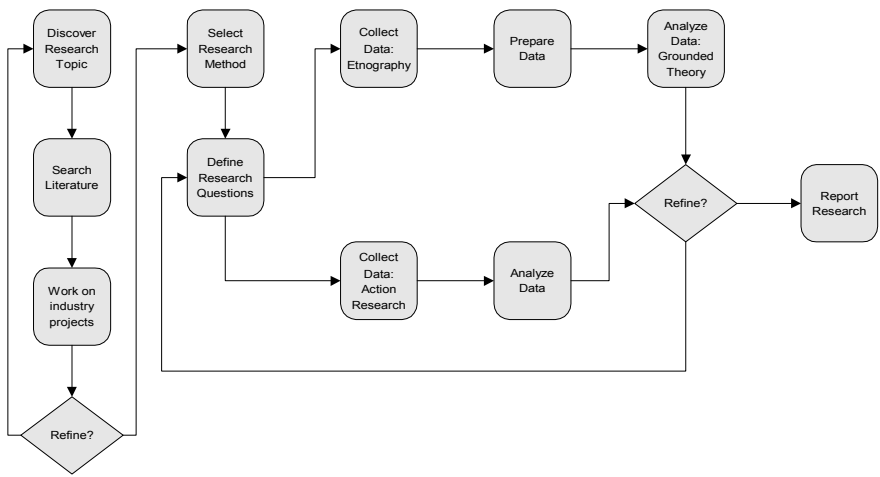

*Figure 4.2: The Research Process that was used for this work.*

So, what kind of research method is appropriate to the research questions we have outlined? The literature study part must obviously be done as a literature study, but we have other options for the two later phases.

In the prestudy, we benefited from working in a research project that involved several companies. The method applied here was action research - in that we worked together with companies to achieve improvement goals. The research method for this phase was then given by the surrounding project.

In the third phase we wanted to investigate issues that happen in a real organization, so we preferred not to carry out controlled experiments. Also, the amount of work previously published in this area is somewhat limited, so we cannot rely solely on material that is already published. To obtain reliable research information from companies, it is necessary with mutual trust. We therefore decided to actively collect the data for the research from a real organization. The company would then also benefit from improving their understanding of how their knowledge management tools work.

Should we then proceed in participation with the industry using action research as in the prestudy, or rely more on observation? First of all, it is a need to know more about what the practice in knowledge management actually is, and not to introduce more new systems without analysing how the old ones behave. Second, it is difficult within the time frame of a PhD work to set up research collaboration that is supposed to give immediate results to the research partners (this research question was not addressed by the research project as a whole). Therefore we opted for an observational method.

To further specify the research method, we would collect data on issues that organizations themselves normally would not collect, so project monitoring was not a suitable method. Also, using assertion was out of the question, as we would have to do the data collection ourselves. Then the issue was to choose whether we would like to do a broad survey in several organizations or to do a deeper survey of a limited set of organizations. If we did a broad survey, of course the results that we would get would be more general than from a limited study. But on the other hand, the issues we are interested in are not fully developed in all organizations, and thus most of the cases might be giving little or no new information.

Also, we had relatively little industrial experience, and we felt that doing a deeper study in a limited number of organizations would give us more insight into the problems that the industry is having. We therefore chose to do case studies, and then the next question was to decide the number of cases. We wanted to limit the number as to get as deeply into the organization as possible, but on the other hand, we would like to have more cases in order to see differences between organizations. The result was then to concentrate on one «deep» case and use another «lighter» case for contrasts.

Which method should we apply to investigate the phenomenon and to examine the boundaries between the phenomena and the context? In general, we wanted to get a good insight into the real-life situation, and to use this for building theories. We also wanted to rely on different data sources in order to make our analysis more reliable. We were inspired by methods from ethnography in the collection of data, and grounded theory in the analysis. These methods were introduced in chapter 2. Why did we not choose to rely solely on questionnaires and make a more qualitative approach? There are several reasons for this. In order to make a good questionnaire you need to know about the domain that you are going to investigate. We had not been working in «real life» in this domain, and we then needed more background information to focus questionnaires. As discussed in chapter 3, there are not many deep case studies available already that would give us such an insight. We therefore would have to use a more qualitative approach anyway, and decided rather to do this in a thorough manner than to do a light study first, and then maybe ending up with questionnaires that are a bit irrelevant for industry practice. So we decided to use mostly a qualitative approach, but supported with quantitative data as well.

## 4.2.1 Selecting Cases for the Literature Study

In the literature study which is reported in section 3.4, we selected a set of papers that were found through searches in databases such as Inspec, Science Citation Index, ACM Digital Library and the IEEE Computer Digital Library. We also searched through proceedings from the last five years of conferences like the International Conference on Software Engineering, The Software Engineering and Knowledge Engineering conference, the International Conference on Product Focused Software

Process Improvement and the International Conference on Case-Based Reasoning manually. Some papers were found after suggestions from others, or from references from other papers. We used keywords as «knowledge management», «corporate memory» and «experience factory» together with keywords like «software engineering» and «software process improvement» in searches. Also, we used a list of 20 knowledge management tools (Goodall, 1999), like «grapeVINE», «KnowMan», and «SemioMap» to see if we could find articles reporting experience with those tools in companies that develop software.

Limitations of this strategy are that we rely on the same understanding of keywords - if there are other papers describing the same topic but using a different vocabulary we would not find them. Another limitation is that it is very common to publish success stories, and not so common to publish results that would either compromise a method, a firm or an organisation. Some researchers might also not consider studies of failures as interesting for the scientific community.

Of course, all the papers were written for some purpose, which does not necessarily correspond with the purpose we had for analysis. Therefore, the papers may contain incomplete information, or the information might be reported using other terminology than we expect.

## 4.2.2 Selecting Cases for the Prestudy

In the prestudy, we were working with four companies that were interested in knowledge management through codification strategies. We therefore had companies that were willing to participate in the project, that the project as a whole picked, and we did not influence ourselves. Did this lead to any special bias? The aim of the project was to work with improvement projects in small- and medium-sized companies, and all of the four companies can be described as medium-sized. Before starting to work with them, we did not know if these were «typical» in their Knowledge Management initiatives, or were special in any kind. When drawing conclusions from this study, we should then be careful when generalising.

### 4.2.3 Selecting the Main Case and a Contrast Case

The two cases were selected on the basis of earlier collaboration in research projects, and on their geographic location. One company had been working with the issues we wanted to investigate for a longer period of time, and this was chosen to be the «deep» case. Another company was just starting up work on the selected issues, and this was chosen as a contrast case.

Reasons for choosing these two was both because they were actively working on the issues we were interested in, and they had an Intranet system working for some time. They were in reach geographically, and willing to let a researcher work in one of their projects to collect data. In both companies, we asked to follow a project that was «typical» for them, and involved a high degree of software development. It was then the companies that suggested the projects that were chosen. To get more into the internal life in the projects, we asked to be assigned roles that we would be able to fill, for example to participate in testing of the product they were to deliver, or acting as a project assistant. In the «deep» case we helped with testing the final project.

Note that the company that is referred to as Alpha in the main study is the same company as Company Three in the prestudy. But these two investigations was done at different times, and involved different people in the company, so we have kept them separate in the following.

### 4.2.4 Collecting Data

We got an office in each of the companies for the time we stayed there, for 4 weeks with Alpha and 2 weeks with Beta. We also got access to their Intranet systems and attended all meetings where all the employees were invited as well as project meetings. In each company, we were allowed to introduce ourselves to the project staff, and explained the purpose of our stay as to «collect data for a thesis on knowledge management in software engineering companies».

We wanted to use multiple methods to collect data, and decided to use the following data sources:

- Interviews - we used semi-structured interviews with open-ended questions to be sure that we covered all the issues of study, to allow the respondents to speak more freely of related issues they felt were important, and to let the interviewer develop the interviews over time. We deliberately created stops in the interviews when writing down what the respondents were saying in order to give the respondent the possibility to elaborate further. The interviews lasted from 25 to 90 minutes, with the typical length being around 40 minutes. At Alpha we interviewed the project manager and the three people participating in the project we followed, the person responsible for knowledge management, three other management representatives: persons responsible for business consulting, international operations and the internal competency centre. We also interviewed six people who had won an internal prize for sharing their knowledge with others, making it a total of 14 interviews. At Beta we interviewed the project manager and two people working on the project, the responsible for knowledge management in the company, and the former responsible for an internal discussion group on software development. The interviews were recorded on MiniDisc and then transcribed. When transcribing, we tried to stay as close to the original material as possible, using italics for words that was emphasized, using exclamation marks for sentences that was said in a humorous tone, and using three dots to indicate that there was a short pause in the interview when the respondent was thinking. The transcripts were sent back to the people who had been interviewed for corrections or further comments. We got feedback from around 50% of the people interviewed, mostly minor corrections. The interviews were done in Norwegian, and the citations that are used in this thesis is our translation of what was said. The interview-guides are given in Appendix B. In total, we got around 150 pages of transcripts of interviews for analysis.
- Usage Logs - we collected logs from the usage of the Knowledge Management system on the Intranet www-pages (from Alpha only).
- Documents - we gathered documents about the design and intent of the Knowledge Management tools. We also collected documents from the projects where we were participating.

- Screenshots - we gathered screen-shots from different areas of the knowledge management system.
- Pictures - we took pictures of people in normal work-situations to get a better understanding of the workplace and work processes.
- Logbook - we wrote down observations from everyday life in the company in a logbook, together with memorandums from conversations we had, meetings and presentations we attended.

## 4.2.5 Analysing the Data

We analysed the qualitative data using principles from grounded theory. We also had some quantitative data in logs, which we first had to pre-process before we could plot them for analysis, as described below. Note that not all categories that are presented later were derived using action research. Some have been taken from the literature on knowledge management, and some from the organisations where the data was collected.

### «Coding»

How did we organise the analysis of the data that was collected? First, we gathered the qualitative material that was collected for each of the main topics of interest that we had. We constructed a database with information from the interviews, documents, and our own logbook observations. We tagged the information to show what kind of source it came from, and applied a simple categorisation of the people that was interviewed: managers, project managers, developers, and people responsible for knowledge management. In the Alpha case we also had one category for the «knowledge sharers of the month».

We searched in this database for areas of interest, and got the information from the different sources. For example, searching this database for the keyword «skill» would result in 43 occurrences in 10 documents.

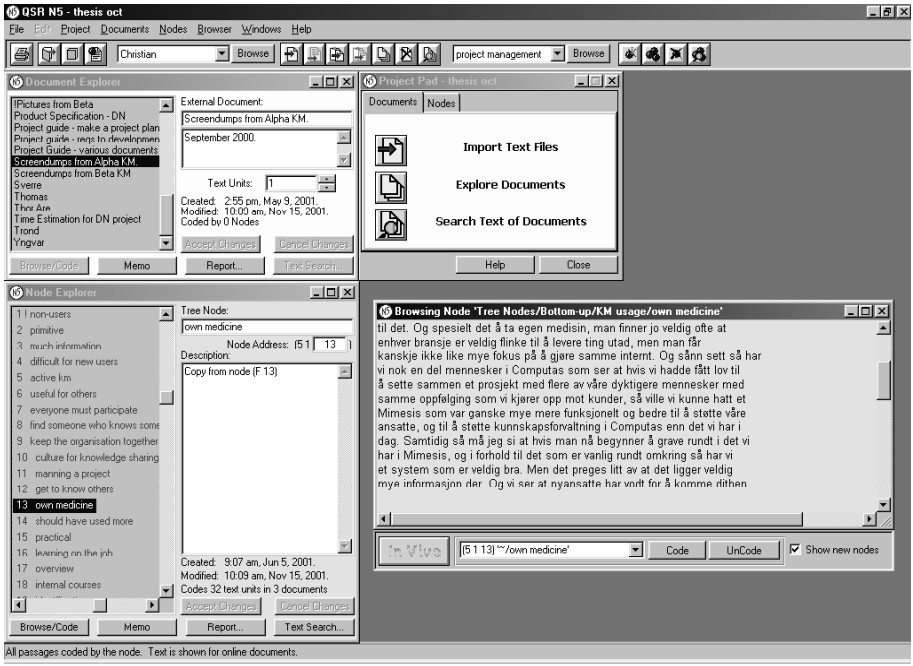

*Figure 4.3: A Screenshot of N5 - a Tool for Analysis of Non-Numerical Data.*

After that, we analysed (and «coded») these chunks of information to find interesting categories that would be usable to build theory later. Would there be any special patterns in what the people were saying? We applied triangulation to see if there were differences between groups of people or between what people were saying and logs or collected documents.

To structure the data, we used a tool for analysis of non-numerical data: N5[2]. In Figure 4.3 we see a screenshot of this tool, with a window in the lower left corner with a text where we can «code» each separate line. On the lower right, we see a group of «Nodes» - concepts that point to lines in the text that are «coded». In the screenshot, the text is about how people in a company described internal knowledge management use as a form of «taking their own medicine».

---

[2] N5 is available from QSR International (http://www.qsr.com.au)

## Usage logs: Preprocessing and Plotting

The usage logs were taken from a web server, and showed all look-ups to the www-pages that was available on the Intranet of Alpha. The logs were on a format like this:

```
194.19.98.236, USER, 4/5/00, 3:06:08, W3SVC1, MIMESIS,
194.19.98.20, 0, 414, 142, 304, 0, GET, /FILE

194.19.98.236, USER, 4/5/00, 3:06:08, W3SVC1, MIMESIS,
194.19.98.20, 0, 406, 142, 304, 0, GET, /FILE
```

In these logs, «user», is the user that requested a download of a webpage, file is the file containing the web-page, and we can also see the date and time when the file was downloaded as well as the IP-address of the computer where the file was downloaded from. We had almost 600 Megabytes of log data, and could not analyse the whole of it. We chose to remove all data except what was relevant to the most interesting tools. Then, we wanted to count how many times a tool was «used». We sorted the filtered logs according to user, and removed duplicate accesses from the same user within eleven minutes. This, to prevent counting an automatic «reload» of a web-page as a «use situation» of a tool. We also produced data without this filtering mechanism to study the effect of the filtering, and give this data in Appendix B. After filtering the information in this manner in a script, we imported the data in Excel and plotted how many times a tool was accessed per week during the time period where we had logs available.

Note that we define «usage» as a «look-up» of a web-page - the front page of a «tool». It does not necessarily imply that the knowledge on that page has been applied.

Also note that we only got logs from one company, Alpha, and did not get logs for all tools. The logs we have are from weeks 14 - 30 and 36 - 39 in year 2000.

## 4.3   Validity Considerations

Here we discuss the limitations in this study. What can we say about the conclusions that we make in the coming chapters? How reliable will they be? We first discuss the action research in the prestudy, and then evaluate the main study according to principles defined in (Yin, 1994).

## 4.3.1 Action Research in the Prestudy

Potential problems with this kind of research is that it can easily be biased, in that everyone is interested in reaching the goals that are set up. Thus, we do not know if the same results would be achieved with another set of researchers, or with other people from the company, or with another company in the same situation. But this kind of research allows interaction with companies in a way that would not be possible if it was not so much in the company's interest.

## 4.3.2 Main Study: Construct Validity

How do we ensure that we really measure the concepts that we intend to study? We have defined a set of areas of interest in the previous chapter, and how can we be sure that the data we collected really say something about what we intended to study?

First of all, we spent some time at each of the fieldwork sites before starting data collection. Thereby we got into the normal work procedures in the company and got a better understanding of which data sources to select and which questions to ask in interviews.

To increase the construct validity, we decided to take two actions: First, to use several data sources so that a bias in one source would not necessarily lead us to wrong conclusions. Second, we had both some of the data (transcripts of interviews) and the final conclusions reviewed by key informants.

## 4.3.3 Main Study: Internal Validity

Our intention in this study, is to explain and develop theory about the use of certain computer systems in some companies. It is therefore crucial to ask: how do we ensure that our explanations are really rooted in the issue that we intend to study, and not in other schemes or concepts in the companies. To handle this, we have done a very data-near analysis, using grounded theory as explained above. When we go on to discuss the material from the case, we will make extensive use of the data sources to show how we have constructed theoretical positions.

### 4.3.4 Main Study: External Validity

To which domains can the results from this study be generalised? We did fieldwork in two companies that develop software in Norway, that are quite small in an international context. The work culture is probably different in a Norwegian setting from for example an American one. It is hard to generalise the results from this type of a study, but companies that are about the same size, produce software, and have a quite flat structure (a «process organisation») might experience similar situations. The topics of study should also be interesting for companies in other domains that are working with knowledge management.

We decided to use a contrast case to increase the external validity of this study, but chose a limited number of cases because we think the research field has a higher need for thorough cases than more wide studies.

### 4.3.5 Main Study: Reliability

Could this study have been repeated with similar results? The procedure for collecting data for the study has been described in this chapter, and the questions for the semi-structured interviews can be found in Appendix A. Deviations from the interview structures can be studied in the transcripts of the interviews. We have also described which considerations we have taken in selecting the cases. For the analysis part, we have constructed a project in the N5 tool which contain processing of the research data. We therefore claim that this study has a high reliability.

## 4.4  Ethical considerations

What about the ethics in doing this type of research? One danger would be to expose material that was meant for research purposes to others. We have always explained to the respondents that the material from the interviews only would be used for writing this thesis, if we wanted to use it for other purposes, we would contact them again. When we have used the material here, we have used quotations anonymously, so it would be difficult to trace them back to persons or companies.

The logs that we gathered have only been analysed at a high level - no analysis has been done on individuals.

The project that we followed in each of the companies are of course known to management, and thus the information from them would not be completely anonymous. However, we always interview several persons in a project, so for example criticism of a company policy by a developer is at least it is not traceable back to an individual.

We have decided to keep the companies and all contact persons in the companies anonymous, not because this thesis contains information that is particularly sensitive for the companies, but because it is not so relevant from a research perspective.

# 5 Empirical Investigation

We first present findings from the prestudy on four codification initiatives, then we present the companies where we did our main study: Alpha and Beta. We also describe a project at each company to give a better impression of what the companies are doing. Finally, we present knowledge management tools at Alpha and Beta and how they are used.

## 5.1   Prestudy: Four Codification Initiatives

We go through the cases, and present some background on each company, what kind of software platform they use for development and then what improvement goal(s) and what codification initiative they introduced. We have kept the company names anonymous, and refer to them as companies One to Four.

The initiatives will be further discussed in chapter 6.2.

### 5.1.1   Company One

Background: Company One is a telecom software house, with 600 developers. It is ISO-9001 certified, and is owned by Norway's largest national telecom provider. Its main profile is administrative support systems for telecom, like logistics, personnel, and billing. It has developed and operates a dozen large information systems, e.g. developed in Oracle 2000-Designer. Company One introduced a web-based quality system in 1995, mainly a bought-in, «canned» process.

Software platform: COBOL, C++, Java, 4GLs. Mainframe, Unix, PC.

Improvement goals: Improve estimation accuracy by 10%.

Codification initiative: A project database and an associated estimation tool were made, using spreadsheet technology. This was linked to the existing, web-based quality system. The estimation tool offers seven different algorithms, mostly based on Function Points, and is based on data from 50 previous projects. It is aimed at project managers, who have been given a one-day course in the tool. A central method group of 4 persons maintain the project database and the estimation tool (Jørgensen et al., 1998).

Results/experience: First, the quality system was mostly introduced «over the head» of people. For example, people were required to write final project reports that were hardly ever used later - a rather demotivating fact. Further, even though the estimation tool gives 10% better accuracy than manual, ad-hoc estimation, and even though project managers have been trained in it, this has not been in extensive use. However, the majority of project managers are positive to start using the tool, according to an internal poll. In all, much synthesized knowledge has been collected and made easily available to key persons, but actual use and reuse of this information has been meagre. On the other hand, all improvement efforts in Company One have been hampered by major reorganizations, and that the key estimation guru resigned.

## 5.1.2 Company Two

Background: Company Two produces software for the bank/finance market, and has around 250 software developers. It is ISO-9001 certified. The software development is organized in large projects that are monitored by a project office. This office is responsible for collecting progress reports, updating process models that are in use, and for collecting experience from projects after completion. The project office is also in charge of the quality system and for resource allocation.

Software platform: COBOL, C, Java, 4GLs. Mainframe, Unix, PC.

Improvement goals: Reduce overruns in projects by better estimation/planning techniques, using a project database.

Codification initiative: The project database was designed and implemented in Oracle 2000-Designer by an undergraduate computer science student, based on requirements from the company. Central data were project profile, project size and function point data. Estimation assistance would be by analogy, looking up similar domains or tool platforms, previous budget/schedule overruns (cf. case-based reasoning). Also, information on risk analysis, estimation and general experiences could be stored.

Results/experience: The experience database was never put into use, because of reorganization in the company and because of financial problems.

## 5.1.3 Company Three

Background: Company Three is a consultancy company with 150 developers, mostly with MSc/PhD degrees in computer science. It focuses strongly on object-oriented, user interaction, and artificial intelligence technologies, and uses the DSDM[3] method for incremental development. It has a flat and «process-oriented" organization.

Software platform: C++, Java, SmallTalk, 4GLs. Unix and PC.

Improvement goals: Make relevant company information more accessible to support the business.

Codification initiative: Company Three has developed a web-based corporate memory tool. This stores administrative information, experience notes, personnel competence profiles, overall project routines (not a full quality system), and day-to-day news and events. It includes a competence base, where all employees are listed, and their present and desired competence areas are indicated. This information is used to allocate people to projects. Very few hard data are collected and stored, except major project data. Processes and roles have also been defined.

Results/experience: For its limited ambition, the tool is functioning fine, and is well received by developers and management. It is an advantage that the company has a flat organization and has already much insight in knowledge engineering, even though the corporate memory presently is very low-tech.

## 5.1.4 Company Four

Background: Company Four is a consultancy company with more than 400 developers, being Norway's third largest and with five branch offices. It has a central method department, with consultants responsible for different technology areas and business domains.

---

[3]www.dsdm.org

Software platform: C++, Java, COBOL, 4GLs, web-tools. Mainframe, Unix, and PC.

Improvement goals: Increase its competitiveness, by making updated methods and related experience more easily accessible for the consultants, that often work at customer sites. A subgoal is to improve project estimation by a new tool and related project base.

Codification initiative: Company Four has developed a web-based Information Well using the Microsoft Exchange repository tool (Halvorsen and Nguyen, 1999). This knowledge base stores general company and personnel information, such as strategies, meetings, various documents, and individual CVs. It also stores recommended routines and work methods («best practices»), as well as experience from using the company's methods and tools in different application domains. All personnel are responsible to develop, publish and adapt the stored material, but special method and domain specialists receive feedback on, quality check and revise certain key method documents.

Results/experience: The Information Well has been in use for over two years. Measures regarding enhancement and use are regularly recorded. Annual, internal surveys conclude that the Information Well is increasingly being used and accepted by Company Four consultants. A technical drawback is that the stored documents exist in different document formats (Powerpoint, Word, etc.) and versions of these formats, being incompatible with the tools installed in each consultant's computer. The planned extension with the estimation tool has been stopped, due to company reorganization and resignation of two key persons.

## 5.2   Main Study: Alpha and Beta

The companies where we collected our material are both software consulting companies that we can call «medium-sized». Alpha has around 150 employees, and Beta around 200 in Norway, and around 40 at the site where we did fieldwork. Both of the companies offer consulting services, and both of them do most of the work «in-house», and not at their customers' offices.

We first present Alpha Consulting and a project we followed when doing fieldwork there, to give a further impression of the company. Then, we present Beta and one of their projects.

## 5.2.1 Alpha Consulting

Alpha Consulting («Alpha») is a consulting company based in Norway, developing knowledge-based systems for a variety of customers. When it was founded in 1985, it was a spin-off of a larger, more general consulting company, and according to a Norwegian newspaper[4], «an international staff of specialists will develop expert systems that above all will cover the needs of the demanding oil industry». In the same article, the newspaper continues: the company shall «offer services in industrial use of knowledge-based expert systems, and software in the field of artificial intelligence».

Since then, the company has grown organically, from just a few employees in the beginning, to around 150 in year 2000 (A small company in another city than where Alpha has offices was bought in 2000). The company has also extended their services and market.

In the annual report for 1999, they state that their vision is to «make knowledge sharing in organizations more effective, and thereby contribute so that knowledge is refined and used to achieve the organization's goal». Their mission is to «deliver services, solutions and products to organizations and individuals who wish to make their business more effective through innovative use of information technology. The company's core competence is knowledge management, process-support and implementation of intelligent systems for knowledge-based behavior and knowledge processes. Within this business area, Alpha will seek international activity based on their role as a leading vendor in Norway». In July 2001, the company was in discussions with Boeing about delivering a system for modelling software and organisations[5].

Important technology for delivering these solutions, are «network and database technology, document management and search, web technology, work process support, coordination technology, artificial intelligence and data mining». The underlying technology for this again is Java, Microsoft and SmallTalk technology.

---

[4]Aftenposten, 7th of March, 1985.

[5]Dagens Næringsliv, 2nd of July, 2001.

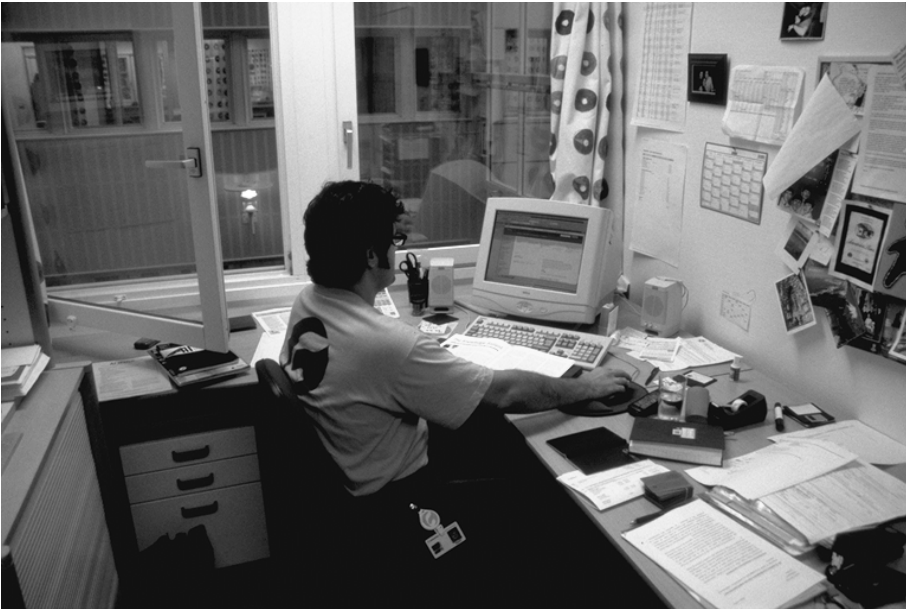

*Figure 5.1: A manager at Alpha working in his office. Most of the employees sit in offices like this, and around 20% work at a customers' offices.*

Customers come from three main groups, the public sector, the marine sector and industry. Projects for these customers typically include 3-10 people working for at least half a year, and in some cases for several years. In projects, the participants take on different roles, as «project manager», «technical manager», and «customer contact». In addition to these projects, the company has a record of participating in different research projects, from highly applied research, like in the Eureka program, to more advanced research in EU- and Norwegian Research Council-funded projects.

The company is organized around «processes» and «projects». The «process organisation» means that they have defined important areas for the company, which has one «process manager», usually with support from a small team. Examples of processes are «Management», «Delivery» and «Support», and also «Knowledge Management». Many employees in the company are responsible for some process issue while working on a project. Most employees have a university degree in Computer Science, and many have a PhD degree, especially in Artificial Intelligence.

The Knowledge Management Process at Alpha is handing out a prize to the «knowledge sharer of the month» in order to promote knowledge management. This prize has been given to people who share their knowledge through Alpha's knowledge management tools, or through oral communication.

On first sight, the organisation seems very «flat» - with people rotating between different «process manager» positions. But as one employee told us, «of course, there is a hierarcy here as well, it is just not written down any place».

When working in projects, most of the development has traditionally been done «in-house», and not at the customers site. But it is now getting more frequent that employees work in the customer companies. When we were visiting the company, around 20% of the staff was working somewhere else than in the main company building.

The building where the people work is located in an «engineering valley» outside a major city in Norway. Most employees have their own office, but some have chosen to sit together because they work on the same project. In general, people have offices located close to other people that work on the same project. Informal discussions in the corridors are promoted; in all floors in the office building, we find coffee-brewers, water dispensers as well as tables and chairs. In some floors, we also find space for leisure-activities such as dart. Another striking thing about the company is that they have focused on getting everyone out of their offices for lunch. A chef is preparing warm and cold meals that the employees pay a small amount for each month. Virtually everyone comes for lunch, which is very rare compared to other companies, where time pressure is strict and it is common to eat lunch while working in the office.

## 5.2.2 The project: «ExWorks»

We spent four weeks in this company, and followed a project that was developing a web-based system for executive work for a government department in Norway. We participated in all meetings in this period, and also helped with testing a part of the system. In the following, we will refer to this project as ExWorks.

Now, we will first describe the contents of the project in more detail, and then the people who participated in the project.

The government department that wanted ExWorks, has many offices, that geographically are very dispersed. But all offices handle a type of applications from «customers», where some can be treated locally, but many will involve the central department, as well as other public departments. The software system that was constructed allowed remote offices of the department to submit applications via a web-interface, and this application would then be treated according to several rules before the «customer» would get feedback, either automatically, or if required from an executive officer.

The size of the project was around 8000 work-hours. 2000 hours was due to changes in the project.

Central technology in the project was first Websphere from IBM, as a server platform. But due to some problems with this type of technology it was changed to Gemstone, a tool for «building large-scale, distributed application architectures for the enterprise». Much of the code for the project was written in Visual Age for Java.

Four people were working in the project when we visited the company; one project manager that had worked at Alpha for five years, ant three developers that had worked there for four, three and one year respectively.

This project started in January 2000, and we followed it through October 2000. It finished with a «final project review» in July 2001.

### 5.2.3  Beta Consulting

Beta Consulting («Beta») is an Internet consulting company, listed on the stock exchange in Stockholm. The company is a result of a large merger in 1999 of 10 companies in Sweden, Norway and Finland, and has two offices in Norway. Their mission statement is: «Beta is an Internet consulting firm that wants to take an active role in the global restructuring of the consulting market. Our mission is to become one of the top five consulting firms in Europe».

In their annual report from 1999, they describe their «business concept» as «with a focus on the Internet, (Beta Consulting) helps innovative companies to identify and exploit new business opportunities by developing and implementing comprehensive IT-based communication solutions and business systems».

Their focus areas are «strategy, communication strategy, design and user interface, on-line applications, back-office applications, system integration and hosting». If we concentrate on the areas involving software development we find e-commerce systems, information portals, internet banks and publishing systems as parts of the on-line applications. As back-office applications they work with knowledge management, Intranets, contact centres, and e-learning.

Underlying technology is a web-publishing system, «Beta Publishing», that the company has developed internally. Another technology that is important is active server pages which are supported by script-languages, where the company is using their own language in addition to JavaScript. Earlier, languages such as Perl were used, as well as C. Most of the software runs on a Microsoft platform.

The customers are Banks, Telecom Companies and other large industrial groups as well as the public sector.

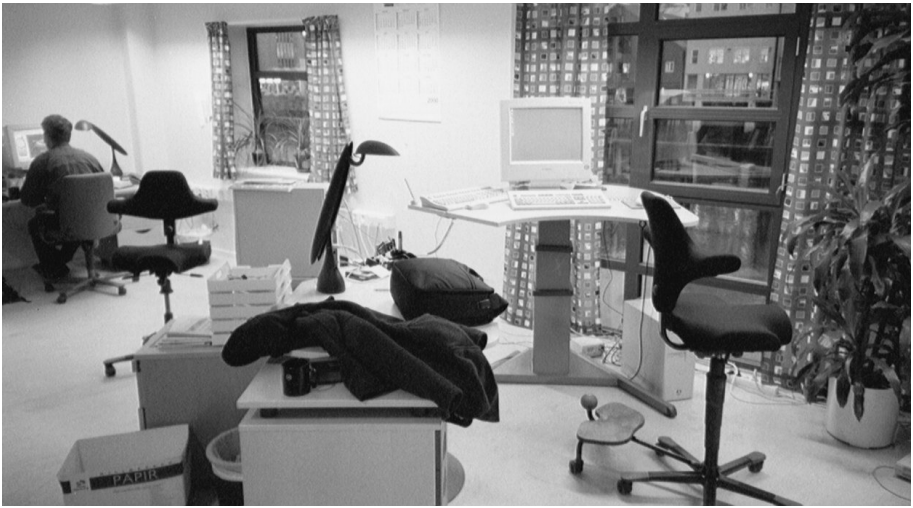

*Figure 5.2: An office room at Beta, where two people are working. Most of the people work with software development work in an environment like this.*

The Company has around 200 employees in Norway, and about 40 of them work at the place we visited. This part of the company was the result of a merger of two companies in 1999, and one of the mergers was also the result of a previous merger some years earlier. When Beta con-sulting was formed at this place, a newspaper wrote:[6] «The Common denominator for the two companies is development of business-critical Internet-solutions. But the two mergers have different starting points: <one of them> has first of all a heavy technological profile, whilst <the other> has had a focus on interfaces, business strategy and design of web-solutions. Now, the engineers in <the first company> and the designers/ advertising people in <the second company> will work under the same roof».

## 5.2.4 The Project: «IntraWeb»

We spent two weeks in this company, and followed a project for a public department in Norway. The aim of the project was to re-design the interface and technical solution of this department's external web-portal as well as their Intranet. The department had several solutions already, that were running on several servers. Some of the software was written in PHP.

The delivered system was an «administration-, editing-, archival-, and publishing-tool for publishing information on Internet, Intranet and Extranet». This meant customising an installation of Beta Publishing for this customer, and creating a new graphical design. It also meant to convert data from existing systems to the new one. About 50% of the work was development, and 50% design. Some work was also devoted to training people from the department to use the system.

The system included search possibility, and the possibility to publish information with or without the approval of an editor. It also supported different «thematical web-pages» for issues of interest, press releases and publications from the department. The whole system was to handle two versions of Norwegian (bokmål and nynorsk) as well as English.

The people participating in the project were two software developers, one who was new to the firm, and another who had worked there for several years. Also, a project manager and a graphical designer was involved. It was the first time that Beta Publishing was deployed in this section of Beta.

---

[6]Dagens Næringsliv, 5th of May, 1999.

We visited the company in early January 2001, and the project had then been running since January 2000. The final delivery was on the 1st of February 2001.

## 5.3   Usage of Knowledge Management Tools in Alpha and Beta

Now we will present our empirical material from Alpha and Beta according to the structure given by the research questions in the previous chapter. In question two, we wanted to know more about how tools are used, and will present each tool in the following with descriptons of usage situations (from the grounded theory analysis) and user groups (taken from the companies organisation), like in the Figure:

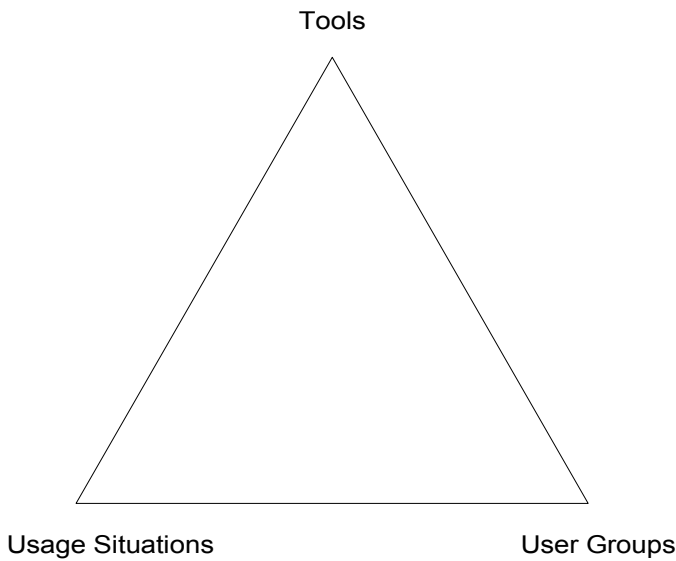

*Figure 5.3: We will use Tools, Usage Situations and User Groups to organise our empirical material from Alpha and Beta.*

Note that we do not follow this structure for all the tools, because we asked people an open question «how do you assess the tools you have available», and there are some tools that either very few people or none at all would mention. But we present all tools anyway, to give a complete overview of what is available in the companies, and do not write about usage situations or user groups where we do not have sufficient information.

We now present the tools that exist in the companies, and have divided them into three groups, if they are Knowledge Repositories and Libraries, Knowledge Cartography Tools, or are not so clearly linked to any of these two groups. This last category which consists of overviews of people in the companies, office locations and so on, we call «community building», and do not present them in such detail, because it is not so relevant for knowledge management. But for the other tools, we will present which usage situations we have found for each tool, as well as what user groups. We also give usage statistics for some of the tools where this was available.

We start by giving a general overview of the front page on the Intranet systems, and what people say about the tools in general. Then we present the groups of tools, and have divided this presentation into a «tool presentation», «usage situation» and «usage groups».

When we describe tools and usage in the following, most of the information is taken from our main case, Alpha. When we present information about Beta, we specifically say so, the other material is taken from Alpha. Note that we did not find as much use of the available tools at Beta as at Alpha, so in many cases information about usage situations and user groups are missing.

### 5.3.1 The Tools in General

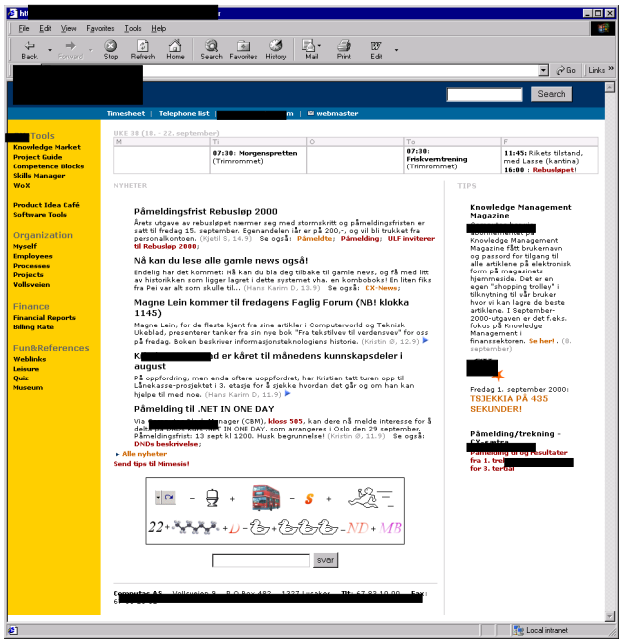

*Figure 5.4: A Screen-shot from the front page of the Intranet at Alpha.*

Now, we first present the tools in general, their usage and a general assessment, as well as statistics from tool usage for some tools at Alpha.

## Tool Presentation

When we enter the main www-page of the knowledge management system at Alpha, we have links to several different subsystems. The first thing we see is the company-internal news, which is placed in the middle of the screen. Above that, we see a calendar, which shows this week's events. On the left, we have links to several other www-pages: The Skills Manager, Competence Blocks, and the Knowledge Repository «WoX»,...

On the top of the page, we have links to each employee's timesheet, a telephone list, the external web pages, and the possibility to send an e-mail to the webmaster. On the right side, we have a «tip» about a knowledge management magazine, and a link to an informal «newspaper» which covers social events in the company. On the bottom of the screen, we see a «quiz of the day» - and we have the possibility to answer this quiz in the box below.

At Beta, we find a similar frontpage of the system. It contains four cate-
gories of news: «project news», «corporate news», «local news» and «staff
news» as well as a column from the director of the company.

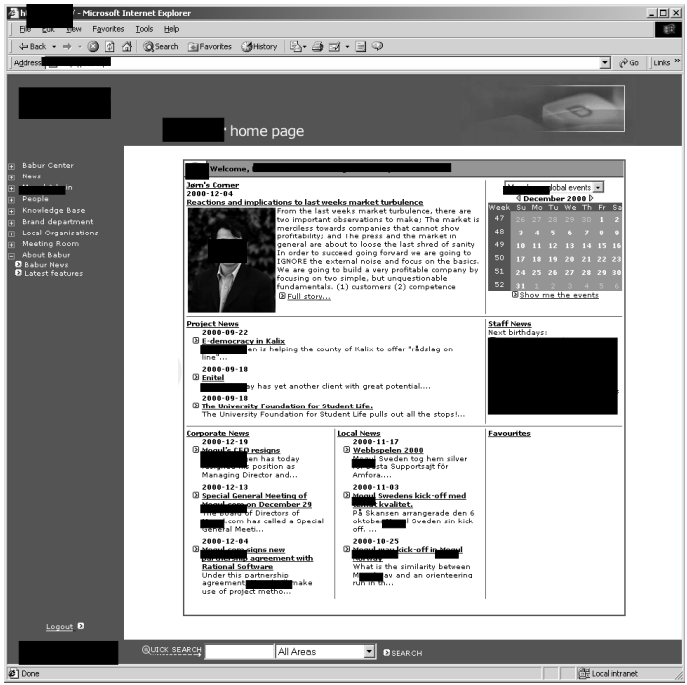

*Figure 5.5: The Front Page of the Intranet at Beta.*

We also find a calendar for the current month that shows upcoming
events in the company. On the bottom of the page we find a search
mechanism, and on the left a set of links to the knowledge management
tools such as the «knowledge base», list of people and their skills and
information on recommended work processes in the company. It also
contains a lot of «corporate information» on organisation and presenta-
tions of the visions and goals of the company.

## Usage

When we asked employees in the companies about how often they
would use the tools for knowledge management, most of the employees
from Alpha said they were using it several times a day. A developer said
«between 5 to 10 times a day», another said «some times a week to write
hours, it is always something you must do... look at news. If you want to
follow what is happening in the company, you have to look at it a couple of times a day. When I open Internet Explorer, it is the first page I get». Of other people we spoke to at Alpha, it seemed that most were using the tools «several times a day», some «daily» and a few «weekly». At Beta, however, all the people we spoke to in the project we followed said they were using it «weekly». «I admit that I just look at it once a week, to check if there is anything new», one developer said.

## General Assessment of Tools

When we asked people to assess the different tools that they have available for knowledge management in their daily work, we got a variety of answers. Some said that the tools that exist now are «primitive», and far from what the company thinks should be possible to use. Others said they worked fine, while others again think that they were «unpractical».

Several people in the company believe in more technically advanced knowledge management tools. One manger said «if we were allowed to set up a project with more of our skilled people, and followed up in the same way as we do against customers, then we would have had a (set of knowledge management tools) that was much more functional and supported our employees better, and supported knowledge management at Alpha better than what we have today». Another manager said: «it (the knowledge management system) is characterized by when it was made, and the need that has been in the organisation at different times. That is, it has been developed once, and has been patched-up a bit afterwards.» So, the technical condition of the system is not something that the company would sell to an external customer. This view is also supported by a developer, who said: «We have a number of tools that represent some good ideas, but the tools' condition today is not the ultimate. We see a lot of possibilities for improvement, especially on technology. What really could have done a difference is that we could have had much better integration between the tools». An example of tools that could be integrated better was the Skills Manager and the WoX (these will be discussed later). Other possible integrations could be between the Skills Manager and the Competence Blocks.

Other people emphasized that the tools are under constant development. A manager said: «it is under constant development, really - and when you get something new, you discover at once the need for something more.»

Several people mention that they would appreciate a more «active» kind of knowledge management, like one manager who said: «The problem is not that we do not document enough experience - but to make the experience appear when it is needed. It is ok in those situations when an employee experience that `now I need knowledge about something' - we could have improved the indexing possibilities (...) But if we had done so, it would be like that if I was thrown into a new project - or a newly employed was - and you get to know that you should do a relatively specific thing, then it could happen that you do some searches for knowledge on the essence on the job, but all the side-experience you have, you would not search for. I see it like the essence of the border of a bit passive form of knowledge management that (the knowledge management system supports).»

One developer said: «I only use the knowledge management system for writing hours, and doing smaller stuff. I do not think it is easy to find information there.» This was because this developer would normally need information whilst working on software development, and she felt it was time consuming to start a browser and look up a web-page for the internally developed framework she was mostly working with. Also, she meant that these web-pages were usually not updated, so she preferred to read code to find answers to problems.

Another user said: «I think the knowledge management system is a bit messy - I do not really know what is in there, because I have never had the time to go through everything».

Other were critical to an extensive use of tools: «Some people talk warmly about `taking our own medicine' by using work processes in development and things like that. That is just bullshit! Maybe it is a good thing for in-house training, but work processes is not the most effective way of working». This developer said that if you are an «expert» user, you have your own way of working that is «probably much better». Work processes would force you into a work pattern that does not suit you, because the way the company is modelling work patterns is «extremely static».

Another developer said that the contents of the tools are «much more up to date than you would expect». He thinks this is because much of the information is generated from databases that are easier to maintain than webpages.

# Statistics

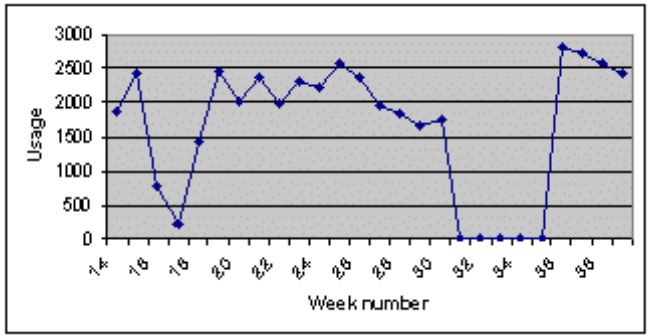

*Figure 5.6: Usage of the Front Page of the Intranet at Alpha over Time.*

The front page of the Intranet at Alpha was accessed as in Figure 5.6. Over the time period shown, it was used an average of 2032 times per week, which is approximately 14 times per week per employee. Note that we did not obtain logs for weeks 31-35. The decrease in usage in weeks 16 - 18 is due to public holidays and company-internal courses.

## 5.3.2 Knowledge Repository and Library Tools

In this group, we have found the following tools at Alpha: The Project Guide, The Well of Experience, The Knowledge Market, Overview of Processes and Overview of Projects. When we have found similar tools at Beta, we give a description of these in the same section as the Alpha tool.

## Project Guide

### Tool Presentation

This is a practical guide to assist project work, which contains descriptions of different processes that are common, such as project start-up and closure, how to do testing and so on. It contains templates for documents that are normally produced during project execution, as well as examples. Different company roles, such as developer, manager and customer contact have different views to the guide.

A manager describes this tool as «it has a form that is very nice - initiatives on peptalks when projects start and such... it is really a step in the right direction, that things are triggered by the system, and that people do not just know how to do things». Another manager described it as «a result of a lot of projects, and some routines and terms around it is an indirect result».

At Beta, we find a similar tool called the «Beta Way», which describes work processes and roles and gives templates and checklists.

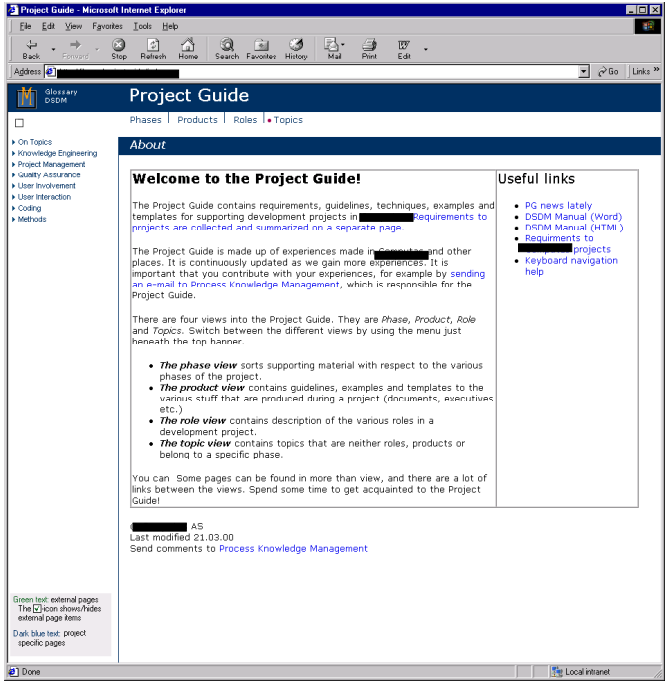

*Figure 5.7: The front of the project guide - where you can choose a «phase», «product», «role» or «topic» view.*

### Usage Groups

Many people at Alpha say that they do not use this tool very often. One manager said «I must say that this is a tool that I might have used more. And when I say that, I suppose there are other people as well that could have used it more». A developer said that «no, I do not use that... or at least not deliberately, but I suppose that there are many things that we do that you can find in the project guide». Another developer said «no, there is no need for me to use it. It is maybe aimed more towards project managers, but to be honest I have not used it as project manager either. Maybe because the projects have been too small. Or that it has been clever people on the projects that have not needed any training». An-other developer had problems with the form of the project guide: «I do not like it a lot, maybe because it is available electronically». This developer felt that he lost overview when reading hypertext documents: For example when investigating about «acceptance tests», «it was a long list of subpoints that you could click on. But you never get through such a list - it is too much! And I am a bit uncertain because it looks like a whole book, and if I pick out a piece to read it, do I have to read everything before it?» A third developer said she got «angry when using it», because it did not contain a complete set of information, and is difficult to navigate in.

### Usage Situations

We found that people in the company had one way of using this tool:

*       Tips and advice in project start-up and execution

A manager said that he «use it as a daily support - how to solve projects in general, and when we needed an acceptance test earlier in the project, we had a look there to see what tips and advice we could find».

# Knowledge Repository: The WoX

### Tool Presentation

*Figure 5.8: The «Well of experience» (WoX) search interface for the knowledge repository of «experience notes».*

The WoX, or «Well of experience», is a small tool for capturing knowledge that would normally be written on yellow stickers, what the company calls «collective yellow stickers». It contains everything from the phone-number to the pizza-restaurant on the corner, and to «how you set up SmallTalk on a special platform». You find information by searching an unstructured database, and can give «credits» to notes that you find useful. Notes with more credits about an issue show up before notes with less. It contains a mechanism to give feedback to the person who wrote the note, and there has been a kind of competition in the company to get the most credits. One developer described this module as «quite useful - it is simple enough to be used in practise». When we visited the company, it contained around 600 «experience notes».

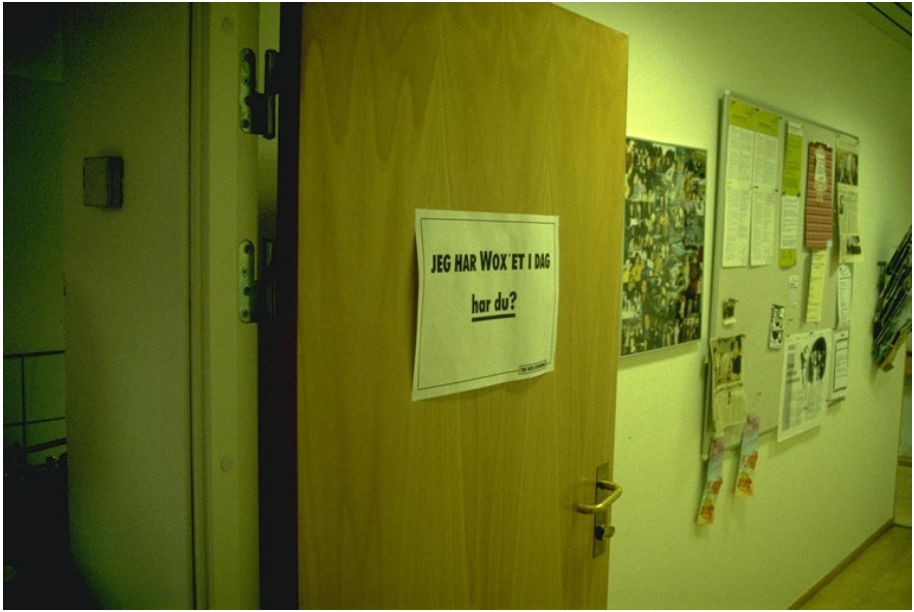

*Figure 5.9: «I've been WoX'ing today, have you?». One of several posters promoting the use of the WoX knowledge repository at Alpha.*

Examples of such notes are «how to reduce the size of your profile in Windows NT», «How to remove garbage from an image in SmallTalk», «Technical problems with cookies» and «An implementation of the soundex algorithm in Java».

According to one developer «people are very good at submitting notes when they think that something can be useful for others». A manager described it as «a behavioural arena that people use in different ways, that is creating a culture of knowledge sharing, and even creates expectations and lets people experience that others make use of their knowledge». The tool is promoted by posters which can be found on places that people visit a lot, like the one in Figure 5.9 which was located just outside the staff restaurant.

At Beta, a tool that covered the same purpose was the «Knowledge Base». In this tool, everyone could post experience without any structure on the company's competency areas.

## Usage Groups

When we asked people to describe what kind of tools they were using in their work, almost all of the developers mentioned that they were using WoX. All developers but one (seven out of eight) say that they have written experience notes, and all of them have tried to search for experience notes. Among the mangers, much fewer were using it actively. Three out of six did not mention WoX when we asked about knowledge management tools in the company.

## Usage Situations

We found five different types of usage of the knowledge repository:

- Solve a specific technical problem.
- Getting an overview of problem areas.
- Avoiding rework in having to explain the same solution to different people.
- Improve individual work situation by adjusting technical tools.
- Finding who has a specific competence in the company.

## Solve a specific technical problem

The most prominent use of this tool seemed to be in «problem solving». As one developer put it «if you run into a problem, then you can use WoX to see if anyone else in the company has had a similar problem», or «when you sit with a problem that you can't solve, or a strange bug, or if you do not understand why the computer does not behave the way it should».

Another developer says: «It happens that I have been searching and have found things in WoX. And then you do not have to search in other places, and maybe spend two or three days».

A problem with the notes that one developer mentioned is that «the person that writes something has a certain background, and with that background they presume that when they write `first you do this, then that...'

- that the others also know what to do». Which is not always the case for complicated matters.

### Getting an overview of problem areas

One said: «if I am stuck and wonder about something: usually, I remember that it was written somewhere in WoX, in fact, and then I go back and find it». An example is some notes about project-startup that this developer will usually go back to when being in that phase, which happens every 6 months or so. Another developer and another manager also said that they would see almost every day what was new «so I know what is in there, and do not have to search for things».

But people do not write about all types of problems as experience notes. Issues that are more «unofficial knowledge» - as one developer put it: «not things that are unethical, but things that you do that could easily be interpreted wrongly by customers, even though I mean we can stand for it» - that kind of issues you do not find any notes about, and that knowledge is transferred through informal oral communication.

### Avoiding rework in having to explain the same solution to different people

One developer said: «when the third person comes and asks about the same thing - then you realize that it is about time to document it». He would then later tell people who were asking about the new topic to look it up in WoX.

### Improve individual work situation by adjusting technical tools

Some said that they would find information on how to improve the tools that they use in their daily work, like Outlook, to make them more easy to use. Another example would be to get to know «how to reduce your profile in Windows NT» - which reduces the booting-time of you operating system quite a bit. A third example of a small improvement is a note on how to burn CDs for customers; which explained how to design covers for the CD so that they look more professional when delivering a final software product.

*Finding who has a specific competence in the company*

«Newbies get a short-cut to discover things that I have spent some time to build up. If they browse WoX a bit, they can find that `this person knows a lot about low-level Windows-patching' and that `this person is good at Apache webserver set up'«, one developer said.

## Knowledge Market

*Tool Presentation*

Here you find links to many other knowledge resources in the company; like company-internal links to information about Java, SmallTalk and other technical information that is relevant to software development.

## Overview of Processes

*Tool Presentation*

This is a list of all processes in the company, and each process has a www-page, which is maintained by the «process owner». It is usually a description of what responsibilities that process has, and how you can «use» it. Examples of processes are consulting, products, sales, knowledge management, mentoring, reuse and administration. A manager says that he likes to look up «information that is relevant to my behaviour, the reports from the resource process. I think the financial process has been very good at publishing reports with simple parameters which makes it possible to follow our billing rate for example».

## Overview of Projects

*Tool Presentation*

Here, we find a list of ongoing and completed projects, with some key information like project name, customer, project manager, which process it belongs to, and status. You also find a link to a project web with all documents from that project for the recent projects.

## Statistics

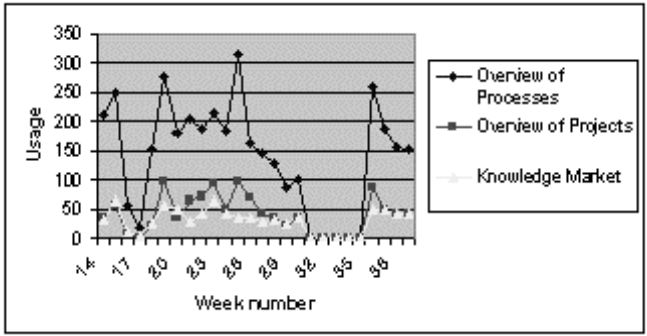

*Figure 5.10: The Usage of the Knowledge Repository and Library Tools over Time.*

We see from Figure 5.10 that the Overview of Processes tool seems to be the tool mostly used of the three we have usage logs from. This tool was used an average of 172 times per week, the Overview of Projects 51 times, and the Knowledge Market 39.

### 5.3.3 Knowledge Cartography Tools

At Alpha we found three cartography tools: Competence Blocks, Skills Manager and Software Tools. At Beta, we found one tool: People.

## Competence Blocks

### *Tool Presentation*

The «competence blocks» is a list of company-internal courses, where you also have the possibility to enter them, and evaluate them after completion. A brief description of each course is given, together with schedule information, and who is responsible. Most of the courses are given in a day of less. Sometimes, courses from other suppliers are also offered through this system. A manager described it as a «very valuable supplement (to normal «on-the-job-training»), with blocks that can be composed specifically». According to a developer, the management «encourage people to hold competence blocks».

*Figure 5.11: A list of «competence blocks» that is available in the competence block manager.*

### Usage Groups

We found six people that mentioned this tool when we interviewed them. This is a tool that people do not use very often - but you have to use it if you want to participate in a course. A developer said that this tool «suits me very well - I prefer oral communication to written».

### Usage Situations

This tool is used when someone wants to participate in a course, or plan a course (or «competence block»).

## Skills Manager

### Tool Presentation

This is a system where all employees can state which level of knowledge they have in different areas that are of interest to the company, like «ob-ject-oriented technology» or the ability to program in Visual Basic. It can be used to indicate which level you want to be at, so if you are interested in learning more about Visual Basic, you can state it in this tool. The tool is used for staffing projects, and many people in the company also use it to find someone who can help them to solve a problem. As one developer said: «I can say that I need a person that `knows HTML', and then I will get a list of people, and see what level of knowledge they have». For a wider discussion of this tool, see (Dingsøyr and Røyrvik, 2001).

At Beta, a similar tool called «People» was available, where people would rate their knowledge in various areas, and where you can search for knowledge.

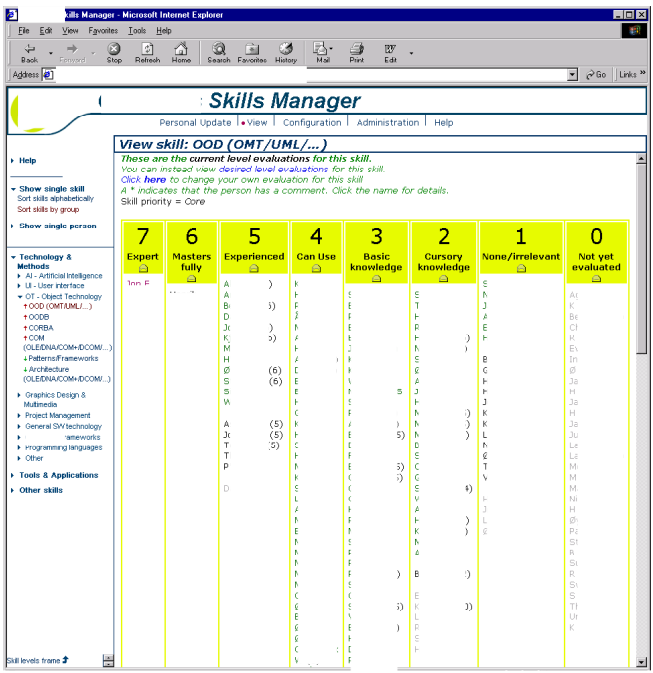

*Figure 5.12: An Example of a result after querying for competence on «object-oriented development» in the Skills Manger. The names of people have been removed.*

## Usage Groups

Project Managers, Managers as well as developers said in our interviews that they were using this tool.

*Usage Situations*

When we now go on to discuss the usage of the skills management tool. From the interviews, we have divided the usage into four categories, some with sub categories:

- Searching for competence to solve problems
- Resource allocation
- Finding projects and external marketing
- Competence development

We discuss each of these uses below:

*Searching for competence to solve problems*

The developers often need to know something about a topic they are not very skilled in themselves. We can then distinguish between two types of usage of the skills management system. First, people use it to find other people in the company who have knowledge about a specific problem that they have to solve - «short term usage». Second, people increase their overall insight in what the core competencies in the company are, what we can call more «long term» usage.

Let us look at the short term usage first: One developer says: «it happens (that I use it), if I suddenly have a specific problem in an area that I do not know much about. Then it sometimes helps to go in there and find someone who knows about it. I have in fact done that once...» Another developer seems to use it more often: «Of course, when I wonder if there is anyone who can help me with something, I look up in the skills management system to see if anyone has the knowledge that I need». In Fig. 5.12, we show a screenshot of the skills management system, which gives an overview of skills in object-oriented development. Here, you can also e-mail people who have a required competence in a specific area. Or you can just print a list of people and ask them yourself, as another developer is usually doing: «Then I find a list, and look at what level they have (...) and then I go around in the building and ask them». Of course, this depends on people to rate themselves in a honest way. One developer used the skills management system to find people, but after asking the believed «experts» found that «I did not get the answers that I needed, so I had to go to someone else. So, it is very dependent on that people update it correctly. And to describe a level is not that easy, so some overrate themselves and others underrate themselves strongly».

Another developer is critical to the categories of competence in the skills management system: «what you can get information about now is if someone knows about web - and that contains quite a lot!...maybe it is not that general, but not too far off. It is based on the core competency areas of the company, but when it comes to more detailed things, like who in fact can write a computer program, and who that can find a solution - you do not find that there».

When we look at more long-term usage, we do not find so much material in our interviews. One developer, however, often finds a group that knows something about a subject on the skills management system, and asks them questions by e-mail. But «if it then happens that you have asked questions about SQL to ten guru's, and it is always the same two that answers, then you start to go to them and talk. You learn after a while who it is any use to attempt to get anything out of».

### Resource allocation

In our empirical material from Alpha, we can see some patterns of the practical uses of the skills management system, in terms of resource allocation.

As one newly employed said: «Contrary to a lot of other companies that uses such a system, here at Alpha we really use the system for resource planning.» Another comment is on the same track: «I think that the skills manager is a useful tool, but a tool that still has got a lot of potential when it comes to practical use. Those who do the resource-management already use the tool a lot in the daily resource allocation work.»

A third Alpha employee comments on the Skills Manager as both an important tool for resource allocation, but also for the strategic development of the company:

«The tools I use the most I think are (...) the competence block manager and the skills manager. Definitely! I'm responsible for the content in many databases, and partly the skills-management base. And the skills manager is a tool that is very important for the resource allocation process (...) Therefore, many employees come up with suggestions to new content, new elements, in the skills database.»

## *Finding projects and external marketing*

Another usage of the system is for the sales department. One manager said that «Even sales can use it (the skills management system), to think out new directions to go in». That is, to find what types of projects that suits the company well. We can also think of another usage that we did not hear from anyone (probably because we did not talk to people in the sales department) - namely to use the system as external marketing; as «proof» of a highly skilled workforce.

## *Competence development*

Concerning the development of competencies at Alpha, the skills manager also seems to play a part. «The problem with all of our systems is that they function only to the degree that they are used. (Systems) like the Skills Manager depends on everybody to update it often and objectively. That could be solved by work-process support. Skills update could be a natural part of the closing of a project, for example by updating the involved competencies - those that have been in use during the project. You are today allocated to projects on the basis of what you have in the Skills Manager. There we have views devoted to people with free time and the competence required in the project. When you are allocated to a project on the basis of a competence profile, then there is also knowledge in the system about which competencies it is expected to be used in the project, and therefore it would be natural to ask for an update on those competencies when the project is finished.»

Another employee sees the Skills manager in light of intellectual capital. «Such tools are very good indicators for accounting intellectual capital. You are able to see in the long term what kind of competencies we will need, evaluate it, and compare it to what competence we already have in the firm, and then say that we have that many man months with C++ competence, or Java, and we see that there is an increase in this competence, and then we can evaluate that.»

In the skills management system at Alpha, the employees can use this tool to state what they want to learn about in the future, not only what they know now. In that way, people can develop their competence by working on relevant projects.

# Software Tools

## *Tool Presentation*

Here you find a list of the software that the company is using, together with information about it and about which person is the contact person for this software.

## *Statistics*

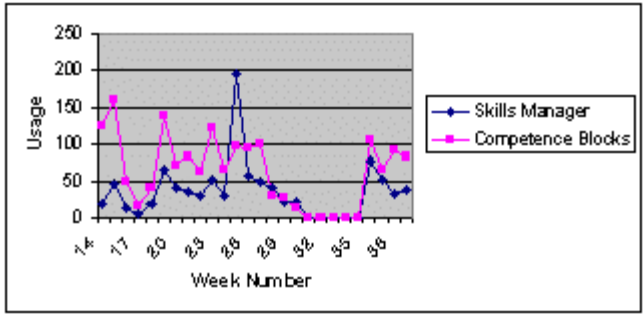

*Figure 5.13: The Usage of The Knowledge Cartography Tools over weeks.*

We see from Figure 5.13 that the Competence Blocks tool seems to be used the most. This tool was used an average of 78 times per week, the Skills Manager was used an average of 55 times.

## 5.3.4  Tools that support Community Building

## Product Idea Cafe

This is a discussion forum devoted to finding new products that the company can sell. It is an on-going brainstorm that has been running at least half a year, with many proposals for possible products.

## Myself

This is a page where you can edit information about yourself, such as contact information and which skill levels you have.

## Employees

This is a list of all employees in the company, with contact information, where their office is, and a picture. You can also see the skills of each person and a brief CV.

## Fun and reference

Here, you find different links, as well as a «museum» of old external www-pages the company has had. In addition you find a link to a quiz with questions and answers from people in the company.

## Finance

Under finance, you find information about the current billing rate of the company, and financial information about projects.

# 6 Discussion and Analysis

We will now discuss our findings from the different case studies in light of our research questions. Our main motive with this study is to investigate: How can Intranet-based knowledge management tools be used in medium-sized software consulting companies to facilitate a «learning software organisation»?

When we go on to discuss the material we have presented in the earlier chapters, we have several axes to structure the discussion. We organise the discussion in three main parts:

- The Knowledge Management initiatives from the literature.
- The Knowledge Management initiatives for codification in the prestudy-companies.
- The usage of Knowledge Management tools in the companies in the main study.

Thereafter, we compare the studies in the different parts. Then, we discuss what implications our findings have for the theory of knowledge management in software engineering that we described in Chapter 3. Finally, we discuss which special conditions that apply for medium-sized companies, and what might be valid also for larger and smaller companies.

## 6.1 Knowledge Management Case Studies from the Literature

In Chapter 3, we described eight case studies of knowledge management initiatives in software engineering. Now, we first discuss what different approaches the companies had to knowledge management, and then what they claim to have achieved.

## 6.1.1 What the Companies Did

First, let us discuss what kind of goals the case study companies had with their knowledge management initiatives, that is their «strategy». Then we will discuss which «processes» and «tools» they made use of to achieve this.

Note that when we use tables with indications of «yes» and «no» in the following, it indicates if we have found evidence in the literature on the question. A «no» means that we do not find evidence for a «yes» from the papers we have found. This can mean either that the topic was not reported in the paper(s) describing the case, or that the answer was really «no».

### Strategy

We find several companies that wanted to improve the situation for their software developers, but did not have clear goals with respect to quality or development costs. Daimler Chrysler, Ericsson Software Technology, sd&m as well as both departments of ICL would come in this category, although ICL High Performance Systems also wanted to «improve the predictability of costs and delivery». At NASA, The Australian Telecom Company, and Telenor Telecom Software they had cost reduction and quality improvement as primary goals for their knowledge management activity. If we categorise the cases according to which type of strategy that was chosen, either to support «personalization» or «codification», we find that all of the companies had a codification strategy, and six of eight also support the personalization strategy (see Table 6.1). The type of knowledge to be collected and distributed in these companies was both qualitative (like descriptions of experience) and quantitative (like measurements on the size of code). Three companies had a focus on quantitative knowledge, whilst seven were focusing on qualitative knowledge.

*Table 6.1: A List of what Companies did, and what Knowledge Management Approach they Chose.*

| Company | Knowledge Management Approach | | | |
|---|---|---|---|---|
| | Personalization? | Codification? | | Reorganisation? |
| | | Quantitative? | Qualitative? | |
| NASA SEL | Yes | Yes | No | Yes |
| Daimler Chrysler | Yes | Yes | Yes | Yes |
| Telenor Telecom Software | Yes | Yes | Yes | Yes |
| Ericsson Software Technology | Yes | No | Yes | Yes |
| Australian Teleom Company | No | No | Yes | No |
| ICL High Performance Systems | No | No | Yes | No |
| ICL Finland | Yes | No | Yes | No |
| sd&m | Yes | No | Yes | Yes |

## Processes

When looking at what kind of processes that are present in each of the cases, we find that many emphasize that developers actively participate in collecting and distributing knowledge. Five out of the eight companies did a reorganisation as a part of the knowledge management initiative, to have a separate part of the organisation responsible for this kind of activities. At sd&m they specifically mention that they organise kick-off and touch-down meetings in the beginning and end of projects.

## Tools

Finally, let us discuss what kind of tools that the companies were using: Intranet systems for exchanging knowledge. We find that the systems at ICL and Telenor Telecom Software contain descriptions of work processes, and at sd&m and ICL Finland, we find lists of employees skills, as well as lists of customers, partners and projects. Telenor Telecom Software is the only company that developed an Intranet-based expert system for estimation.

## 6.1.2 What were the Results?

Now we would like to discuss the results of the knowledge management initiatives mentioned in the case studies. We have listed the case studies as well as reported benefits in Table 6.2.

*Table 6.2: A List of Effects of Knowledge Management in the Companies.*

| Company | What was the effect? | Benefit | | |
|---------|---------------------|---------|---|---|
| | | Developer satisfaction? | Lower cost? | Higher quality? |
| NASA SEL | Reduced number of defects, reduced Sof tware production costs, increased | No | Yes | Yes |
| Daimler Chrysler | The case gives no information on the effect for the company. | No | No | No |
| Telenor Telecom Sof tware | The study indicates that estimation accuracy was improved, and focus on risk management was increased. | No | Yes | No |
| Ericsson Sof tware Technology | The study claims that the initiative was "more valuable" than a database and measurement approach. | No | No | No |
| Australian Telecom Company | Good acceptance of product among users. | Yes | No | No |
| ICL H igh Performance Systems | A perception that it has facilitate a "new mode of working". | Yes | No | No |
| ICL Finland | Saved time, bacause it is easier to find documents. Easier to learn new project members about project work. | Yes | Yes | No |
| sd&m | Previous problems due to rapid growth have diminished. | Yes | No | No |

If we look at whether the introduction of a knowledge management system improves the quality of software, we only find an answer to that in the first article from NASA Software Engineering Laboratory. Although it is mentioned in the article from sd&m that employees now do not make the same mistakes again so often, it is not directly said that the software now has higher quality than it used to be.

Then, does the introduction of a knowledge management system lower the cost of developing software? We find evidence for this in three of the cases. Again, this is documented with measurements from the NASA Software Engineering Laboratory. At ICL Finland, it is claimed that project managers and other project members «save about 30 percent in time, when making a new project member familiar with the system under development». At Telenor Telecom Software, the paper authors believe that the work has resulted in improved estimation accuracy.

So three out of the eight companies say that the cost in some way is lower after introducing the knowledge management system.

If we ask how the introduction of a knowledge management system influences the work of employees in an organisation - we find more in the cases: ICL Finland conducted an internal survey amongst employees that showed that the initiative «supported project work and saved time», and «made it easier to find documents». Another benefit is described as «better visibility through knowing what kind of projects are going on». sd&m claims that their problems due to rapid growth «do not occur nearly as often as before». And ICL High Performance Systems claim that there is a «perception by project members» that the company is in a «new mode of working». Also at the Australian Telecom Company, employees said in a survey that the product «was good». So in all, in four of the eight cases, we find some evidence for improved developer or employee satisfaction.

## 6.2   Success Factors in Codification Initiatives

In our prestudy, working with Companies One to Four, we were interested in success criteria for codification initiatives. Our research question was:

*R1)      Which factors influence the success of codification strategies for Intranet-based knowledge management tools in medium-sized companies that develop software?*

What can we learn from the experience with these four companies? We have divided the discussion of these cases into four subchapters: First, we characterize each of the initiatives, then the results, before discussing what «success criteria» we think we can find. Finally, we discuss our suggested success criteria with respect to other studies.

## 6.2.1 Characteristics of the Improvement Initiatives

All four companies had an improvement initiative to codify knowledge from software projects, but they differed in purpose, technical platform and what type of content the system should transfer between projects. We have summarized the different approaches in Table 6.3:

*Table 6.3: Some Characteristics of the four Initiatives.*

| | Company | | | |
|---|---|---|---|---|
| | **One** | **Two** | **Three** | **Four** |
| *Purpose* | Better estimation. | Better estimation. | Assist software development. | Assist software development and improve estimation. |
| *Platform* | Web, spreadsheets. | Orcale w/ 4GL. | Web, files. | MS Exchange Repository. |
| *Contents* | QA System (process models), estimation tool. | Project Experience reports, simple estimation models. | QA system, company info, CVs, experience notes. | Guidelines, course material, project experience reports, CVs. |

We see that Company One and Company Two have the same purpose, and Company Three and Four also have similar, but more widely focused purpose of the initiative. If we look at the technical platform, the common denominator is that all systems were available for all employees in the companies through a web-interface - on the companies' Intranets. Apart from that, many different technologies were used. The contents that was distributed through these systems were knowledge from previous projects in the form of several tailored estimation models at Company One, various project data in Company Two. In Three and Four, we find much more material, from personnel CVs to experience reports from projects.

## 6.2.2 Results from the Improvement Initiatives

Earlier, we said that the results of the companies' efforts varied a lot. Let us examine this in further detail:

First, what kind of impact is it meaningful to discuss from such initiatives? Is it possible to do a cost/benefit analysis? This is a very difficult question. Making codified knowledge available for new projects can only give a long-term effect, and an effect that one would expect to grow over time as more and more knowledge is available. We did not work with the companies long enough to study such effects. Also, if we had done so, it is difficult to measure such effects, other than by indirect measures like asking employees in the companies about what they think themselves.

In order to determine some kind of «success» of these initiatives, we have to rely on indirect measures, or notions. What should these be? One factor that comes to mind is to ask employees if the tools have any «business value». That is - do employees have a subjective feeling that the efforts they put into using the tools are worth it? Another factor could be to look at the culture for sharing knowledge in the company. To be willing to share knowledge is crucial for the tools to have a lasting effect. If employees are not willing to share their experience, tools will quickly be outdated and irrelevant. We will call this factor «cultural changes». A third factor that we could observe is, of course, if the tools that were developed were actually in daily use or not in the companies. This should also be a good indication on if this would be a lasting improvement or an initiative that would quickly die.

We have summarized what we observed in the different companies in Table 6.4:

*Table 6.4: Results of the Companies' Efforts in Codification Initiatives.*

| | Company | | | |
|---|---|---|---|---|
| | **One** | **Two** | **Three** | **Four** |
| *Business Value* | Some | Low | High | High |
| *Cultural Changes* | Some | Low | High | Some |
| *Current Use* | Low | Low | High | High |

As we see from the table, is seems that only the two last companies have a system that we can describe as having a business value, which is in current use, and where the organisation have achieved cultural changes.

## 6.2.3 Success Criteria

Now, why is it that Company Three and Four ended up with systems that seems to be working, whilst company One and Two had initiatives that does not seem to work, although these companies had a more focused approach to what they were doing?

When working with these companies, the main difference was the degree of turbulence. This was a time when many companies merged with others, which usually meant a large reorganisation process. Many employees also changed their job often, which meant that it was difficult to work with stable «champions» for improvement initiatives. We have listed how this affected our four companies in Table 6.5, listing stability both in the companies - if they had a stable improvement strategy or not, and in the «champions» that we worked with. Other noticeable differences were in the relevance of the tools that were developed, and how fast the improvement initiative was introduced in the company. Some of the companies had a more incremental approach, whilst some other went more for a «big bang» approach.

*Table 6.5: Some Influential Factors for the Codification Initiatives.*

|  | Company | | | |
|---|---|---|---|---|
|  | **One** | **Two** | **Three** | **Four** |
| *Stable Strategy?* | Until 1998 | Until 1998 | Yes | Yes |
| *Stable Champion?* | Until 1998 | Until 1998 | Yes | Until 1999 |
| *Relevance?* | High | Low | High | High |
| *Introduction?* | Big Bang | Big Bang | Incremental | Incremental |

From the table, we see that Company One and Two had large changes in their improvement strategy. In both cases this was due to major reorganisations in the companies. These two companies also suffered the most from key employees leaving the company, which meant that many improvement activities had to start almost from scratch.

Further, we see that the relevance of the initiative was lower for Company Two than for the three other ones - mainly because the company did not invest much in development, and the resulting system was therefore more general.

We also found that the companies differed in their focus on supporting knowledge sharing. In Company Three a part of the knowledge management system was an incentive system that promoted knowledge sharing by giving employees who contributed knowledge feedback from the employees who were actually applying the knowledge later. This company also marketed themselves to other companies as a «knowledge management» company; so this issue was considered very important for them. We also observed that the management supported the initiatives at Company Four, where major resources was put into this type of work. Also, in Company One, some attention was given to developing a more «knowledge sharing culture» by giving courses to employees who would use the tool that was developed. But this effort was not sustained when the company went through a reorganisation.

If we look at how the new tool was introduced, we also find differences. Company One and Two wanted a more «big bang» approach where a working system would be introduced in the company. Companies Three and Four first introduced a small system that was gradually expanded over several years.

Another difference that we can note is the domain of the companies: The two last ones are consulting companies that are much more dependent on being updated on different topics than the more traditional Companies One and Two. We could say the two last companies are more «knowledge intensive» in their work, and thus more dependent on having good internal systems.

If we sum up what we can learn from this, by dividing the companies into groups of «unsuccessful» and «successful», we may state the following:

From the «unsuccessful» attempts, there were problems with changing strategies and changing «champions» that provided major set-backs in the improvement efforts. Also, the importance of a «culture» for sharing knowledge was not so much emphasized in these companies. Another issue is the coupling to business goals: it was not crucial for these companies to succeeded in their attempts, because they were operating in business domains that were less knowledge intensive. A last issue was the «big bang» approach that required many resources, and did not produce such an «immediate» result as was hoped for.

If we look at the «successful» companies, we see that they had more stable strategies and champions, were better at working with cultural aspects, and were much more dependent on being «successful». They also proceeded in a more incremental fashion. An interesting aspect is that these companies chose a more «holistic» approach, and did not have such a focused strategy.

Again, to sum up, we think that the following factors are important:

- A culture for sharing knowledge (f1).
- Stable focus on knowledge management (f2).
- Incremental development; show benefits during development (f3).
- Good coupling to business goals (f4).

We label the factors f1-f4 to make it easier to compare with other factors in the following.

### 6.2.4 Success Criteria from other Studies

So how do our findings relate with those of others? In the software engineering domain, we have not been able to find similar studies of «success factors». But if we turn to the general literature on knowledge management, we find several other results, what we described in Chapter 3.

We see that the set of factors from the 31 knowledge management projects in the study by Davenport, Long and Beers include more about technology issues than we experienced in our setting - the factors «technical and organisational infrastructure» (d1) and «standard, flexible knowledge structure» (d3). Another noticeable difference is the criteria for «multiple channels for knowledge transfer» (d6). This is in a way similar to having a broader focus on what to transfer - like in Company Three and Four, who focused on knowledge in different forms, from competence profiles to experience reports. What this first study refers to as «clear purpose» (d5) and «senior management support» (d7) might be seen as our criteria on stable focus (f2). What we did not find here is a criteria for incremental development (f3), which was one of the major factors in our environments. The second and third studies emphasise on having a culture for sharing knowledge, which corresponds very well with our findings. The McKinsey study is more detailed in their important factors, but at least we recognise that the knowledge management efforts should be well linked to the business goals, for example to improve development and process efficiency. An interesting point here is the degree of involvement of the end users in developing the knowledge management systems. Coming from a Scandinavian work environment, this is a factor that is easy to forget, as it is so common to focus on employee participation. We think the end users - developers and project managers - was involved to a large degree in Companies One, Three and Four, but not very much in Company Two. But we think that companies will not be able to create a «knowledge sharing culture» (f1) without involving employees in discussing how to achieve it.

## 6.3   Knowledge Transfer by Intranet-Tools

In our main study, we were interested in what kind of knowledge management tools that were available in medium-sized software consultancy companies, and how these tools were used. Our research question was:

*R2)      How do different groups of users in medium-sized consultancy organisations use Intranet-based knowledge management tools to transfer knowledge between software development projects?*

We will now structure the discussion of this question after the types of tools and strategies we described in Chapter 3. There, we divided between two strategies for knowledge management: codification and personalization. We also divided between three types of tools: Knowledge Repositories and Libraries, Knowledge Cartography and tools that support Communities of knowledge workers. In Alpha and Beta, we only examined the first two types of tools, and we will now discuss how these different tools were used for codification and personalization in the companies. Then, we discuss what kind of learning that takes place as a result of these tools.

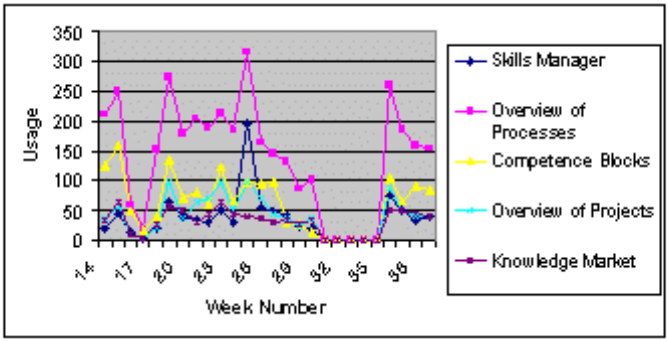

*Figure 6.1: Usage of Knowledge Repository and Knowledge Cartography Tools over Time.*

But first, let us look at to what degree different tools are used at Alpha, where we have logs of usage over time for some tools, as shown in Figure 6.1. We see that The overview of processes is the tool that is used the most, followed by the Competence Blocks and the Skills Manager. It is a bit surprising that the process overview was accessed this much, as none of the employees we interviewed said they were actively using this tool. This is probably because the webpage with this tool is used to access other tools. It might also be that people did not mention this tool because they did not consider it as a proper tool.

As we wrote in the Empirical Investigation chapter, some employees were not using the knowledge management tools actively. And some were critical to the contents of the various tools. But we will focus on the users in the following, and not on the non-users.

### 6.3.1 Knowledge Repositories and Libraries

Before discussing the different usage of the tools, let us quickly summarize which tools we found at Alpha and Beta. We have made a list of the knowledge repository and library tools in Table 6.6.

*Table 6.6: List of Knowledge Repositories and Libraries at Alpha and Beta.*

| Tool | Description | Company |
|---|---|---|
| Project Guide | Description of common processes and work roles in project work, with templates, checklists and examples. | Alpha |
| Handbooks and policies | Descriptions of common processes and work roles in the company. | Beta |
| Well of Experience | A knowledge repository ("collective yellow stickers"). Contains everything from bugfixes to telephone numbers. | Alpha |
| Knowledge Base | A repository of knowledge on competence areas, methods, customers and company-internal courses. | Beta |
| Knowledge Market | Links to knowledge resources like company-internal information on Java, SmallTalk and in-house libraries. | Alpha |
| Overview of Processes | Lists all the processes in the company, like consulting, products, sales. | Alpha |
| Overview of Projects | Gives an overview of ongoing and completed projects, with key information like project name, customer, project manager and status. | Alpha |

We see that Alpha has more tools than Beta, and that the Knowledge Base of Beta seems to cover the same type of knowledge as the Well of Experience at Alpha. Further, the «Handbooks and policies» tool at Beta contains descriptions of work processes much in the same way as the Project Guide at Alpha. The «extra» tools we find at Alpha are overviews of projects, processes and knowledge reserves in the company.

When we go on to ask about how these knowledge repository and library tools are used for transferring knowledge between development projects, we divide the usage in two types. First, we look at usage of codified knowledge from the tools - what corresponds to the codification strategy that we have presented in Chapter 3. Second, we also have found some types of usage that suits better in the personalization strategy.

## Codification strategy

In the empirical investigation chapter, we listed a number of usages of the tools from different groups. Of the knowledge repository and library tools, we found the following usage situations (with corresponding tool in brackets):

- Getting tips and advice in project start-up and execution (Project guide)
- Solve a specific technical problem (Well of Experience)
- Avoid rework in having to explain the same solution to different people (Well of Experience)
- Improve work situation by adjusting technical tools (Well of Experience)

From the interviews it seemed that the Project Guide was in use by different employees and with a different frequency than the Well of Experience. Very few employees mentioned any use of the overviews of knowledge areas. But from Figure 6.1 we see that these tools do not seem to be used less than others, in fact the overview of processes is in wide use.

The Project Guide seemed to be mostly in use by some project managers, and not very much in use by developers. The Well of Experience on the other hand, seems to be used by many employees, and at a much higher frequency. We note that it was mainly developers who said that they actively contributed to the contents of the Well of Experience, and not employees who acted as project managers or managers. If we divide between employees who use tools to «contribute» and «use» knowledge from the two main repository-tools, and divide employees in groups of manager, project managers and developers, we get the following table:

*Table 6.7: Groups of Contributors and Users of Knowledge in the most used Knowledge Repositories/Libraries.*

| | | Tool | |
|---|---|---|---|
| | | Project Guide | Well of Experience |
| Employee Groups | Contributors | *Mostly Project Managers* | *Mostly Developers* |
| | Users | *Mostly Project Managers* | *All Groups* |

Why do we see this difference between the usage of these tools? Is it because of the intended focus of the knowledge in the tools, or the way the tools can be used? The Project Guide is intended as a support in project work, and contains abstracted knowledge from previous projects. The Well of Experience has no structure and could contain any type of information. Yet, it seems that it is the developers that use the tool, and fill it with technical information - either in order make it easier for others to solve a problem, or to avoid rework oneself by having to explain the same thing several times. Or: adjusting technical tools to increase performance.

The «user interfaces» of the tools are quite different: The Project Guide can display knowledge according to different roles in a development project, and is browsable. The Well of Experience is a small search engine containing company-relevant information.

It might be that developers require more specific information in order to solve most of their daily problems; when they have a specific problem, the solution is often in a «bug fix», or a technical description on how to change something. The solution is not found in an abstract way to reason on such problems - which is what you might expect from the Project Guide. Maybe this type abstract knowledge that you can find there is better suited in situations when you have to decide on some overall structures, but not in concrete problem situations.

## Personalization strategy

When asking employees about usage, we found two uses of Knowledge Repositories/ Libraries that we can say is a part of the personalization strategy, namely:

- Getting an overview of problem areas (Well of Experience)
- Finding who has a specific competence in the company (Well of Experience)

Here, the employees did not use the knowledge found in the Well of Experience directly. They saw the available knowledge, who made it, and used that information for getting an overview of problem areas the company faced often, and of who was frequently posting tips on topics, and could then be considered some kind of «expert». It is an interesting point that the tools with codified knowledge can be seen as having an additional purpose than pure «codification» and «distribution».

## 6.3.2 Knowledge Cartography

We have listed the Knowledge Cartography tools in Table 6.8, and indicated what tools we found at each company. At Alpha we fund three tools, and at Beta we found one. The Skills Manager and the People tools cover the same needs, but at Alpha we also found tools that organise company-internal courses and list which software tools that are available for development as well as contact persons for these tools.

*Table 6.8: List of Knowledge Cartography Tools at Alpha and Beta.*

| Tool | Description | Company |
|---|---|---|
| Competence Blocks | A list of company-internal courses with brief descriptions, schedule information and the possibility to sign on and evaluate courses. | Alpha |
| Skills Manager | An overview of the skill level of all employees on about 250 different skills that are considered important for the company. | Alpha |
| People | An overview of the skills of all employees, in categories like "Java programming". | Beta |
| Software Tools | A list of the software that the company is using for software development, with a contact person for each tool. | Alpha |

## Codification strategy

All the Knowledge Cartography tools we found at Alpha and Beta are overviews of knowledge sources, and we therefore did not find any use that we can classify as a codification strategy.

## Personalization strategy

Of the cartography tools, we found the Skills Manager to be in use for four different purposes:

- Searching for competence to solve problems (Skills Manager)
- Resource allocation (Skills Manager)
- Finding projects, and external marketing (Skills Manager)
- Competence development (Skills Manager)

Only two employees mentioned that they were using the Competence Blocks, and we did not hear anyone mentioning the Software Tools. From the interviews it seems that these tools are used much less than the Skills Manager that almost everyone mentioned, and where most employees had updated their skill levels. But from Figure 6.1 we see that the Skills Manager is used less than the Competence Blocks.

It was developers that said they were using the Skills Manager for solving problems and competence development, while managers and administration used it for resource allocation and to find external projects and market the company externally.

### 6.3.3 Learning at Alpha

We now go on to discuss what kind of learning the different usage types we found in Alpha support. We do not discuss this at Beta, as we did not find many descriptions of tool use in our interviews from that company.

We found some use in solving problems, namely:

- Solve a specific technical problem (Well of Experience)
- Searching for competence to solve problems (Skills Manager)

We also found use in avoiding rework and improving the work situation:

- Avoid rework in having to explain the same solution to different people (Well of Experience)
- Improve work situation by adjusting technical tools (Well of Experience)

Other types of use were for getting an orientation in the company, and for making some work processes more effective:

- Getting an overview of problem areas (Well of Experience)
- Finding who has certain competence in the company (Well of Experience).
- Resource allocation (Skills Manager)
- Finding projects, and external marketing (Skills Manager)
- Competence development (Skills Manager)
- Getting tips and advice in project start-up and execution (Project guide)

If we describe these forms of usage in relation to the theories about learning in Chapter 3, we see that all these usages can be said to be what we called single loop learning. It is either transferring knowledge in order to solve a problem or perform a work task more efficient in the future, or it is knowledge which is necessary for automating a work task that would otherwise require manual effort, such as resource allocation. We do not find knowledge that we can say facilitates double loop learning.

At Alpha, people who had the same position in the company would sometimes use different tools. Some preferred to use the Skills Manager to find experts in order to solve a technical problem, while others would search in the knowledge repository WoX. This might be an indication that it is not only what knowledge you can expect to find through a tool that decides what tool you use, it is also the way the knowledge is presented. In Kolb's theory of experiential learning we mentioned that people have different learning styles, and this might then affect what tools they prefer to use.

## 6.4   Comparison of the Different Studies

Now, we would like to compare the findings from the three previous subchapters, and see how we can place the findings from Company One to Four, Alpha and Beta in comparison to the case studies from the literature.

We proceed by describing Company One to Four, Alpha and Beta in the general framework we used for the case companies. That is, we ask, what did the companies do, and what was the result of these actions.

### 6.4.1  What the Companies did

As we see from Table 6.9, we were concerned with codification initiatives in the prestudy, and did not look at mechanisms that supported the personalization strategy. All companies One to Four were using knowledge repositories with qualitative knowledge, and Companies One and Two were also collecting quantitative data from completed research projects.

So, could it be that the problems that we discussed in Companies One and Two were because it is more difficult to gather quantitative data than qualitative data? We think that this is not the case, because the initiative in Company One showed promising results in the start, that actually gave a better precision in estimation than what was normal with previous methods. The problems lie more in sustaining such an improvement project as we described in the success factors.

At Alpha and Beta, we did not find any knowledge repositories for quantitative knowledge, only for qualitative knowledge. But we also found systems that supported the personalization strategy. From the case studies in the literature, ICL Finland and sd&m are the companies that are mostly similar to Alpha and Beta, in that they both have a personalization strategy, a qualitative codification strategy, and no quantitative codification. Unfortunately, these studies were done early after «implementing» a new knowledge management initiative, so we know little of the actual usage over time of the tools that are described.

But if we look at the tools that we find at ICL Finland and at sd&m, we see several similarities to Alpha and Beta.

ICL Finland had three groups of «knowledge resources», what they called external knowledge, structured internal knowledge and informal internal knowledge. The external knowledge is similar to the lists of customers, technical tools and the list of company-internal knowledge resources that we found at Alpha. The structured internal knowledge contains what we found in the Skills Manager and the Project Guide, as well as databases for marketing and sales information (although not so much related to software development), and what the company describes as research reports. In the informal internal knowledge category, we find electronic discussion forums, news, and «project folders». We also found similar resources at Alpha and Beta, although we did not discuss news, folders and discussion boards as tools. From the description of tools and resources at ICL Finland, it seems that they do not have a search interface to unstructured knowledge such as the Well of Experience Tool at Alpha. But this kind of knowledge would probably be found in the discussion boards.

sd&m had several databases with employees, customers, partners, projects, a skill database, and also some Intranet web-pages about «core topics» of the company, that was edited by knowledge brokers. If we compare this to what we found in Alpha and Beta, we see that the databases contain the same material, except a «method handbook» with templates and tips, such as the Project Guide at Alpha and the «Beta Way» at Beta.

We also do not find such a tool as the Well of Experience knowledge repository in this company.

In all, it seems that Alpha and Beta are fairly well equipped with knowledge management tools in comparison to what we have found in the literature.

*Table 6.9: What was done in Company One - Four, Alpha and Beta.*

| Company | Knowledge Management | | | |
|---|---|---|---|---|
| | Personalization? | Codification? | | Reorganisation? |
| | | Quantitative? | Qualitative? | |
| Company One | Yes | Yes | Yes | Yes |
| Company Two | No | Yes | Yes | No |
| Company Three | Yes | No | Yes | No |
| Company Four | Yes | No | Yes | No |
| Alpha Consulting | Yes | No | Yes | No |
| Beta Consulting | Yes | No | Yes | No |

If we look at the issue or reorganisation, we see from Table 6.9 that Company One added new roles for editing and maintaining the knowledge repository that was built up. In the other companies, existing roles were redefined.

## 6.4.2 What were the Results?

If we describe the companies from our case studies in the same way that we described the cases from the literature, can we say anything about the results of the knowledge management initiatives?

As we discussed in the section about success criteria, it is very difficult to determine the «effect» of a knowledge management initiative. But a start is to look at the different approaches and see whether they sustained over time or not. As we know, the initiatives as Company One and Two were abandoned after some time, and the tools that were developed went out of use. In company Three and at Alpha, the tools seemed to be in extensive use, whilst in Company Four and at Beta, the tools were in use, but not in frequent use («weekly» instead of «daily»).

But if we return to the «results», although being unable to point at direct results, we can describe what we observed in the companies - were they measuring whether they produced software of higher quality? At lower cost? Or did they improve the work situation for the employees? We have described this in Table 6.10, where we see that we found employees in the companies that expressed that the work conditions had improved as a result of the knowledge management initiatives. None of the companies tried to measure reduced costs or higher quality, although they probably also expected this a result. A reason for focusing on employee satisfaction at the time when we were working with the companies, could be that it was difficult to get enough employees, and it was crucial to keep the ones the companies had.

*Table 6.10: The Effect of the Knowledge Management Initiatives in Companies One - Four, Alpha and Beta.*

| Company | What was the | Benefit | | |
|---|---|---|---|---|
| | | Developer satisfaction? | Lower cost? | Higher quality? |
| Company One | The Knowledge Management system was abandoned. | No | No | No |
| Company Two | The Knowledge Management system was abandoned. | No | No | No |
| Company Three | The Knowledge Management system in extensive use. | Yes | No | No |
| Company Four | The Knowledge Management system in use. | Yes | No | No |
| Alpha Consulting | The Knowledge Management system in extensive use. | Yes | No | No |
| Beta Consulting | The Knowledge Management system in use. | Yes | No | No |

If we compare the results from these companies with the ones we found in the literature in Table 6.2, we can group our cases with the Australian Telecom Company, ICL High Performance Systems, ICL Finland, and sd&m.

## 6.5    Empirical Investigations in Relation to Theory

How does our empirical studies from companies that develop software compare to what we know from theory about knowledge management in software engineering?

The Experience Factory concept places much emphasis on having a separate part of the organisation that develops software to work with experience (or knowledge) management. In our cases, we found that very few companies had a separate organisation for this purpose. Most companies had distributed the responsibility for knowledge management throughout the organisation, with some management function in charge. This was possible because the extensive use of Intranet tools made it easy to communicate various knowledge. We did not find much focus on actively «packing» knowledge in the organisations we studied. Some tools like the Project Guide and the Handbooks and Policies contained processed knowledge, but most relied on normal employees to edit knowledge themselves. This form of self-administration was also applied in the skills management systems, where the employees would rate their own skills. In all, we can say that the medium-sized organisations were spending less human effort in maintaining a knowledge management infrastructure than we would think from reading the Experience Factory literature. Also, we found more focus on computer-support.

If we look at the type of knowledge we find in the tools that we examined closely, namely at Alpha and Beta, we see that the tools contain much of the knowledge that is described in the Experience Factory literature. We find knowledge about:

- Products - in the knowledge repositories such as the Well of Experience. We also find this type of knowledge in the Knowledge Market at Alpha and Betas Knowledge Base.
- Processes - in the Project Guide at Alpha and Handbooks and Policies at Beta.
- Tools - in the «Tools» overview at Alpha.
- Management - in the Project Guide and Handbooks and Policies.

What is mentioned in the Experience Factory literature that we do not find in our main case study are relationship packages for prediction in cost, defect and resource models. This is because neither Alpha nor Beta focused on collecting quantitative knowledge. The same applies to the data packages mentioned in the Experience Factory.

We note that the focus in Experience Factory is primarily on codification, and not on personalization. Therefore, we do not find references to packages or tools for skills management or for organising internal courses in the Experience Factory organisation.

In Chapter 3, we distinguished between active and passive tools - or tools that operate without the influence of a user, and tools that users have to apply themselves. All the tools we found in use at Alpha and Beta were tools that required an active effort from the users - passive tools. Also, the tools we found in the case studies from literature required active user participation. In all, it seems that active tools are not applied. Why is this the case? This might be because it is more expensive to develop active tools, because they require more technology, and have to be more tailored to usage situations in companies. It might also be because the existing passive tools fulfil the role of knowledge management - that the overhead of having to become aware of a «knowledge need» and actively having to search for knowledge is not large.

When we presented the two strategies, codification and personlization in Chapter 3, we mentioned that Hansen et al. argue that companies should focus on just one of these strategies. We found many successfull combinations of both. People at Alpha also argued for a tighter integration between the Skills Manager and the Well of experience, which seems sound in that the two tools are used for some of the same purposes. We therefore disagree with the view of choosing a single strategy.

## 6.6   What is Special for Medium-Sized Companies?

We have discussed findings from medium-sized companies in the prestudy and main study. In the case studies from the literature, some of the companies can be called «large», namely NASA, Daimler Chrysler and Ericsson. However, the studies from these companies were done only on one or a few departments, which we may say are comparable to a medium-sized company.

A characteristic of medium-sized companies is that they can afford some overhead in projects, for example for quality improvement and knowledge management. This of course also depends of what kind of business domain the company is working in. If the company delivers products with strong quality requirements, the overhead will probably be higher than a company that focuses more on delivering products fast.

For companies to make use of the ideas we discussed in this chapter, they will need to invest time in developing and maintaining different solutions. This might also be valid for some smaller companies, and all larger companies. But larger companies might experience problems that we did not see in medium-sized companies because of scale. To use an unstructured database as a form of knowledge transfer, such as in the Well of Experience, might make it difficult to find relevant experience, for example. We can expect larger companies to have to invest more in organizing and packaging experience, such as in the Experience Factory concept.

# 7 Conclusion and Further Work

To conclude this thesis, we first draw conclusions from the literature study, then the prestudy, and finally the main study. We then discuss what implication these conclusions have, and present possible directions for further work on knowledge management for learning software organisations.

As we discussed in the introduction, we see the field of software engineering as a particularly interesting domain to study the use of knowledge management systems, as the employees in software companies are skilled in using computer tools, usually are motivated to use computer tools, and also spend most of their workday in front of computers. In addition, the need for learning about new technology, and about new markets is large in the software domain. We think that more and more companies will work in a computer-intensive way, and that the conclusions we draw on such tools here might be usable for such companies in the future.

## 7.1 Conclusions from the Literature Study

We analysed eight case studies of knowledge management systems in software engineering companies. We found that five of the companies set up a separate department in the organisation with responsibility for knowledge management. All the companies report that they store experience (codification). Seven companies focused on qualitative knowledge, and three focused on quantitative knowledge. Five of the eight companies also facilitate personalization in the organisation.

Our research question was:

- *How can Intranet-based knowledge management tools be used in medium-sized software consulting companies to facilitate a «learning software organisation»?*

Companies who have made knowledge management initiatives for personalization or codification report on better working conditions for employees, improved software quality, or lower development costs. That is, we see learning effects from both types of strategies for tool-supported knowledge management in software companies, and for codification with a qualitative and quantitative focus. Knowledge management seems to contribute to facilitate learning software organisations. We can therefore conclude:

- Intranet-based knowledge management tools can be used for personalization and codification strategies (the latter with both a quantitative and qualitative focus). The usage of these tools can result in learning effects like improved software quality, reduced development costs and a better working environment for employees.

Concerning the relatively few number of cases on use of knowledge management systems that we found in the literature, and the large work on technology issues, we can conclude: There is a great interest in developing technology to support learning software organisations, but empirical analysis of how experience sharing actually works is lacking.

## 7.2    Conclusions from the Prestudy

In the prestudy, we asked the research question:

*R1)      Which factors influence the success of codification strategies for Intranet-based knowledge management tools in medium-sized companies that develop software?*

In the discussion, we found that it can be difficult to get knowledge management tools for codification into use, but we found four success factors that we believe are important when introducing knowledge management initiatives for codification:

- *A culture for sharing knowledge.* Using knowledge management tools and sharing knowledge with others requires employee motivation. It is easy to postpone such activities because of lack of time, or that employees do not see how others can value their knowledge. Also, if management «require» it and employees are not motivated, it is easy to «fake» reporting of knowledge. A culture for sharing knowledge has to be rooted in all employees. We also find support for this in studies by Orlikowski, Pan and Scarbrough.

- *Stable focus on knowledge management.* Key champions who leave a company, or frequent changes in organisation cause initiatives to be abandoned before they start working. Knowledge management requires stable focus over time. The studies of Davenport, Long and Beers confirm this view of knowledge management.

- *Incremental development.* Show benefits during development. There is no such thing as the «perfect» knowledge management tool. Tools need to be updated as work practise and the organisation's focus change. Incremental deliveries show that the organisation values knowledge management over time, and allows employees to participate in the development process. The other studies we have found do not mention this criteria, which is probably more visible in companies that develop software, as they are more interested and skilled to participate in the development process.

- *Good coupling to business goals.* The company who invests in knowledge management tools must really need them. There has been a lot of hype around the term knowledge management, and many companies have tried to apply it without having a clear idea of which needs they have. If the direct benefit is not obvious, it is harder to motivate employees to contribute, and also harder to get management support over time. This view is also confirmed from other studies.

Other candidate factors from other studies are: Having multiple channels for knowledge transfer, and a good technical and organisational infrastructure. We have not included multiple channels for knowledge transfer because we see it as a part of a good technical infrastructure. Good technical infrastructure is also left out in our study, because the technical dimension is emphasized enough in the companies we have experience with.

Also, other studies focus more on the importance of involving employees in knowledge management programs, but we see this as a part of getting a culture for knowledge management.

For the field of learning software organisations, we have found that organisational factors are very important in getting knowledge management tools in use. This is a point that should be taken into account when new prototypes for knowledge management tools are developed.

## 7.3   Conclusions from the Main Study

Our research question in the main study was:

*R2)      How do different groups of users in medium-sized consultancy organisations use Intranet-based knowledge management tools to transfer knowledge between software development projects?*

We found a variety of specialised tools in the two companies. We found five knowledge repository tools at Alpha and two at Beta. Some contained knowledge that was not structured, like the Well of Experience, and some contained packaged knowledge like the Project Guide (both of these tools were from Alpha). We found three knowledge cartography tools at Alpha, and one such tool at Beta. From the interviews and the usage logs, we see that the use of these tools varied. From this we conclude that:

- We found many different knowledge management tools in medium-sized software companies, and the tools were used to varying degrees.

If we go on to look tool usage, it seems that the repositories that present more «packaged» knowledge are used less than the tools with unstructured knowledge. If we take into account the different groups of employees, it also seems that project managers prefer tools with more abstracted knowledge, and that the developers prefer tools with more specific knowledge. Also, many tools that existed in the two companies were not mentioned by the people we interviewed, although they seem to be in stable use from the usage logs. This might be because employees do not consider some of the tools as knowledge management tools. Further, usage of tools also varied between people in the same group. Some developers preferred oral communication to written, and would then make more use of the personalization tools. Others preferred written communication, and some of these preferred to have it on paper while others preferred to have it electronically. Others again were sceptical to the use of tools in general, because it was hard to find relevant information. In all, we can conclude that:

- The use of knowledge management tools varies both between developers, project managers and managers, and after the personal preferences employees have.

From the examination of the Knowledge Repository and Library tools, we found six types of usage:

–Getting tips and advice in project start-up and execution (Project guide)
– Solve a specific technical problem (Well of Experience)
– Avoid rework in having to explain the same solution to different people (Well of Experience)
– Improve individual work situation by adjusting technical tools (Well of Experience)
– Getting an overview of problem areas (Well of Experience)
– Finding who has a specific competence in the company (Well of Experience).

We see the four first of these usage types as a codification strategy, and the two latter as a personalization strategy.

If we look at the Knowledge Cartography tools, we found the following types of usage of the Skills Manager:

– Searching for competence to solve problems
– Resource allocation
– Finding projects, and external marketing
– Competence development

All theses type of usage fit in a personalization strategy. From this we can conclude that:

- Knowledge management tools are used for a variety of purposes. The practitioners in companies will adapt and use tools to suit their normal work situations.
- Knowledge Repositories can function as a personalization strategy as well as a codification strategy.

For companies that want to develop knowledge management tools, this shows that different groups of users in software companies, such as developers, project mangers, and management benefit from different types of tools. Developers require more detailed knowledge, while the other groups seem to benefit more of abstract knowledge in their tool use.

## 7.4    Implications of our Findings

We think the conclusions of this thesis can have implications for the theory of learning software organisations, as well as for practical work with knowledge management initiatives in companies.

The theory of learning software organisations should acknowledge organisational issues in deploying knowledge management initiatives in companies, and also acknowledge that employees have different working styles that need different types of tool support. We argue that the field of learning software organisations should focus more on tailoring knowledge management tools to users than pursuing technological advancements, like working with active knowledge management tools. There is a large potential in getting existing tools into wider and more effective use.

Companies can benefit from simple knowledge management tools with low development cost. Examples of tools that are simple to develop but give high visibility on knowledge management in an organisation are the Skills Management tool at Alpha and the People tool at Beta. Companies should promote knowledge sharing through flexible tools to suit different working styles. They should also focus on sharing both specific and abstract knowledge so that knowledge relevant for all groups can be shared, and not only focus on knowledge that is important to management.

## 7.5    Evaluation

Although we discussed research methods both in our chapter on Software Development, and Research Goals, Method and Design, we think we should mention some limitations of the studies we have conducted.

In the literature study, we found eight case studies of knowledge management initiatives. We used a set of bibliography databases as well as manual search in proceedings to find these, but we have no guarantee that we have found all relevant case studies. Also, we might expect that there are unsuccessful attempts to do the same as we found in the case studies, that might not have been published because companies would not like to admit failures. It could also be that this kind of results are seen as less interesting by researchers. It is therefore not possible to claim that the cases we have found are «representative» for software engineering companies in general.

The prestudy was conducted as action research in a larger Norwegian research project. We gathered data by observing company representatives in meetings, and participated in improvement projects. In this kind of studies it is a danger of researcher bias, as there are not a variety of data sources, and observations might be interpreted differently by different people.

Finally, in the main study, we had data of higher quality from different sources, but from a limited number of cases. We can therefore not claim that our findings can be generalised directly. However, as with all case studies, the aim is to construct general theory.

## 7.6   Further Work

We now present possible further work in this area, that could contribute to a further understanding of how knowledge management tools function in medium-sized companies that develop software:

- When knowledge is codified and represented in a repository, it always has to be interpreted by a user when the knowledge is applied later. How the knowledge is interpreted will affect how it is used. It would be interesting to study such interpretation processes in companies that develop software to see what factors affect how software engineering knowledge is interpreted. In doing so, it would be more easy to predict the effect of a repository.

- It would also be interesting to study how the knowledge in the repositories and libraries has been codified. Have companies developed methods for externalisation, such as the Postmortem review technique presented in Chapter 3, or is this left to the individual contributor? And what different practises exist, if they are determined by individuals? This could help to get a better understanding of how to increase the quality of codified knowledge in repositories.

- In the discussion chapter, we characterized the knowledge management tool use in Alpha and Beta as support for single loop learning. It would be interesting to see whether we can find any support for more thorough learning processes, what we have referred to as double loop learning. If companies are changing their work processes in a more radical way because they focus on knowledge management - it would be a very good argument to further initiatives in this field.

- A final interesting topic would be to do a quantitative survey of the different purposes for, and frequency of use of knowledge management tools in software engineering companies. This could show whether the use of knowledge management tools we found at Alpha and Beta is similar or different to what we find in other companies. This could help to find more precisely in what domains our conclusions are applicable.

# Appendix A  Interview Guides

## A.1    Questions for developers:

*Background*

1. What kind of work do you do in the company?

2. How long have you been employed?

3. What kind of knowledge is important in your job?

4. How do you take care of this knowledge?

5. Can you give a description of the project you are working on now?

*Knowledge management tools*

1. How do you assess the tools for knowledge management that are available?

2. What knowledge have you had benefit of from these tools?

*Processes around the tools*

1. When do you use the tools?

2. How were the tools for knowledge management introduced to you?

### Attitudes to knowledge management

1. What relationship has the management to knowledge management?

2. What benefit have you had of knowledge from other projects through the tools?

3. How have you found this knowledge?

4. How often do you use the tools?

5. Have you contributed to the contents in the tools? With what?

6. How did you transfer this knowledge?

7. Do you have the impression that your contribution has been used by others?

8. Do you have time enough to use such systems?

9. What type of knowledge from the projects you are on now would you like to transfer to new projects?

10. What type of knowledge has been suitable for reuse from the projects you have been working on earlier?

### The quality of the knowledge in the tools

1. Do you have the impression that more knowledge is explicitly available now then before the tools were introduced?

2. What quality has the knowledge that is in the tools?

*In general*

1. How do you define «knowledge»?

2. How do you define «experience»?

3. How do you define «information»?

4. How do you define «knowledge management»?

## A.2   Questions for process owner for knowledge management:

*Background*

1. What kind of work do you do in the company?

2. How long have you been employed?

3. What kind of knowledge is important in your job?

4. How do you take care of this knowledge?

5. Can you give a description of the project you are working on now?

*Knowledge management tools*

1. How do you assess the tools for knowledge management that are available?

2. What knowledge have you had benefit of from these tools?

### Processes around the tools

1. When do you use the tools?

2. How were the tools for knowledge management introduced to you?

### Attitudes to knowledge management

1. What relationship has the management to knowledge management?

2. What benefit have you had of knowledge from other projects through the tools?

3. How have you found this knowledge?

4. How often do you use the tools?

5. Have you contributed to the contents in the tools? With what?

6. How did you transfer this knowledge?

7. Do you have the impression that your contribution has been used by others?

8. Do you have time enough to use such systems?

9. What type of knowledge from the projects you are on now would you like to transfer to new projects?

10. What type of knowledge has been suitable for reuse from the projects you have been working on earlier?

### The quality of the knowledge in the tools

1. Do you have the impression that more knowledge is explicitly available now then before the tools were introduced?

2. What quality has the knowledge that is in the tools?

### In general

1. How do you define «knowledge»?

2. How do you define «experience»?

3. How do you define «information»?

4. How do you define «knowledge management»?

# A.3 Questions for management:

### Background

1. What kind of work do you do in the company?

2. How long have you been employed?

### Knowledge management tools

1. How do you assess the tools for knowledge management that are available?

### Processes around the tools

1. How were the tools for knowledge management introduced?

### Attitudes to knowledge management

1. How important do you think knowledge management is?

2. What type of knowledge do you think it is most important to preserve?

3. What strategy does the company have for knowledege management?

4. What benefit does the company have of the knowledge management tools today?

### The quality of the knowledge in the tools

1. Do you have the impression that more knowledge is explicitly available now then before the tools were introduced?

2. What quality has the knowledge that is in the tools?

### In general

1. How do you define «knowledge»?

2. How do you define «experience»?

3. How do you define «information»?

4. How do you define «knowledge management»?

## A.4   Questions for knowledge sharers of the month:

### Background

1. What kind of work do you do in the company?

2. How long have you been employed?

3. What kind of knowledge is important in your job?

4. How do you take care of this knowledge?

5. Can you give a description of the project you are working on now?

## *Knowledge management tools*

1. How do you assess the tools for knowledge management that are available?

2. What knowledge have you had benefit of from these tools?

## *Processes around the tools*

1. When do you use the tools?

2. How were the tools for knowledge management introduced to you?

## *Attitudes to knowledge management*

1. What relationship has the management to knowledge management?

2. What benefit have you had of knowledge from other projects through the tools?

3. How have you found this knowledge?

4. How often do you use the tools?

5. Have you contributed to the contents in the tools? With what?

6. How did you transfer this knowledge?

7. Do you have the impression that your contribution has been used by others?

8. Do you have time enough to use such systems?

9. What type of knowledge from the projects you are on now would you like to transfer to new projects?

10. What type of knowledge has been suitable for reuse from the projects you have been working on earlier?

### The quality of the knowledge in the tools

1. Do you have the impression that more knowledge is explicitly available now then before the tools were introduced?

2. What quality has the knowledge that is in the tools?

### In general

1. How do you define «knowledge»?

2. How do you define «experience»?

3. How do you define «information»?

4. How do you define «knowledge management»?

# Appendix B: Processed Usage Logs

| Week | Day | Date | Main Page | Skills | | Competence | Project | Knowledge | News |
| | | | | Manager | Processes | Blocks | Guide | Market | |
| --- | --- | --- | --- | --- | --- | --- | --- | --- | --- |
| | Mon | 3-apr | 507 | 7 | 46 | 16 | 18 | 8 | 7 |
| | Tue | 4-apr | 455 | 5 | 72 | 30 | 8 | 7 | 1 |
| | Wed | 5-apr | 430 | 5 | 48 | 49 | 3 | 11 | 52 |
| | Thu | 6-apr | 359 | 1 | 30 | 29 | 8 | 3 | 109 |
| | Fri | 7-apr | 85 | 1 | 12 | 0 | 0 | 2 | 122 |
| | Sat | 8-apr | 13 | 0 | 0 | 0 | 0 | 0 | 51 |
| 14 | Sun | 9-apr | 15 | 0 | 2 | 0 | 0 | 1 | 16 |
| | Mon | 10-apr | 496 | 11 | 62 | 41 | 17 | 13 | 2 |
| | Tue | 11-apr | 439 | 7 | 52 | 36 | 11 | 15 | 1 |
| | Wed | 12-apr | 420 | 12 | 44 | 32 | 17 | 12 | 119 |
| | Thu | 13-apr | 389 | 8 | 39 | 31 | 7 | 8 | 28 |
| | Fri | 14-apr | 407 | 7 | 42 | 20 | 5 | 16 | 13 |
| | Sat | 15-apr | 143 | 0 | 8 | 0 | 0 | 0 | 50 |
| 15 | Sun | 16-apr | 138 | 0 | 2 | 0 | 0 | 0 | 57 |
| | Mon | 17-apr | 314 | 1 | 29 | 17 | 6 | 1 | 1 |
| | Tue | 18-apr | 274 | 6 | 24 | 21 | 4 | 7 | 0 |
| | Wed | 19-apr | 128 | 5 | 6 | 12 | 0 | 3 | 13 |
| | Thu | 20-apr | 24 | 0 | 0 | 0 | 0 | 0 | 8 |
| | Fri | 21-apr | 14 | 0 | 0 | 0 | 0 | 0 | 1 |
| | Sat | 22-apr | 7 | 0 | 0 | 0 | 0 | 0 | 0 |
| 16 | Sun | 23-apr | 6 | 0 | 0 | 0 | 0 | 0 | 0 |
| | Mon | 24-apr | 0 | 0 | 0 | 0 | 0 | 0 | 0 |
| | Tue | 25-apr | 0 | 0 | 0 | 0 | 0 | 0 | 0 |
| | Wed | 26-apr | 0 | 0 | 0 | 0 | 0 | 0 | 0 |
| | Thu | 27-apr | 183 | 3 | 16 | 17 | 3 | 4 | 0 |
| | Fri | 28-apr | 38 | 0 | 2 | 0 | 0 | 0 | 0 |
| | Sat | 29-apr | 0 | 0 | 0 | 0 | 0 | 0 | 24 |
| 17 | Sun | 30-apr | 0 | 0 | 0 | 0 | 0 | 0 | 7 |
| | Mon | 1-mai | 9 | 0 | 0 | 0 | 0 | 0 | 0 |
| | Tue | 2-mai | 330 | 5 | 40 | 12 | 5 | 7 | 0 |
| | Wed | 3-mai | 435 | 4 | 49 | 9 | 7 | 5 | 0 |
| | Thu | 4-mai | 230 | 0 | 21 | 13 | 0 | 4 | 50 |
| | Fri | 5-mai | 396 | 10 | 40 | 5 | 9 | 7 | 47 |
| | Sat | 6-mai | 6 | 0 | 0 | 0 | 0 | 0 | 22 |
| 18 | Sun | 7-mai | 34 | 0 | 2 | 2 | 0 | 3 | 18 |
| | Mon | 8-mai | 560 | 9 | 59 | 34 | 26 | 13 | 0 |
| | Tue | 9-mai | 471 | 6 | 58 | 6 | 3 | 4 | 0 |
| | Wed | 10-mai | 429 | 8 | 42 | 20 | 14 | 9 | 81 |
| | Thu | 11-mai | 453 | 21 | 58 | 48 | 29 | 12 | 158 |
| | Fri | 12-mai | 440 | 20 | 45 | 24 | 21 | 17 | 44 |
| | Sat | 13-mai | 40 | 1 | 3 | 1 | 1 | 1 | 44 |
| 19 | Sun | 14-mai | 45 | 1 | 10 | 3 | 3 | 2 | 121 |
| | Mon | 15-mai | 488 | 4 | 41 | 5 | 11 | 4 | 5 |
| | Tue | 16-mai | 473 | 14 | 48 | 22 | 11 | 13 | 3 |
| | Wed | 17-mai | 37 | 1 | 2 | 4 | 0 | 0 | 147 |
| | Thu | 18-mai | 545 | 14 | 46 | 20 | 10 | 20 | 129 |
| | Fri | 19-mai | 386 | 8 | 38 | 18 | 5 | 14 | 1 |
| | Sat | 20-mai | 42 | 0 | 2 | 0 | 0 | 0 | 186 |
| 20 | Sun | 21-mai | 29 | 0 | 1 | 2 | 0 | 0 | 66 |
| | Mon | 22-mai | 527 | 7 | 55 | 27 | 15 | 4 | 0 |
| | Tue | 23-mai | 469 | 8 | 37 | 17 | 15 | 6 | 1 |
| | Wed | 24-mai | 412 | 7 | 30 | 5 | 8 | 5 | 141 |
| | Thu | 25-mai | 438 | 8 | 40 | 10 | 12 | 4 | 117 |
| | Fri | 26-mai | 414 | 4 | 31 | 18 | 14 | 9 | 54 |
| | Sat | 27-mai | 42 | 0 | 5 | 2 | 0 | 0 | 67 |
| 21 | Sun | 28-mai | 69 | 0 | 7 | 3 | 0 | 1 | 44 |
| | Mon | 29-mai | 494 | 4 | 53 | 15 | 21 | 16 | 5 |
| | Tue | 30-mai | 423 | 8 | 39 | 16 | 16 | 8 | 10 |
| | Wed | 31-mai | 491 | 11 | 42 | 13 | 9 | 12 | 137 |
| | Thu | 1-jun | 74 | 0 | 6 | 1 | 2 | 0 | 133 |
| | Fri | 2-jun | 413 | 2 | 42 | 15 | 23 | 7 | 178 |
| | Sat | 3-jun | 33 | 1 | 3 | 0 | 0 | 0 | 18 |
| 22 | Sun | 4-jun | 55 | 4 | 3 | 2 | 0 | 0 | 140 |
| | Mon | 5-jun | 455 | 7 | 56 | 21 | 34 | 14 | 1 |
| | Tue | 6-jun | 482 | 21 | 31 | 19 | 28 | 16 | 6 |
| | Wed | 7-jun | 410 | 10 | 47 | 14 | 8 | 8 | 107 |
| | Thu | 8-jun | 426 | 9 | 42 | 42 | 19 | 12 | 67 |
| | Fri | 9-jun | 472 | 5 | 34 | 24 | 7 | 14 | 71 |
| | Sat | 10-jun | 33 | 0 | 1 | 3 | 0 | 0 | 110 |
| 23 | Sun | 11-jun | 28 | 0 | 2 | 0 | 0 | 0 | 103 |
| | Mon | 12-jun | 59 | 2 | 4 | 1 | 2 | 0 | 0 |
| | Tue | 13-jun | 488 | 6 | 52 | 19 | 10 | 15 | 4 |
| | Wed | 14-jun | 574 | 6 | 46 | 17 | 15 | 7 | 8 |
| | Thu | 15-jun | 520 | 6 | 47 | 12 | 11 | 6 | 119 |

| | | | | | | | | |
|---|---|---|---|---|---|---|---|---|
| | Fri | 16-jun | 486 | 10 | 31 | 17 | 13 | 14 | 129 |
| | Sat | 17-jun | 27 | 0 | 1 | 0 | 0 | 0 | 87 |
| 24 | Sun | 18-jun | 68 | 0 | 3 | 0 | 0 | 1 | 49 |
| | Mon | 19-jun | 570 | 28 | 63 | 29 | 12 | 8 | 3 |
| | Tue | 20-jun | 496 | 10 | 69 | 16 | 21 | 9 | 7 |
| | Wed | 21-jun | 487 | 93 | 67 | 19 | 25 | 3 | 61 |
| | Thu | 22-jun | 517 | 39 | 65 | 24 | 26 | 11 | 85 |
| | Fri | 23-jun | 467 | 25 | 52 | 8 | 13 | 7 | 129 |
| | Sat | 24-jun | 17 | 1 | 0 | 0 | 0 | 0 | 118 |
| 25 | Sun | 25-jun | 17 | 0 | 0 | 0 | 0 | 0 | 74 |
| | Mon | 26-jun | 493 | 20 | 41 | 27 | 17 | 6 | 1 |
| | Tue | 27-jun | 439 | 14 | 42 | 13 | 25 | 14 | 0 |
| | Wed | 28-jun | 455 | 4 | 27 | 11 | 11 | 9 | 78 |
| | Thu | 29-jun | 467 | 10 | 27 | 24 | 11 | 1 | 114 |
| | Fri | 30-jun | 424 | 8 | 25 | 19 | 4 | 6 | 78 |
| | Sat | 1-jul | 37 | 0 | 1 | 0 | 0 | 0 | 157 |
| 26 | Sun | 2-jul | 45 | 0 | 0 | 0 | 1 | 1 | 110 |
| | Mon | 3-jul | 453 | 14 | 35 | 20 | 14 | 5 | 6 |
| | Tue | 4-jul | 415 | 10 | 36 | 15 | 11 | 8 | 3 |
| | Wed | 5-jul | 372 | 13 | 34 | 31 | 5 | 8 | 88 |
| | Thu | 6-jul | 384 | 7 | 24 | 26 | 5 | 8 | 129 |
| | Fri | 7-jul | 286 | 3 | 13 | 6 | 8 | 1 | 167 |
| | Sat | 8-jul | 23 | 0 | 1 | 0 | 0 | 0 | 121 |
| 27 | Sun | 9-jul | 40 | 2 | 2 | 1 | 0 | 0 | 85 |
| | Mon | 10-jul | 359 | 7 | 25 | 9 | 13 | 12 | 3 |
| | Tue | 11-jul | 376 | 7 | 23 | 3 | 8 | 3 | 15 |
| | Wed | 12-jul | 398 | 3 | 26 | 4 | 8 | 11 | 89 |

| | | | | | | | | |
|---|---|---|---|---|---|---|---|---|
| | Thu | 13-jul | 377 | 13 | 27 | 1 | 7 | 1 | 153 |
| | Fri | 14-jul | 295 | 11 | 27 | 9 | 0 | 6 | 115 |
| | Sat | 15-jul | 10 | 0 | 0 | 0 | 0 | 0 | 74 |
| 28 | Sun | 16-jul | 17 | 0 | 2 | 3 | 0 | 0 | 53 |
| | Mon | 17-jul | 327 | 10 | 23 | 15 | 4 | 6 | 1 |
| | Tue | 18-jul | 354 | 7 | 27 | 5 | 5 | 3 | 7 |
| | Wed | 19-jul | 368 | 2 | 17 | 0 | 3 | 4 | 40 |
| | Thu | 20-jul | 339 | 1 | 6 | 2 | 5 | 5 | 58 |
| | Fri | 21-jul | 223 | 2 | 9 | 5 | 6 | 7 | 52 |
| | Sat | 22-jul | 25 | 0 | 0 | 0 | 0 | 0 | 36 |
| 29 | Sun | 23-jul | 18 | 0 | 4 | 0 | 0 | 1 | 17 |
| | Mon | 24-jul | 351 | 2 | 16 | 4 | 14 | 5 | 1 |
| | Tue | 25-jul | 352 | 0 | 23 | 4 | 10 | 8 | 6 |
| | Wed | 26-jul | 372 | 10 | 14 | 0 | 7 | 7 | 67 |
| | Thu | 27-jul | 331 | 6 | 23 | 0 | 3 | 7 | 113 |
| | Fri | 28-jul | 297 | 3 | 24 | 4 | 2 | 8 | 48 |
| | Sat | 29-jul | 17 | 0 | 0 | 0 | 0 | 0 | 34 |
| 30 | Sun | 30-jul | 21 | 0 | 1 | 0 | 0 | 0 | 31 |
| | Mon | 31-jul | 0 | 0 | 0 | 0 | 0 | 0 | 0 |
| | Tue | 1-aug | 0 | 0 | 0 | 0 | 0 | 0 | 0 |
| | Wed | 2-aug | 0 | 0 | 0 | 0 | 0 | 0 | 0 |
| | Thu | 3-aug | 0 | 0 | 0 | 0 | 0 | 0 | 0 |
| | Fri | 4-aug | 0 | 0 | 0 | 0 | 0 | 0 | 0 |
| | Sat | 5-aug | 0 | 0 | 0 | 0 | 0 | 0 | 0 |
| 31 | Sun | 6-aug | 0 | 0 | 0 | 0 | 0 | 0 | 0 |
| | Mon | 7-aug | 0 | 0 | 0 | 0 | 0 | 0 | 0 |
| | Tue | 8-aug | 0 | 0 | 0 | 0 | 0 | 0 | 0 |
| | Wed | 9-aug | 0 | 0 | 0 | 0 | 0 | 0 | 0 |
| | Thu | 10-aug | 0 | 0 | 0 | 0 | 0 | 0 | 0 |
| | Fri | 11-aug | 0 | 0 | 0 | 0 | 0 | 0 | 0 |
| | Sat | 12-aug | 0 | 0 | 0 | 0 | 0 | 0 | 0 |
| 32 | Sun | 13-aug | 0 | 0 | 0 | 0 | 0 | 0 | 0 |
| | Mon | 14-aug | 0 | 0 | 0 | 0 | 0 | 0 | 0 |
| | Tue | 15-aug | 0 | 0 | 0 | 0 | 0 | 0 | 0 |
| | Wed | 16-aug | 0 | 0 | 0 | 0 | 0 | 0 | 0 |
| | Thu | 17-aug | 0 | 0 | 0 | 0 | 0 | 0 | 0 |
| | Fri | 18-aug | 0 | 0 | 0 | 0 | 0 | 0 | 0 |
| | Sat | 19-aug | 0 | 0 | 0 | 0 | 0 | 0 | 0 |
| 33 | Sun | 20-aug | 0 | 0 | 0 | 0 | 0 | 0 | 0 |
| | Mon | 21-aug | 0 | 0 | 0 | 0 | 0 | 0 | 0 |
| | Tue | 22-aug | 0 | 0 | 0 | 0 | 0 | 0 | 0 |
| | Wed | 23-aug | 0 | 0 | 0 | 0 | 0 | 0 | 0 |
| | Thu | 24-aug | 0 | 0 | 0 | 0 | 0 | 0 | 0 |
| | Fri | 25-aug | 0 | 0 | 0 | 0 | 0 | 0 | 0 |
| | Sat | 26-aug | 0 | 0 | 0 | 0 | 0 | 0 | 0 |
| 34 | Sun | 27-aug | 0 | 0 | 0 | 0 | 0 | 0 | 0 |
| | Mon | 28-aug | 0 | 0 | 0 | 0 | 0 | 0 | 0 |
| | Tue | 29-aug | 0 | 0 | 0 | 0 | 0 | 0 | 0 |
| | Wed | 30-aug | 0 | 0 | 0 | 0 | 0 | 0 | 0 |
| | Thu | 31-aug | 0 | 0 | 0 | 0 | 0 | 0 | 0 |
| | Fri | 1-sep | 0 | 0 | 0 | 0 | 0 | 0 | 0 |
| | Sat | 2-sep | 0 | 0 | 0 | 0 | 0 | 0 | 0 |
| 35 | Sun | 3-sep | 0 | 0 | 0 | 0 | 0 | 0 | 0 |
| | Mon | 4-sep | 654 | 11 | 80 | 42 | 16 | 9 | 1 |
|    | Tue | 5-sep  | 574 | 35 | 58 | 18 | 13 | 13 | 1   |
|    | Wed | 6-sep  | 600 | 15 | 44 | 5  | 28 | 8  | 138 |
|    | Thu | 7-sep  | 437 | 6  | 41 | 17 | 15 | 5  | 136 |
|    | Fri | 8-sep  | 475 | 9  | 34 | 22 | 13 | 16 | 1   |
|    | Sat | 9-sep  | 25  | 0  | 1  | 0  | 0  | 1  | 22  |
| 36 | Sun | 10-sep | 44  | 0  | 2  | 0  | 0  | 0  | 126 |
|    | Mon | 11-sep | 599 | 5  | 40 | 25 | 18 | 14 | 93  |
|    | Tue | 12-sep | 537 | 10 | 36 | 9  | 2  | 17 | 104 |
|    | Wed | 13-sep | 505 | 13 | 35 | 9  | 19 | 10 | 77  |
|    | Thu | 14-sep | 506 | 8  | 39 | 11 | 9  | 3  | 97  |
|    | Fri | 15-sep | 502 | 16 | 34 | 11 | 3  | 7  | 6   |
|    | Sat | 16-sep | 26  | 0  | 1  | 0  | 0  | 0  | 10  |
| 37 | Sun | 17-sep | 43  | 0  | 1  | 0  | 0  | 0  | 175 |
|    | Mon | 18-sep | 560 | 6  | 33 | 12 | 9  | 14 | 160 |
|    | Tue | 19-sep | 489 | 11 | 36 | 9  | 11 | 16 | 102 |
|    | Wed | 20-sep | 492 | 9  | 41 | 26 | 11 | 4  | 147 |
|    | Thu | 21-sep | 476 | 3  | 25 | 28 | 3  | 1  | 115 |
|    | Fri | 22-sep | 444 | 2  | 19 | 13 | 7  | 10 | 1   |
|    | Sat | 23-sep | 38  | 0  | 1  | 0  | 0  | 0  | 8   |
| 38 | Sun | 24-sep | 57  | 1  | 2  | 3  | 0  | 0  | 100 |
|    | Mon | 25-sep | 561 | 11 | 39 | 34 | 12 | 7  | 106 |
|    | Tue | 26-sep | 450 | 9  | 30 | 10 | 5  | 5  | 74  |
|    | Wed | 27-sep | 410 | 2  | 23 | 15 | 2  | 6  | 127 |
|    | Thu | 28-sep | 481 | 13 | 25 | 17 | 14 | 13 | 90  |
|    | Fri | 29-sep | 429 | 4  | 29 | 7  | 8  | 9  | 6   |
|    | Sat | 30-sep | 32  | 0  | 0  | 0  | 0  | 1  | 10  |
| 39 | Sun | 1-okt  | 53  | 0  | 7  | 0  | 0  | 0  | 114 |

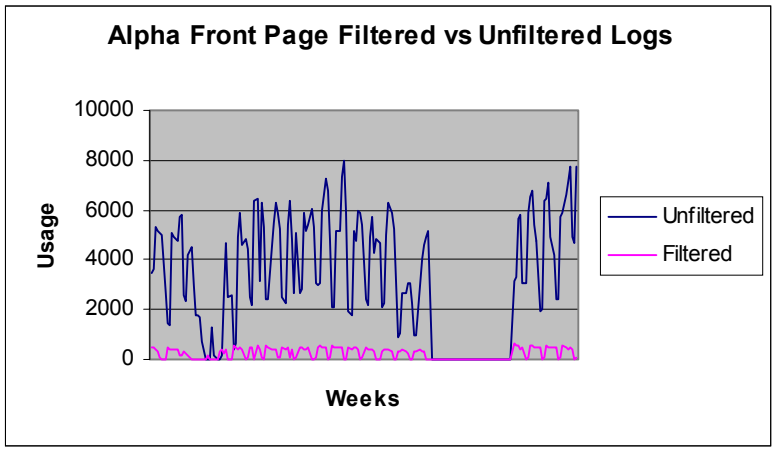

# Index

## R

Research Methods 29
reuse 25
Root Cause Analysis 51

## S

sd&m 63
silver bullet 17
single-loop learning 38
situated learning 39
skills management 105
socialization 42
Software development 13
software engineering 13
software process 22
software process improvement 18
SPIQ 9
Standish report 15
Structured techniques 18
subjectivist 8

## T

tacit knowledge 35
Telenor Telecom Software 59
The flow of knowledge 54
Total Quality Management 21
transformation 40

## W

waterfall model 13

# Utgivelser i Perspektiv: 17 x 24

**Nr. 22 Torgeir Dingsøyr**

Knowledge management in Medium-Sized Software Consulting Companies.

An Investigation of Intranet-based Knowledge Management Tools for Knowledge Cartography and Knowledge Repositories for Learning Software Organisations